\documentclass[fleqn,usenatbib]{mnras}

\usepackage{newtxtext,newtxmath}

\usepackage[T1]{fontenc}
\usepackage{ae,aecompl}

\DeclareRobustCommand{\VAN}[3]{#2}
\let\VANthebibliography\thebibliography
\def\thebibliography{\DeclareRobustCommand{\VAN}[3]{##3}\VANthebibliography}

\usepackage{graphicx}	
\usepackage{amsmath}	
\usepackage[normalem]{ulem} 
\usepackage{hyperref}

\newcommand{\app}[1]{Appendix~\ref{sec:#1}}
\newcommand{\eq}[1]{Equation~\ref{eq:#1}}
\newcommand{\fig}[1]{Figure~\ref{fig:#1}}
\renewcommand{\sec}[1]{Section~\ref{sec:#1}}

\newcommand{\App}[1]{Appendix~\ref{sec:#1}}

\newcommand{\Fig}[1]{Figure~\ref{fig:#1}}
\newcommand{\Sec}[1]{Section~\ref{sec:#1}}
\newcommand{\Tab}[1]{Table~\ref{tab:#1}}

\newcommand{\ceagle}{\mbox{\sc{C-Eagle}}}
\newcommand{\eagle}{\mbox{\sc{Eagle}}}

\newcommand{\flares}{\mbox{\sc Flares}}

\newcommand{\jwst}{\mbox{\it JWST}}

\newcommand{\Msun}{\mbox{M$_\odot$}}

\newenvironment{shortitem}
{\begin{list}{$\bullet$}{\topsep=0pt\itemsep=0pt\parsep=0pt\parskip=0pt\leftmargin=12pt}}
{\end{list}}

\newcommand{\flaresone}{\mbox{\sc Flares-I}}
\defcitealias{Flares1}{\mbox{\sc Flares-I}}

\title[\flares\ X: Bias and survey variance]{First Light and Reionisation Epoch Simulations (\flares) X:\\ Environmental Galaxy Bias and Survey Variance at High Redshift}

\author[P.~A.~Thomas et al.]{Peter A.~Thomas$^{1},$\thanks{E-mail: p.a.thomas@sussex.ac.uk}
Christopher C. Lovell$^{2}$,
Maxwell G.~A.~Maltz$^{1}$,
\newauthor
Aswin P.~Vijayan$^{3}$,
Stephen M.~Wilkins$^{1}$,
Dimitrios Irodotou$^{4}$,
\newauthor
William J.~Roper$^{1}$,
Louise Seeyave$^{1}$
\\
$^{1}$Astronomy Centre, University of Sussex, Falmer, Brighton BN1 9QH, UK\\
$^{2}$Institute of Cosmology and Gravitation, University of Portsmouth, Burnaby Road, Portsmouth, PO1 3FX, UK\\
$^{3}$Cosmic Dawn Center (DAWN), DTU-Space, Technical University of Denmark, Elektrovej 327, DK-2800 Kgs. Lyngby, Denmark \\
$^{4}$Department of Physics, University of Helsinki, Gustaf Hällströmin katu 2, FI-00014, Helsinki, Finland
}

\date{Accepted XXX. Received YYY; in original form ZZZ}

\pubyear{2023}

\begin{document}
\label{firstpage}
\pagerange{\pageref{firstpage}--\pageref{lastpage}}
\maketitle

\begin{abstract}
  
Upcoming deep galaxy surveys with \textit{JWST} will probe galaxy evolution during the epoch of reionisation (EoR, $5\leq z\leq10$) over relatively compact areas (e.g.~$\sim$ 300\,arcmin$^2$ for the JADES GTO survey).
It is therefore imperative that we understand the degree of survey variance, to evaluate how representative the galaxy populations in these studies will be.
We use the First Light And Reionisation Epoch Simulations (\flares) to measure the galaxy bias of various tracers over an unprecedentedly large range in overdensity for a hydrodynamic simulation, and use these relations to assess the impact of bias and clustering on survey variance in the EoR. 
Star formation is highly biased relative to the underlying dark matter distribution, with the mean ratio of the stellar to dark matter density varying by a factor of 100 between regions of low and high matter overdensity (smoothed on a scale of 14\,$h^{-1}$cMpc). 
This is reflected in the galaxy distribution -- the most massive galaxies are found solely in regions of high overdensity.
As a consequence of the above, galaxies in the EoR are highly clustered, which can lead to large variance in survey number counts. 
For mean number counts $N\lesssim 100$ (1000), in a unit redshift slice of angular area 300\,arcmin$^2$ (1.4\,deg$^2$), the 2-sigma range in $N$ is roughly a factor of four (two).
We present relations between the expected variance and survey area for different survey geometries; these relations will be of use to observers wishing to understand the impact of survey variance on their results.
  
\end{abstract}

\begin{keywords}
galaxies: high-redshift -- galaxies: luminosity function, mass function 
\end{keywords}



\section{Introduction}
\label{sec:intro}

This paper investigates the clustering and bias of galaxies in the Epoch of Reionisation (EoR), $5\lesssim z\lesssim10$ using the First Light and Reionisation Epoch Simulations \citep[hereafter \citetalias{Flares1}]{Flares1}.  This can lead to variations in the  number counts of upcoming galaxy surveys in the EoR (95 percentile range) of factors of around 2$-$4.

Galaxies form within dark matter haloes, which themselves form at the peaks of the density field (smoothed on the halo mass scale) and which are overdense with respect to the background \citep{Zeldovich82,Kaiser84}.  In the early Universe especially, those peaks rely on contributions from a wide range of scales \citep[e.g.~][]{Bardeen86} and can therefore only be properly represented in a region of very large extent.  The non-linear relationship between galaxies and the underlying matter distribution is known as \emph{galaxy bias}, a term which is also used more generally to describe the relation between a range of different galaxy tracers and the underlying matter distribution \citep[see a review by][]{Desjacques18}. 

\textit{Survey variance}\footnote{We avoid the oft-used term \emph{cosmic variance} which more accurately describes the uncertainty from having a single observable universe.} describes the uncertainty in observed estimates of galaxy number densities that arises from spatial variation within different survey volumes: both clustering of dark matter and galaxy bias contribute to this effect, in addition to stochasticity in the galaxy formation process itself.  The choice of survey area and geometry is closely linked to the amplitude of these fluctuations and can give rise to significant variation in the measured number counts.  Any actual survey will also be subject to sample variance arising from Poisson counting statistics, and this is likely to dominate at small number counts.  We do not include that in our analysis as the survey and sample variances, as described above, can simply be added in quadrature to find the total variance in the observations.

A combination of dark matter only (DMO) simulations and analytic models are a computationally efficient means of assessing the variance over large volumes. 
These tend to connect haloes to galaxies given some mass -- luminosity relation, or using some abundance matching prescription \citep[e.g.~][]{Newman02,Somerville04,Stark07,Trenti08,Moster11}.  Ideally, however, one would use a more astrophysical, semi-analytic model \citep[SAM, for a comparative review see][]{Knebe18} for which simulations with sufficient resolution are limited in size.  The most well-known and well-used of these is the Millennium Simulation \citep{Springel05} which has a volume of just (500\,$h^{-1}$cMpc)$^3$, for which estimates of survey variance out to $z=5$ were undertaken by \citet{Kitzbichler07}.  Larger volumes are available at lower resolution \citep[see, e.g.,][]{Kim09,Angulo12,Maksimova21}, more suitable for use with (sub)halo abundance matching.

To more accurately model galaxies, hydrodynamic simulations are required and these have even more limited extent.  The first to be widely used were {\sc Illustris} and \eagle\ \citep[respectively]{Genel14, Eagle15}, both of order (70\,$h^{-1}$cMpc)$^3$, followed by {\sc Simba}\ \citep{Dave19} at (100\,$h^{-1}$cMpc)$^3$ and {\sc Illustris-TNG} \citep{Nelson17,Pillepich17} at (200\,$h^{-1}$cMpc)$^3$.  Perhaps the most ambitious in this respect is the large-scale simulation {\sc Bluetides} \citep{Feng15} which simulated (400\,$h^{-1}$cMpc)$^3$ -- still less than the Millennium Simulation -- but only down to $z\approx8$.

To overcome this limitation requires new approaches.  One such is zoom simulations which run hydrodynamics at high resolution in selected regions of very large, low resolution, DMO simulations \citep[e.g.~][which all concentrated on massive clusters]{Katz93,Bahe17,Barnes17}.  \flares\ built on this approach to simulate galaxy formation in a wide range of environments \citep[following the approach adopted in][]{Crain09} within a (2.2\,$h^{-1}$cGpc)$^3$ box.  It resimulates 40 regions with a wide range of overdensities, allowing us both to capture the very high overdensity environments within which the first galaxies will form, but also to investigate in detail the dependence of galaxy formation upon environment.


In recent years a number of multiwavelength surveys have measured the abundances and properties of galaxies at high redshift \citep[e.g.~][]{Gonzalez11,Duncan14,Song16,Stefanon17,Bhatawdekar19}.
Deep galaxy surveys using \jwst\ over the coming years will measure many of these functions to much greater depth, increasing the redshift and dynamic range probed, e.g.~: CEERS \citep{Ceers22}, COSMOS-Web \citep{CosmosWeb22}, GLASS-ERS \citep{Treu22}, JADES \citep{Jades20} and PRIMER \citet{Primer21}. 
These surveys will cover areas in the range $100-2000$\,arcmin$^2$ and one of the purposes of this paper is to estimate the effect of survey variance on the expected number counts.
This is particularly pertinent given the recent discovery of massive galaxy candidates at very high redshifts in relatively small early fields \citep[e.g.][]{Donnan22,Labbe22,Adams22,Harikane22,Rodighiero22,Naidu22}.

A number of studies have used analytic methods to estimate survey variance at these high redshift \citep{Somerville04, Trenti08, Trapp20,Trapp21,Einasto23}, and have shown that the normalisation and slope of measured luminosity functions can be significantly affected.
However, these studies use simplified models to map galaxies onto dark matter halos, which have not been tested in this regime.  Also, they are presented in a way that is hard to relate to the population of galaxies likely to be observed in deep surveys.  \citet{Trenti08} and \citet{Moster11} provide cosmic variance calculators (CVCs) that can be used to estimate the survey variance for given field sizes and depths.  They each reach similar conclusions to this paper, in that they show that the variance can be significant for small areas and number counts, but we find a larger magnitude for the effect than they do.  The former provide an online CVC and we contrast our results with theirs in \sec{variance} below.

One purpose of this paper is to investigate the relationship between galaxies and the underlying dark matter distribution in the EoR in a much more direct way than previous studies, using a hydrodynamic method \citep[\eagle,][]{Eagle15} that has been shown to reproduce the galaxy population extremely well in the current day Universe and which provides a good match to the observed luminosity functions of galaxies in the EoR \citep{Vijayan21}. 
We measure the galaxy bias of various components over an unprecendented range in overdensity for a hydrodynamic simulation, and provide new estimates of the effect of survey variance on high redshift galaxy number counts.

\Sec{method} briefly describes \flares, the method that we use to define large-scale overdensity, and to map stars and galaxies onto the dark matter distribution.  \Sec{bias} presents results for the biasing of the smooth stellar distribution and of galaxies relative to that of the dark matter.  \Sec{variance} then explores the clustering of those galaxies in areas typical of those of deep surveys.  Finally, \Sec{conc} summarises our conclusions.

\section{Method}
\label{sec:method}

\subsection{\flares}

The First Light And Reionisation Epoch Simulations \citep[\flares][]{Flares1,Vijayan21} are a series of 40 large zoom simulations selected at $z = 4.69$.
\flares\ uses the same hydrodynamics code, \textsc{Anarchy}, as the \eagle\ simulation, described in detail in \citep{Eagle15,Schaller15}.
It employs the AGNdT9 parameter configuration, which leads to a closer match with observational constraints on the hot gas properties in groups and clusters \citep{Barnes17} than does the standard configuration, although in \flares\ these changes should have little effect, since the number of such massive halos is very low at $z = 5$.

\flares\ uses an identical resolution to the fiducial \eagle\ simulation, with gas particle mass $m_\mathrm{g} = 1.8 \times 10^6 \, \mathrm{M_{\odot}}$, and a softening length of $2.66\,\mathrm{ckpc}$.  Resimulation regions are selected from the same (3.2\,cGpc)$^3$ dark matter-only parent simulation as that used in the \ceagle\ simulations \citep{Barnes17}.  The highest redshift snapshot available for this simulation is at $z = 4.69$, which was used to select spherical volumes that sample a range of overdensities.  The size of the resimulation regions (radius 14\,$h^{-1}\,$cMpc) was chosen such that density fluctuations averaged on that scale are linear: then the distortion in the shape of the Lagrangian volume during the simulation is relatively small, and the ordering of the density fluctuations is preserved.
Full details on the 40 selected regions and their overdensities are provided in \citetalias{Flares1}.

As shown in \citetalias{Flares1}, the galactic stellar mass functions from the \flares\ simulations agree with those from \eagle\ at $z=5-10$ in the mass range within which they overlap, but with those from \flares\ extending to higher masses that are not accessible within the limited \eagle\ volume.  As \eagle\ has been shown to agree well with observations of galaxies in the low-redshift universe \citep{Eagle15} that gives us confidence that our galaxies will also provide a reasonable match to the real galaxy population in the EoR.  This is reflected in the success of \flares\ in matching existing observations in this regime.  For example, the \flares\ galactic stellar mass function \citep[][Figure~8]{Flares1} is in good agreement with observational constraints at all redshifts up to $z=8$, beyond which it is slightly {\it lower} than the observations; on the other hand, the UV luminosity function is in excellent agreement with observations at all redshifts up to $z=9$  \citep[][Figure~7]{Vijayan21}, beyond which slightly {\it over-predicts} the number counts.  The complex astrophysics of star formation and feedback means that the physical nature of high redshift galaxies is quite different from those at low redshift, for example they are much more compact in size \citep{Roper22}, which renders suspect any extrapolation from (semi)-analytic models constrained by observations at lower redshift.

\subsection{Determination of overdensities}

It is useful to be able to relate the density of stars, galaxies, or other observable quantities, to the overdensity of matter smoothed on a scale for which fluctutations are still linear and hence deducable from the initial density field.  In \flares, we do this at a redshift of 4.69, shortly after the end of reionisation.\footnote{With that particular redshift being chosen because it was the highest snapshot available for the underlying dark matter simulation.}

The  \emph{parent simulation} has a volume of (3.2\,cGpc)$^3$.  We divide that up into 1200$^3$ \emph{grid cells} each of side 2.67\,cMpc.  We use nearest grid point assignment to associate simulation particles with grid cells.  We then determine the mean overdensity of those regions, smoothed using top hat filters of three radii: 10, 14 and 20\,$h^{-1}$cMpc: we will call these $\delta_{10}$, $\delta_{14}$ and $\delta_{20}$, where $1+\delta$ is the ratio of the density of matter within a smoothing sphere to the mean density of matter within the simulation.

\begin{figure*}
  \includegraphics[width=17.4cm]{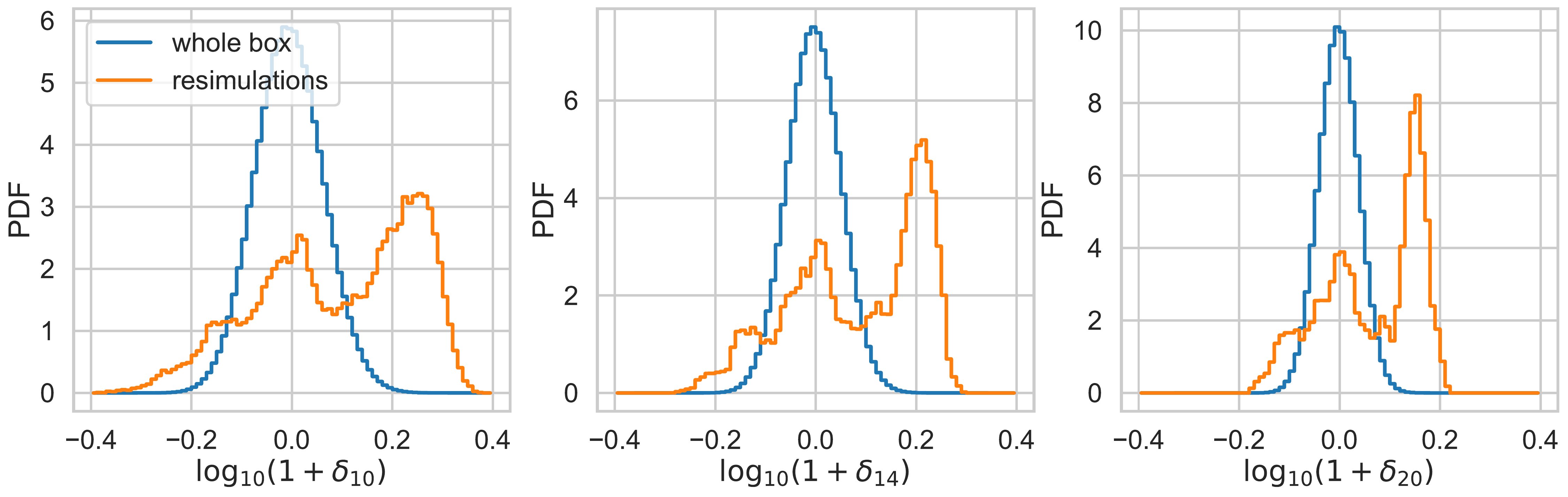}
  \caption{The PDF of overdensities smoothed within top-hat windows of different radii: 10, 14 and 20\,$h^{-1}$\,cMpc for the left, centre and right panels, respectively.  Blue is the entire simulation box; orange is the regions that we resimulate.}
  \label{fig:weights}
\end{figure*}

\Fig{weights} shows the probablity densty functions (PDFs) for these three different definitions of overdensity.  Blue shows the PDF for the parent simulation; orange shows the overdensities within the regions that we resimulate.  You can see that we have deliberately chosen to over-sample regions of high density in order to get a significant population of massive galaxies.

To determine the mean (i.e. universal average) of a given quantity, we need to know how to weight the contributions from individual grid cells.  We do this using the procedure described in Section~2.4 of \citet{Flares1}. Essentially, we count the number of grid cells in bins of overdensity, both in the resimulated regions and within the parent simulation as a whole. The ratio of the latter to the former then gives us the relative weighting that needs to be applied to each resimulated grid cell.

\begin{figure}
  \includegraphics[width=8.7cm]{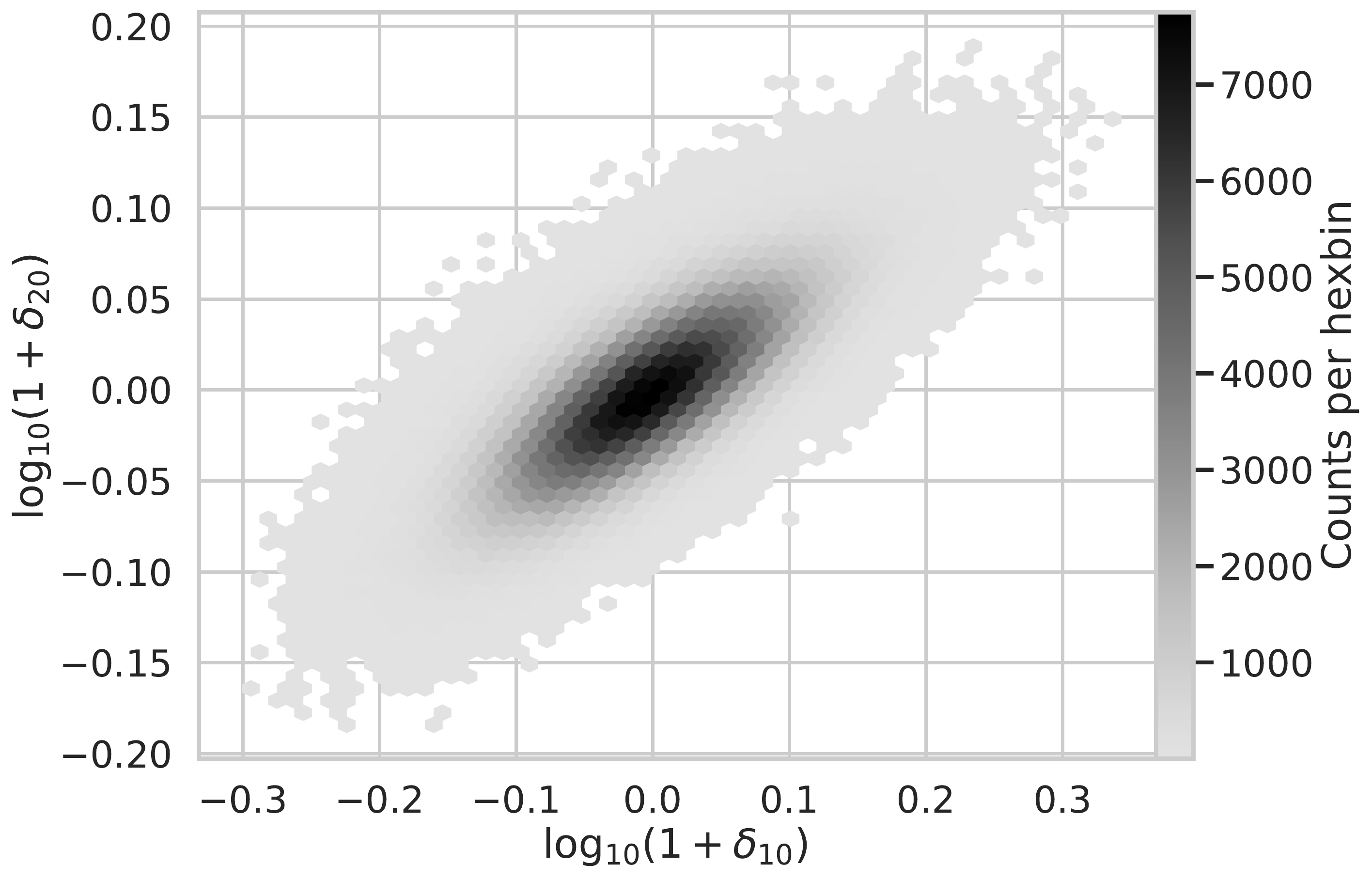}
  \caption{For a subsample of grid cell locations, the relationship between overdensity smoothed with top-hat filters of radii 10\,$h^{-1}$\,cMpc and 20\,$h^{-1}$\,cMpc.}
  \label{fig:delta}
\end{figure}

\Fig{delta} shows the relationship between $\delta_{10}$ and $\delta_{20}$.  While clearly there is a strong correlation between the two, there is also significant scatter.  We have found that all three smoothing radii give very similar results for the quantities that we investigate in this paper and so we stick with the original choice of $\delta_{14}$ used in \citetalias{Flares1} below.

\subsection{Mapping stars and galaxies to dark matter}
\label{sec:method:mock}

Although we resimulate only a small fraction of the parent volume, we sample a wide range of environments that span the whole range of overdensities.  We use this to populate the parent simulation with galaxies in order to create large mock surveys.  To do this, we tabulate galaxy properties\footnote{the galaxy stellar mass function (GSMF), or the star formation rate function, (SFRF).} within overdensity bins, and then use this as a lookup table to populate grid cells that we have not resimulated.

We map individual particles within the simulation (dark matter, gas, stars or black holes) to the grid cell that they occupy at $z=4.69$.  This mapping can then be recovered at higher redshifts using the particle IDs that are preserved during the simulation and when particles transform from gas into stars.\footnote{The exception is merging of black hole particles for which only the ID of the most massive progenitor is stored -- hence we trace the main branch.}

We have tried using each of $\delta_{10}$, $\delta_{14}$, $\delta_{20}$, the grid cell overdensity without smoothing ($\delta_\mathrm{grid}$), and the velocity divergence within a grid cell, both alone and in combination.  Although there is a slight reduction in residual scatter when combining two or more diagnostics, the gain is very slight, and we choose in this paper to stick to the single input of $\delta_{14}$ that was used in \flaresone.

\section{Bias}
\label{sec:bias}

Bias is a measure of how much a quantity is clustered relative to the overall mass density.  \Sec{bias:smooth} looks at bias in the distribution of stars and other smoothed quantities within grid cells, and \Sec{bias:galaxies} in that of the galactic population.  Unless otherwise stated, all results shown here are for a redshift $z=4.69$.

\subsection{Bias in the matter distribution}
\label{sec:bias:smooth}

We first look at the bias in the distribution of different types of matter within grid cells, compared to that of the dark matter.  We plot our results as a function of $\delta_{14}$ in order to investigate how the bias changes with overdensity.

\subsubsection{Dark matter}

\begin{figure}
  \includegraphics[width=8.7cm]{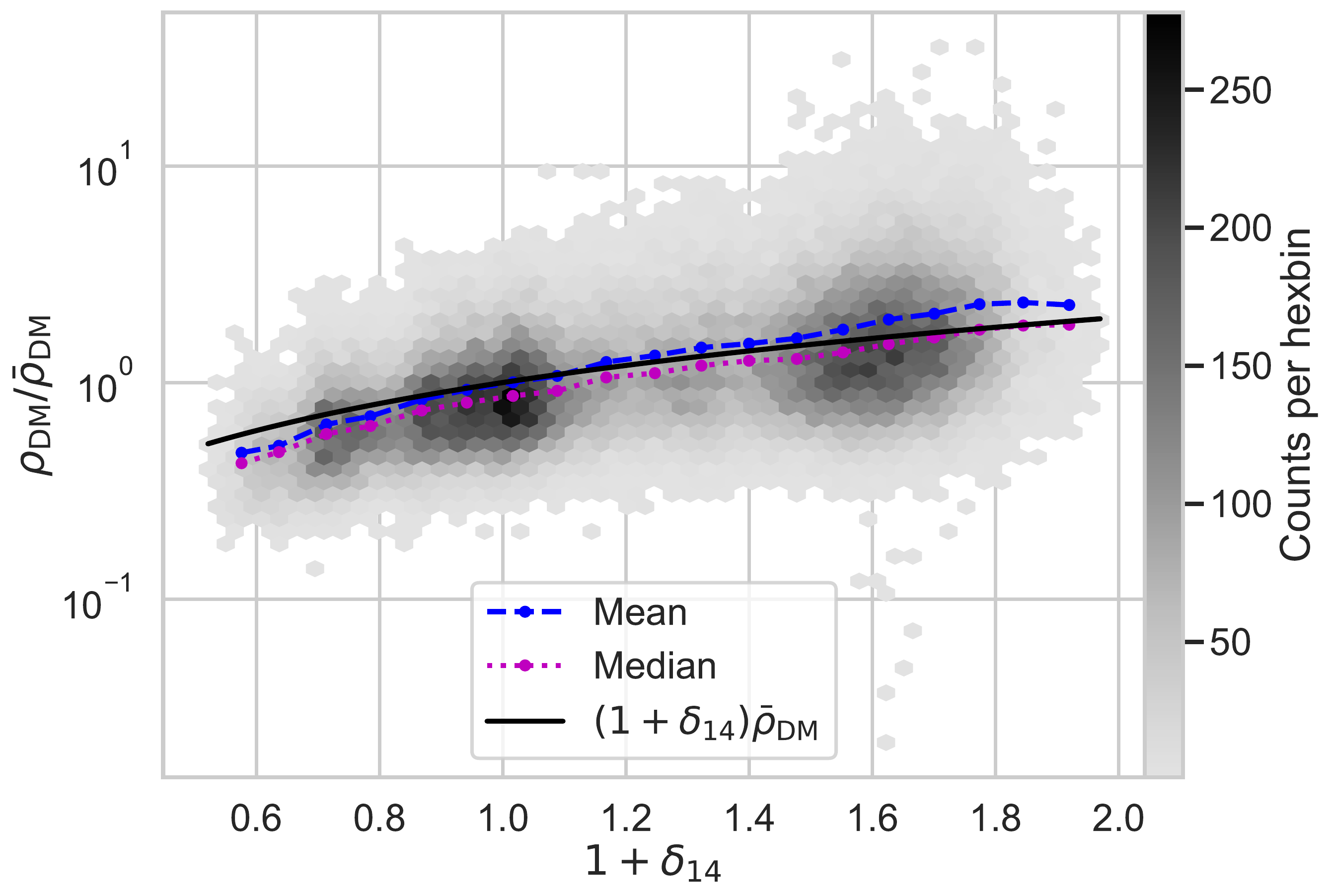}
  \caption{The dark matter density within grid cells plotted as a function of the mean matter density at that location, smoothed with a top-hat window of radius 14\,$h^{-1}$\,cMpc.  Only grid cells within the resimulated regions are plotted and used to calculate the mean and median within bins of $\delta_{14}$.}
  \label{fig:DM_od_raw}
\end{figure}

\Fig{DM_od_raw} shows the measured density of dark matter within each grid cell in resimulated regions, compared to the smoothed matter density, $1+\delta_{14}$ at that location.  The solid black line shows the 1-to-1 relation, i.e.~$y=1+\delta_{14}$; the blue dashed and magenta dotted lines the mean and median, respectively, averaged in bins of $\delta_{14}$.  The mean passes through the point (1,1), as is to be expected, but has a slope greater than that of the 1-1 relation: this is because the density averaged within a sphere of radius 14\,$h^{-1}$\,cMpc will tend to be closer to 1 than when averaged within grid cells.

Note that the horizontal variation in the density of points simply reflects the overdensity of the regions that we have chosen to resimulate: the excess near $\delta_{14}=0$ comes from the fact that this is the peak of overall density field; that at high values of $\delta_{14}$ because we have chosen to simulate a large number of regions of high overdensity.  The vertical variation in the density of points does, however, show the true variation of $\rho_\mathrm{DM}/\bar\rho_\mathrm{DM}$ at a given overdensity.


The scatter in $\rho_\mathrm{DM}/\bar\rho_\mathrm{DM}$, that is the dark matter density within grid cells measured in units of the universal mean, is very large and roughly symmetrical in the log, i.e.~skewed to high values in real space.
This skewness is caused by the non-linear growth of density fluctuations within grid cells whose overdensity approaches or exceeds unity.  This leads to a huge bias in star formation, as we will see in the next section.

\subsubsection{Stellar to dark matter mass ratio}

\begin{figure}
  \includegraphics[width=8.7cm]{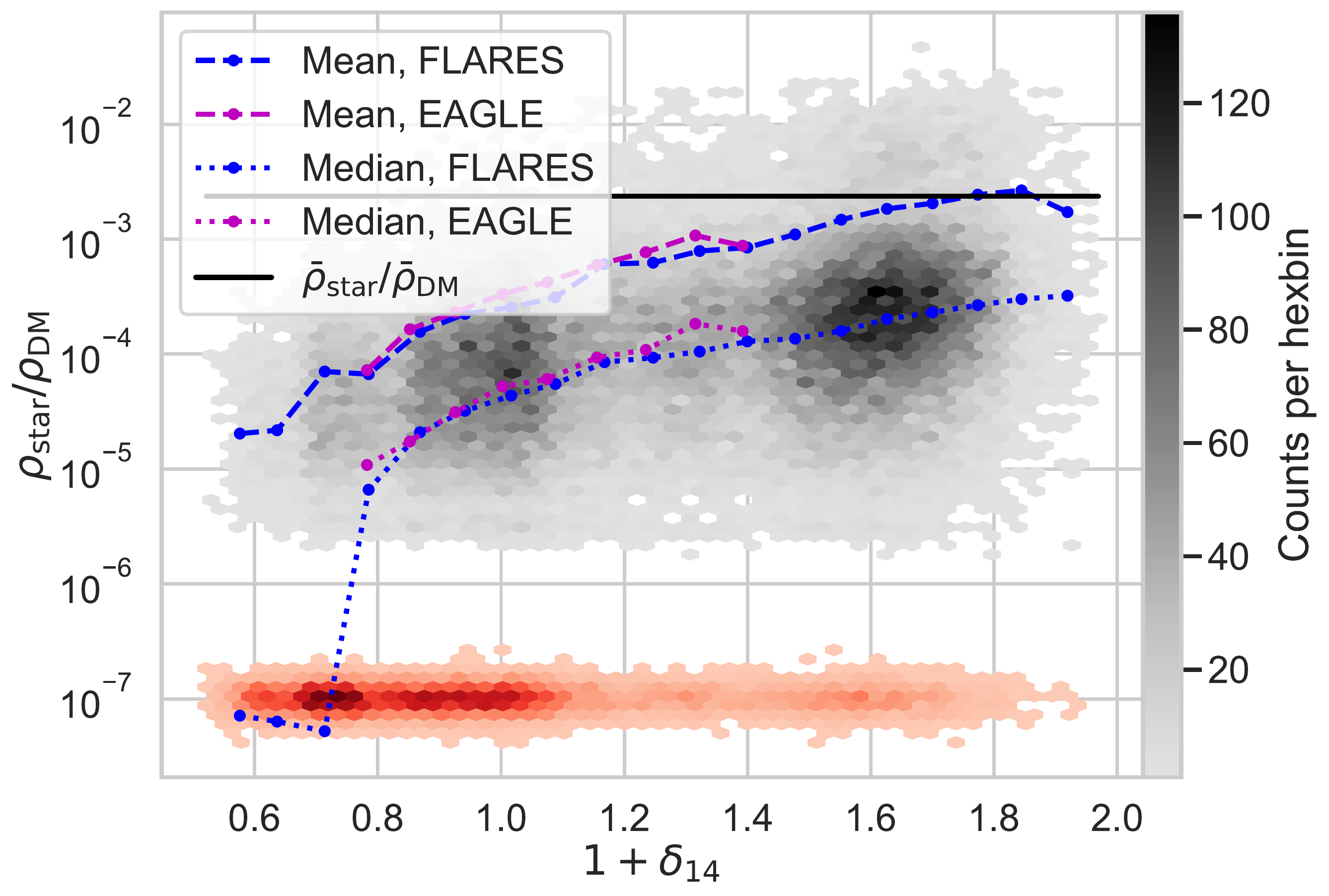}
  \caption{The ratio of stellar to dark matter density within grid cells plotted as a function of the mean matter density at that location, smoothed with a top-hat window of radius 14\,$h^{-1}$\,cMpc.  Only grid cells within the resimulated regions are plotted and used to calculate the mean and median within bins of $\delta_{14}$.  The red hexes correspond to an absence of stars, but have been given a nominal value so that they appear on the plot.}
  \label{fig:stars_DM_od_raw}
\end{figure}

\Fig{stars_DM_od_raw} shows, at $z=4.69$, the ratio of the stellar mass to the dark matter mass in individual grid cells, as a function of the smoothed matter density $1+\delta_{14}$.  The points coloured in red correspond to cells that have zero stars, but which have been given a nominal value so that they appear on the plot.  The blue dashed and dotted lines correspond to the mean and median value, respectively, within overdensity bins.  In magenta, we show the equivalent ratios for the standard \eagle\ 100\,cMpc box, which correspond quite closely to the relation seen in \flares. However, \eagle\ does not extend to the higher or lower values of overdensity sampled by \flares.

One thing that is immediately apparent is that the mean stellar to dark matter mass ratio varies by a factor of 100 between the highest and lowest values of $\delta_{14}$: star formation is thus highly biased towards regions of high overdensity.  Moreover, even at a fixed value of $\delta_{14}$, the scatter is enormous and the distribution is highly skewed such that the mean is 10 times the median.

\begin{figure}
  \includegraphics[width=8.7cm]{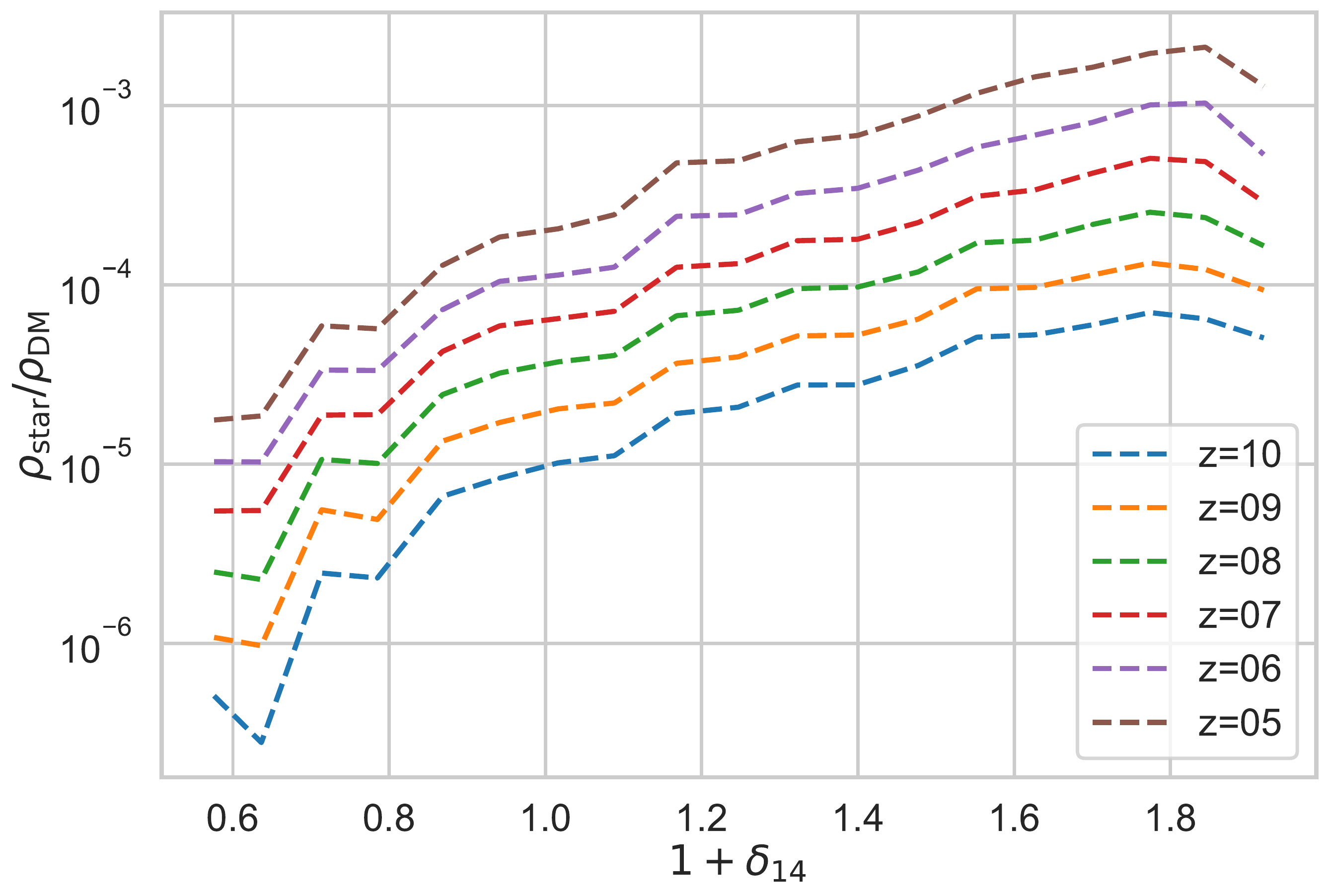}
  \caption{The mean stellar to dark matter density ratio within grid cells, as a function of redshift.}
  \label{fig:stars_DM_od_mean_snaps}
\end{figure}

The variation of the mean stellar to dark matter density ratio as a function of redshift is shown in \Fig{stars_DM_od_mean_snaps}.  Note that the value of $\delta_{14}$ used here is that measured at $z=4.69$ (using the ability to track particles over time), so that the same grid cells contribute to the $x$-axis bins at all redshifts.  The relative bias as a function of $\delta_{14}$ steepens slightly over time, with the overall normalisation rising steadily with decreasing redshift.  

\begin{figure}
  \includegraphics[width=8.7cm]{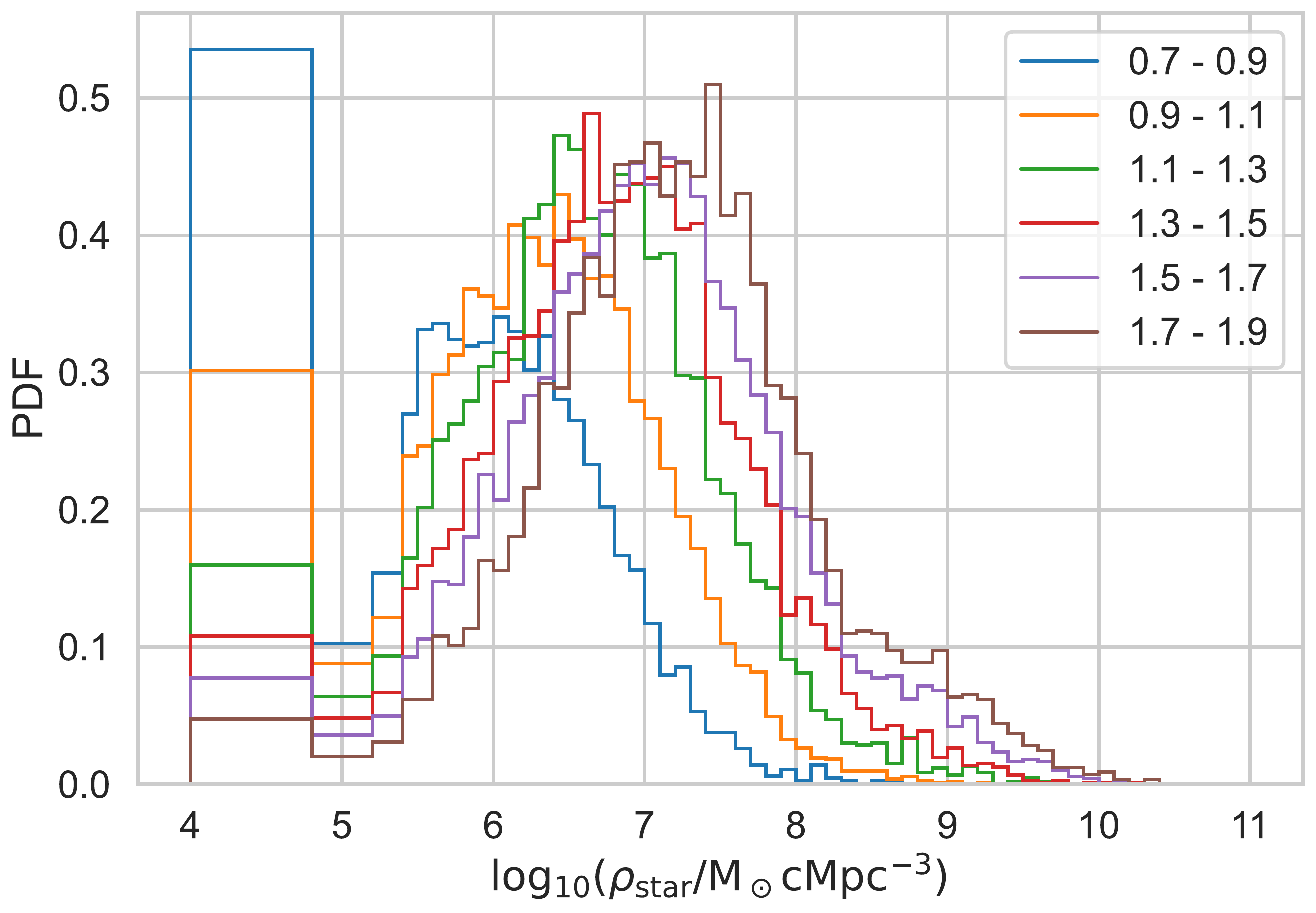}
  \caption{The distribution of stellar mass densities within grid cells split by overdensity.  The large bin on the left captures grid cells that have no stars within them.  The legend shows the range of $1+\delta_{14}$ within each colour bin: the peak stellar density shifts to the right as $\delta_{14}$ increases.}
  \label{fig:stars_skewness_od}
\end{figure}

We show the dispersion of values about the mean in \Fig{stars_skewness_od}.  As can be seen, there is a huge variation in stellar density, even within grid cells with the same value of the smoothed matter density $\delta_{14}$.  For the lowest values of $\delta_{14}\lesssim-0.25$ more than half the grid cells have no stars whatsoever within them.  By contrast, the highest stellar density, within a grid cell with $\delta_{14}=0.76$, is $2.3\times10^9$\,\Msun\,cMpc$^{-3}$, almost 4 times the universal baryon density: within that grid cell approximately 10 per cent of the baryons have been turned into stars. 

\subsubsection{Other properties}

\begin{figure}
  \includegraphics[width=8.7cm]{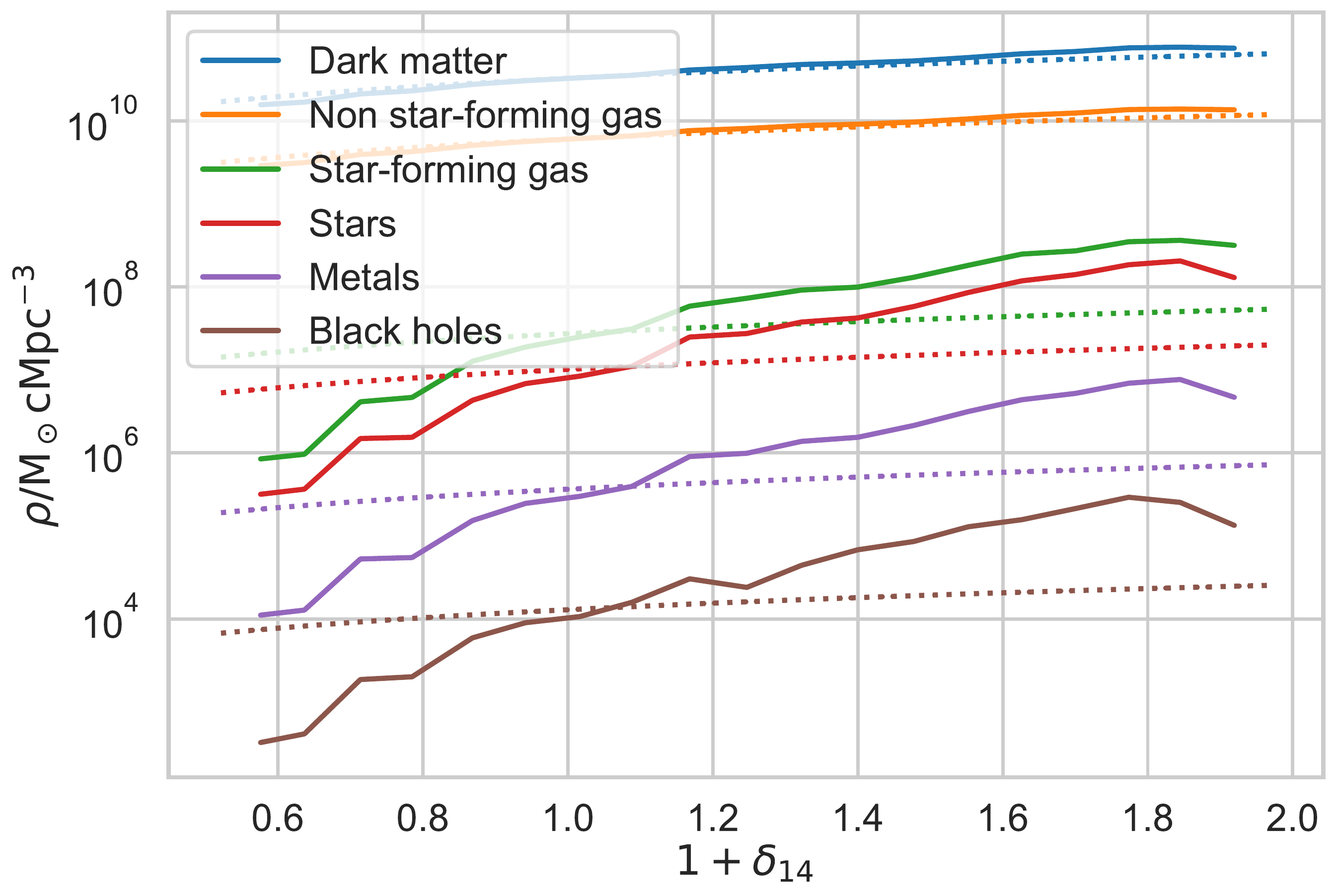}
  \caption{The mean density of various particle types within grid cells, as a function of the smoothed matter density.  The dotted lines show the expected relations if the particles traced the smooth matter distribution.}
  \label{fig:various_od_mean}
\end{figure}

\Fig{various_od_mean} contrasts the density variation of different particle types within grid cells at $z=4.69$.  The dotted lines show the expected relations if the particles traced the smooth matter distribution.  The bias for both dark matter and non star-forming gas is minimal.  However, that of star-forming gas, stars themselves, and the mass of metals produced is significant, varying by more than an order of magnitude above and below the mean in the highest and lowest density regions, respectively.  A similar effect is seen in the distribution of black hole mass.

\begin{figure}
  \includegraphics[width=8.7cm]{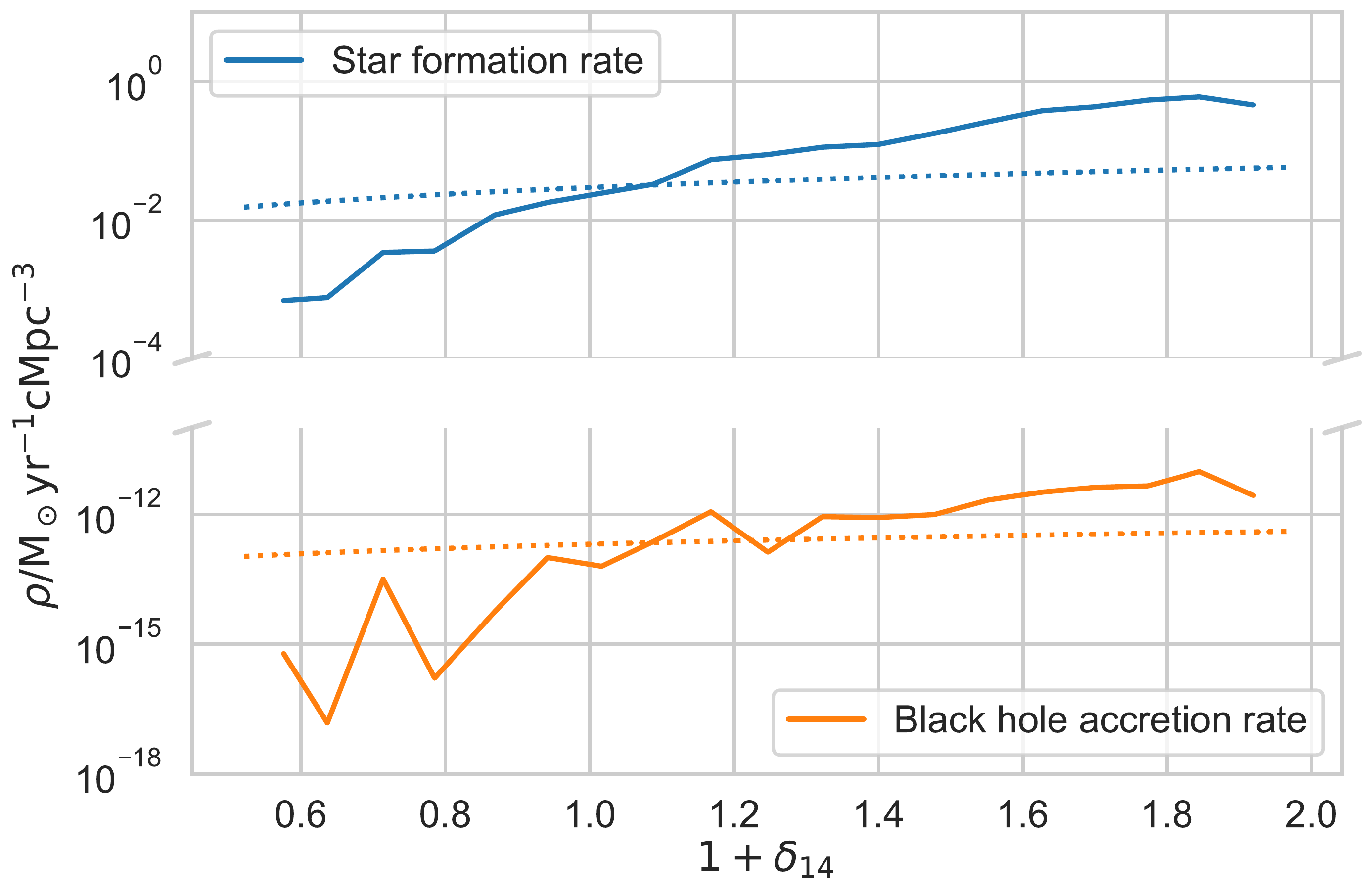}
  \caption{The mean star formation and black hole accretion rate densities, as a function of the smoothed matter density.  The dotted lines show the expected relations if the rates traced the smooth matter distribution.}
  \label{fig:various_od_rates}
\end{figure}

Finally, \Fig{various_od_rates} shows the star formation and black hole accretion rate densities, which roughly track those of the stellar and black hole mass density, respectively.

\subsection{Bias in galaxy properties}
\label{sec:bias:galaxies}

\begin{figure}
  \includegraphics[width=8.7cm]{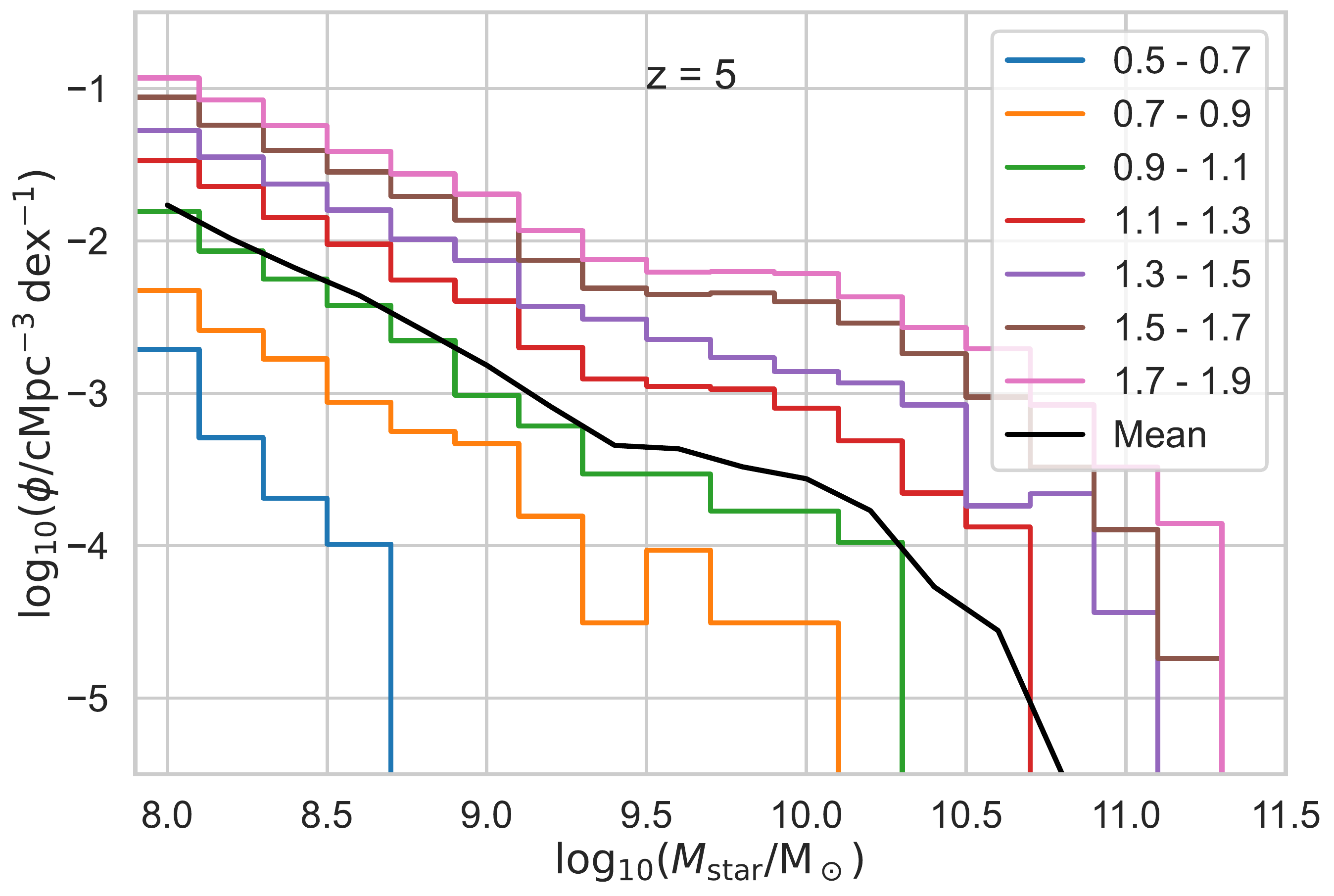}
  \includegraphics[width=8.7cm]{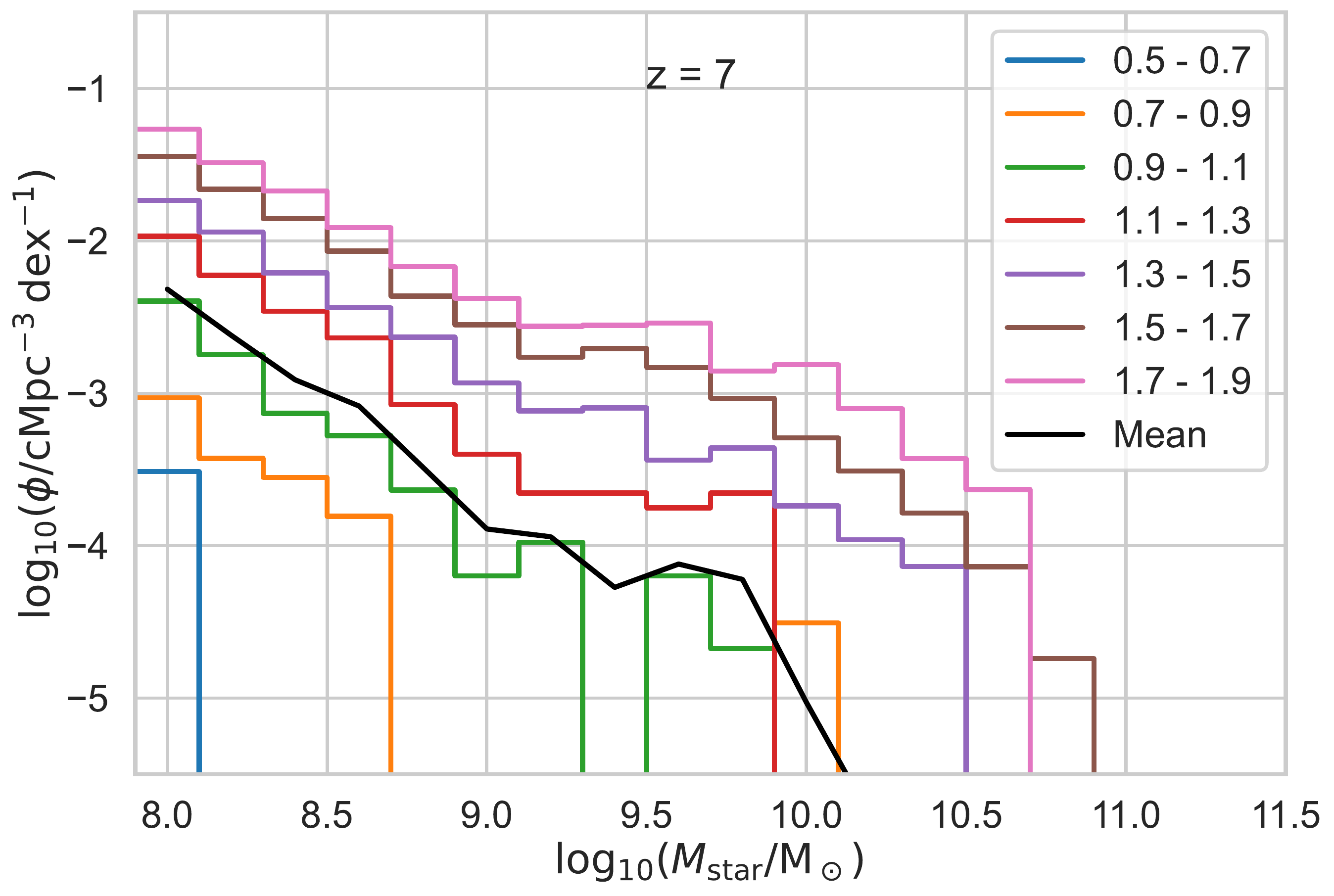}
  \includegraphics[width=8.7cm]{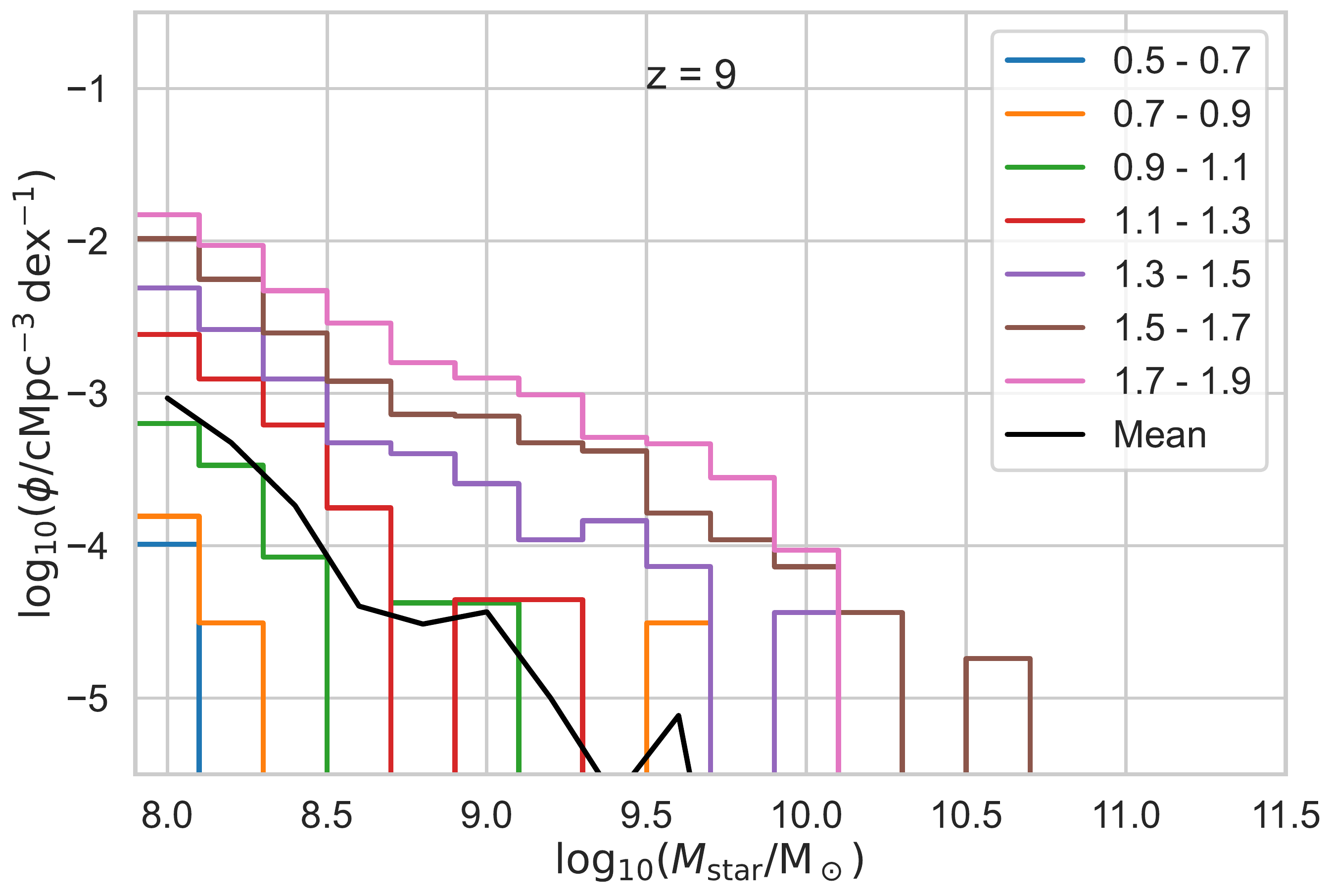}
  \caption{The galactic stellar mass function split by overdensity $1+\delta_{14}$ at three different redshifts.  The legend shows bins of $1+\delta_{14}$.  The universal mean is shown by the solid, black line.}
  \label{fig:GSMF_od_bins}
\end{figure}

We now look at the bias in the distribution of integrated galaxy properties, as a function of the matter overdensity.

\Fig{GSMF_od_bins} shows the galactic stellar mass function (GSMF) as a function of $1+\delta_{14}$.  This can be directly compared to Fig.~9 in \citetalias{Flares1}, which showed a similar plot but with each galaxy being associated with the whole range of overdensities within its resimulation volume, rather than that specific to its individual grid cell.  The two plots are very similar except that the new one better captures the true overdensity local to each galaxy, and so has a slightly larger difference between overdensity bins.

The mean GSMF follows a similar form to that for the grid cells of mean matter density ($\delta_{14}\approx0$) in the mass range where they overlap, but has a slightly higher normalisation due to the strong bias towards extra star formation in overdense regions.  An important thing to note, however, is that, at the high mass end, only the highest overdensity regions contribute to the mass function, increasingly so at higher redshift.  These regions are very rare, which gives rise to the exponential decline in the GSMF at the high mass end.  High mass galaxies are strongly clustered in these high density regions, leading to a large sample variance in observational surveys: this is discussed further in \Sec{variance:results}.

\begin{figure}
  \includegraphics[width=8.7cm]{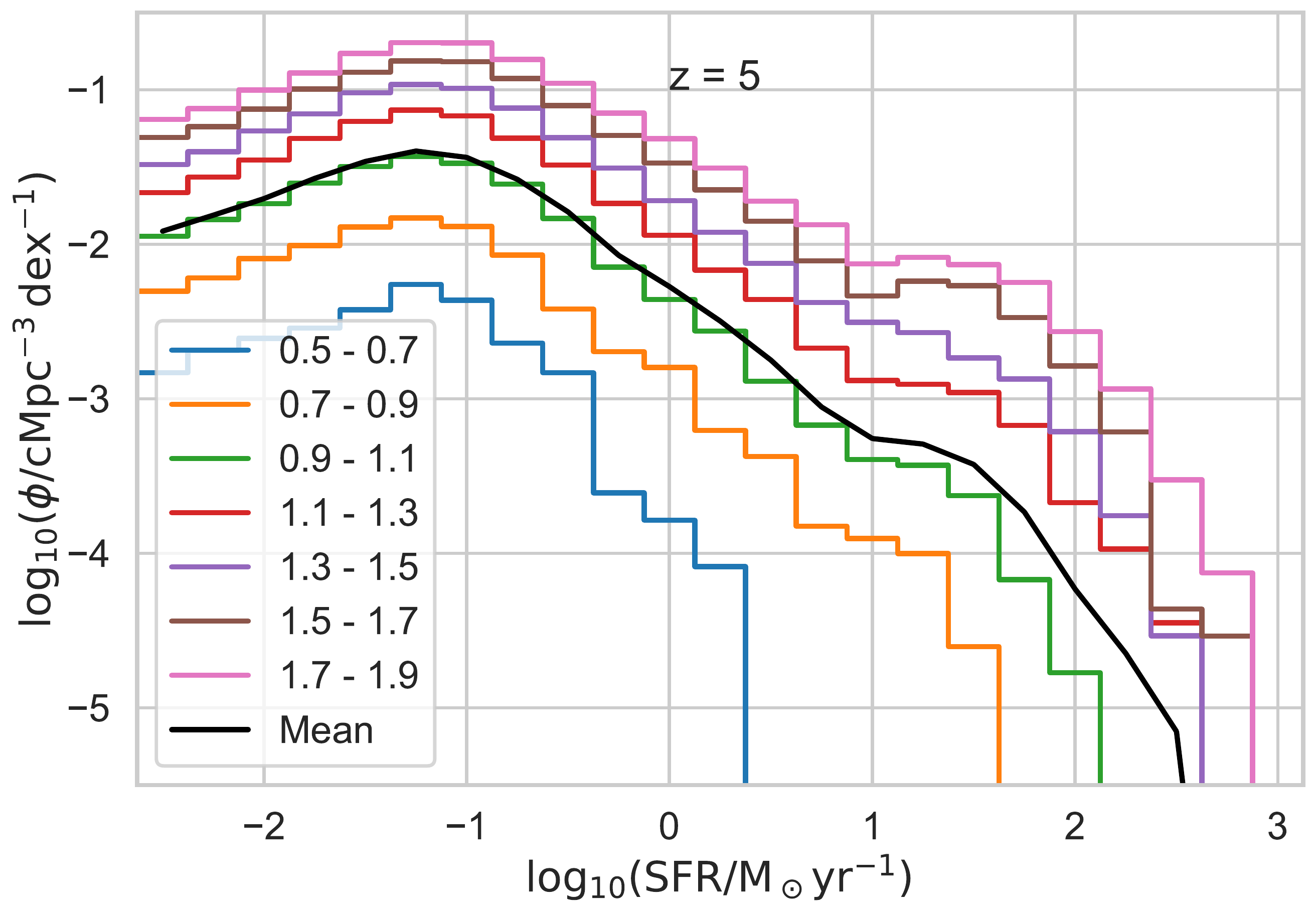}
  \includegraphics[width=8.7cm]{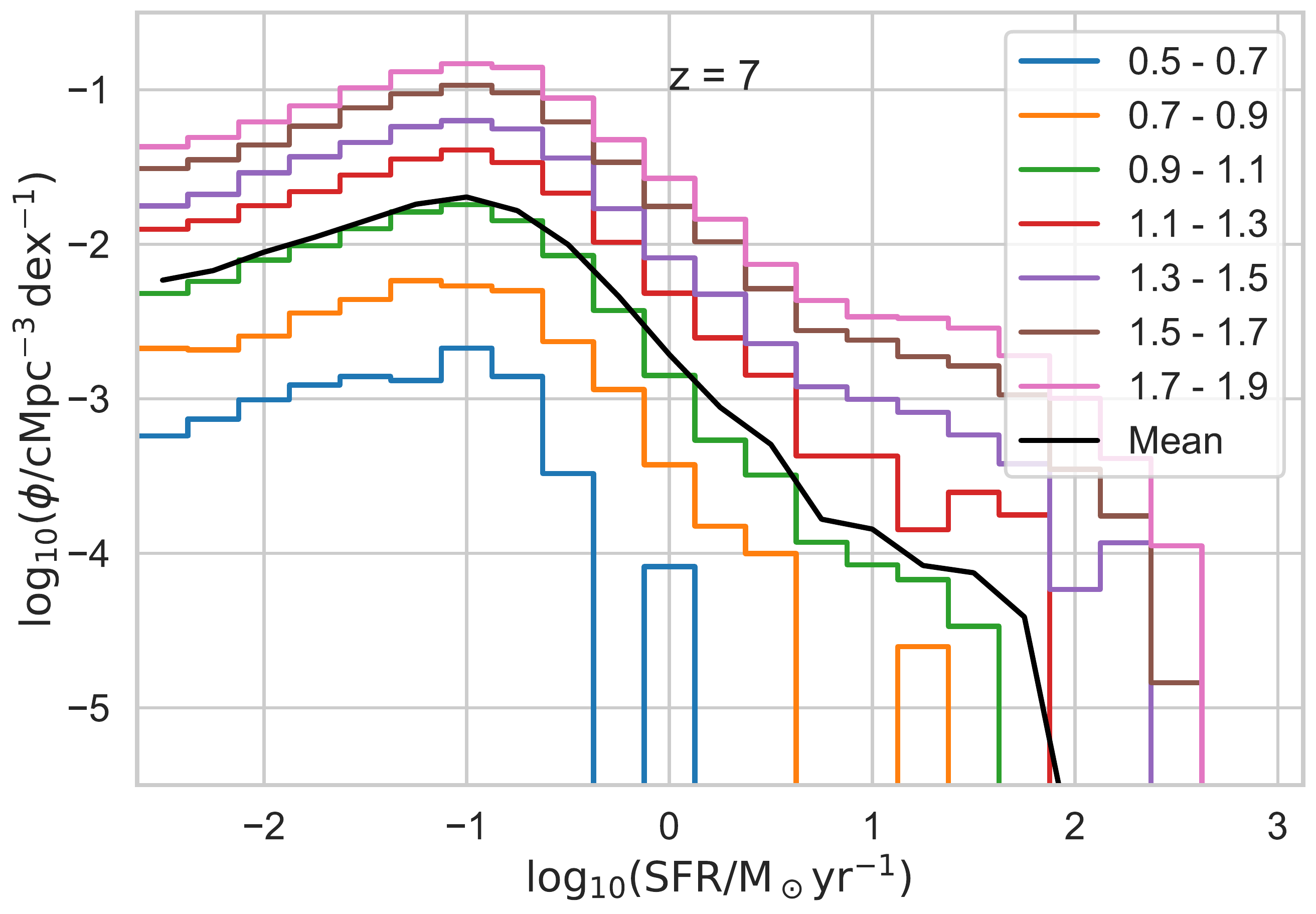}
  \includegraphics[width=8.7cm]{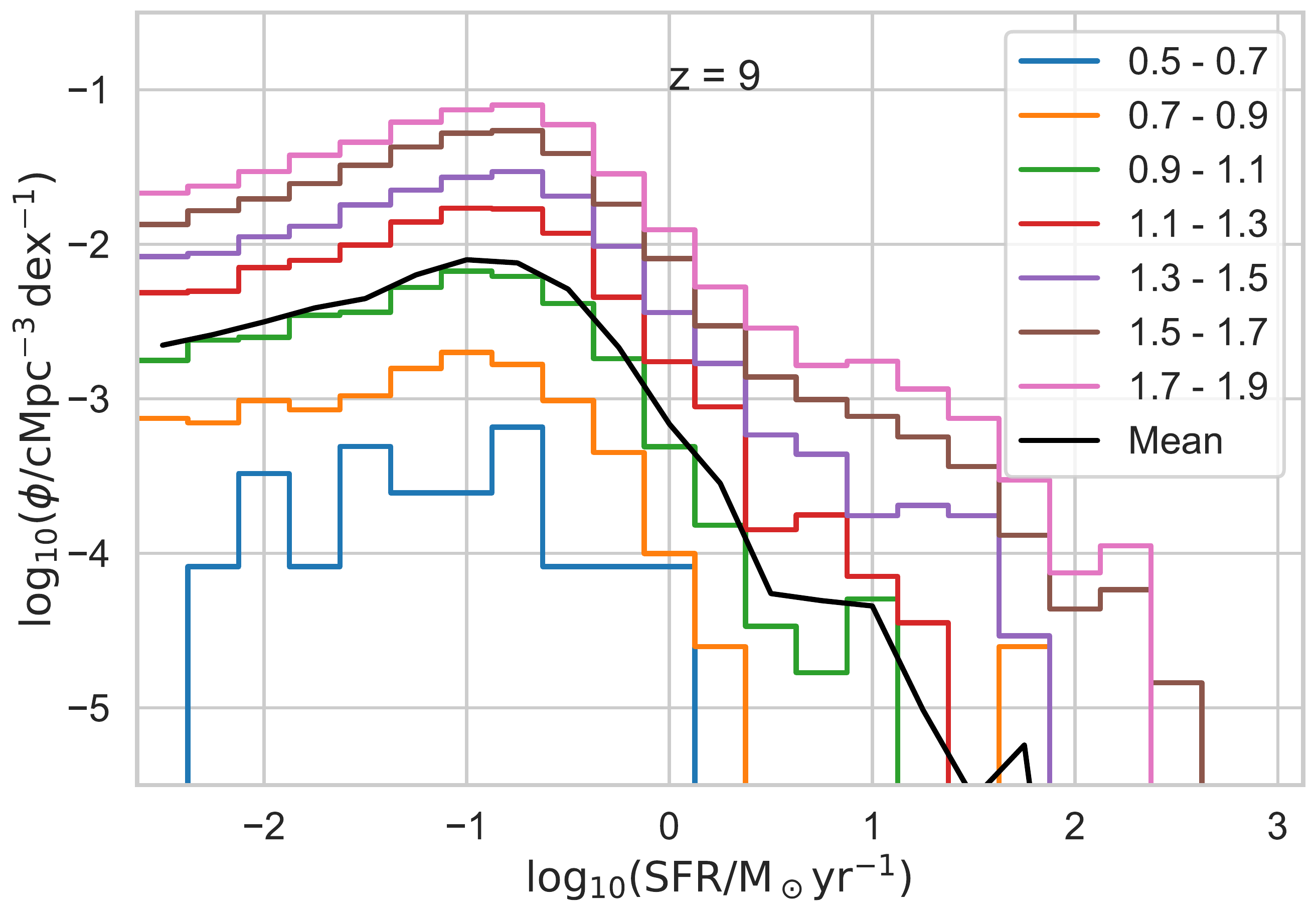}
  \caption{The galactic star formation rate split by overdensity $1+\delta_{14}$ at three different redshifts.  The legend shows bins of $1+\delta_{14}$.  The universal mean is shown by the solid, black line.}
  \label{fig:SFRF_od_bins}
\end{figure}

The star formation rate function (SFRF) for galaxies is shown in \Fig{SFRF_od_bins}, again split by matter overdensity.  It shows a similar behaviour to the GSMF, with the largest star formation rates being dominated by galaxies in the highest overdensity bins, especially at high redshift.  This reflects the strong positive correlation between stellar mass and star formation at high redshift.

\section{Survey variance}
\label{sec:variance}

In this section, we investigate the clustering of galaxies on the sky and discuss the implications for survey design.  This is very much a first look and we make a number of simplifying assumptions.  We show that compact surveys such as those that are expected for deep fields are subject to large variance and we will return to a more detailed study of this in future work.

\subsection{Populating grid cells with galaxies}
\label{sec:variance:model}

We use the information that we have gathered from our high-resolution hydrodynamic simulations to populate the underlying dark-matter-only (DMO) simulation with galaxies.  The mass of the DMO particles is $8.01\times10^{10}\,\Msun$, meaning that Milky Way sized halos would be barely resolvable, hence we choose instead for the purposes of this current paper to use as input the average properties of dark matter within grid cells.  We have investigated a number of different ingredients: as well as the average densities on different smoothing scales, $\delta_{10}$, $\delta_{14}$ and $\delta_{20}$, described above, we have also tried the unsmoothed density within an individual grid cell, $\rho_\mathrm{grid}$, and the divergence of the local velocity field within each cell.  We find that all are highly correlated and have similar predictive power for the determination of the bias, $\rho_\mathrm{star}/\rho_\mathrm{DM}$, within each grid cell.  We have also checked that combining two or more of these inputs provides only a marginal improvement in predictive accuracy.  For that reason, we stick here with the quantity that we have used both in the design of \flares\ and throughout most of this paper, $\delta_{14}$.

We tabulate the galactic stellar mass function (GSMF) as a function of overdensity and redshift; i.e. we determine the expected number of galaxies in each grid cell, which will be a fractional number.  \App{variance:ngal} investigates the variance in this GSMF and shows that it is close to Poisson, even for low galaxy number counts.  As we are interested in the variation in number counts caused by large scale structure, we do {\bf not} include that variance here: hence within each grid cell, we simply assign galaxies according to the corresponding mean GSMF.    

\subsection{Generating maps}
\label{sec:variance:maps}

To generate maps, we project grid cells along one axis of the simulation.  Strictly speaking, we should project the simulation box onto a cone centred on an observer at $z=0$; however, provided that the angular diameter of a grid cell varies by only a small amount within the depth of each map, then the parallel projection is a good approximation and avoids having to smooth over grid cell -- in this paper we are interested in only a rough estimate of the clustering of sources, hence this is sufficient for our purposes.  We summarise the slice properties for unit redshift intervals in \Tab{variance:maps}.

\begin{table}
  \caption{Properties of the redshift slices that we use: first two columns, the redshift and angular diameter of grid cells at the slice edges; third/fourth columns, the thickness of the slice in cMpc and grid cells, respectively.}
  \label{tab:variance:maps}
  \begin{tabular}{cccc}
    \hline
    Redshift& Ang.~diam./arcmin& Thickness/cMpc& Thickness (cells)\\
    \hline
    \begin{tabular}{@{}c@{}}
      4.5 \\ 5.5 \\ 6.5 \\ 7.5 \\ 8.5 \\ 9.5 \\ 10.5
    \end{tabular} &
    \begin{tabular}{@{}c@{}}
      1.20 \\ 1.12 \\ 1.06 \\ 1.02 \\ 0.99 \\ 0.96 \\ 0.94
    \end{tabular} &
    \begin{tabular}{@{}c@{}}
      541 \\ 430 \\ 352 \\ 295 \\ 252 \\ 218
    \end{tabular} &
    \begin{tabular}{@{}c@{}}
      203 \\ 161 \\ 132 \\ 111  \\ 94  \\ 82
    \end{tabular} \\
    \hline
  \end{tabular}
\end{table}

\subsection{Results}
\label{sec:variance:results}

In this section, we will present results for the number of galaxies that exceed a certain mass limit, in different survey areas and redshift slices.  Very similar results are found for galaxies that exceed particular star formation rates, and these are shown in \App{variance:SFR}.

\begin{figure}
  \includegraphics[width=8.7cm]{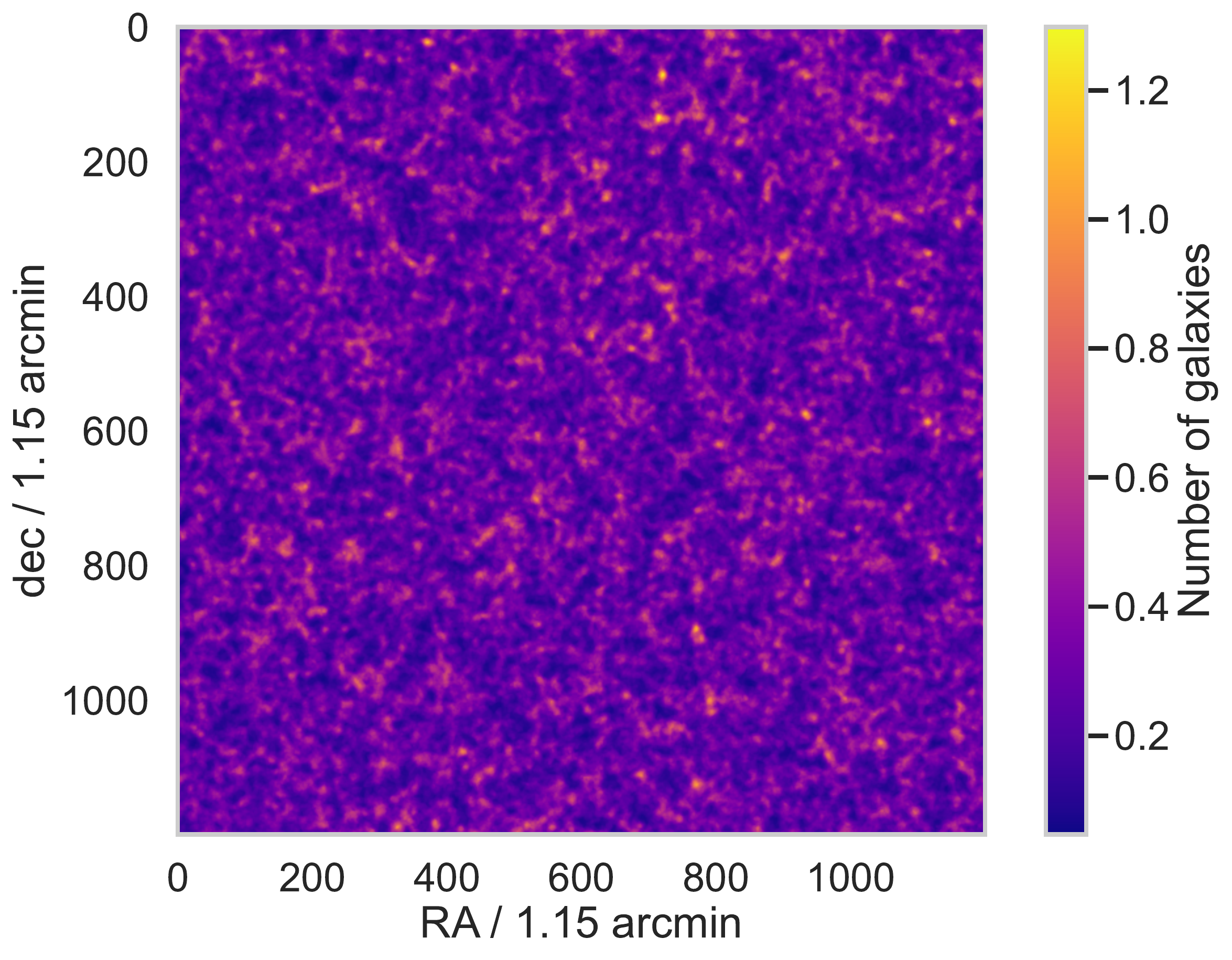}
  \includegraphics[width=8.7cm]{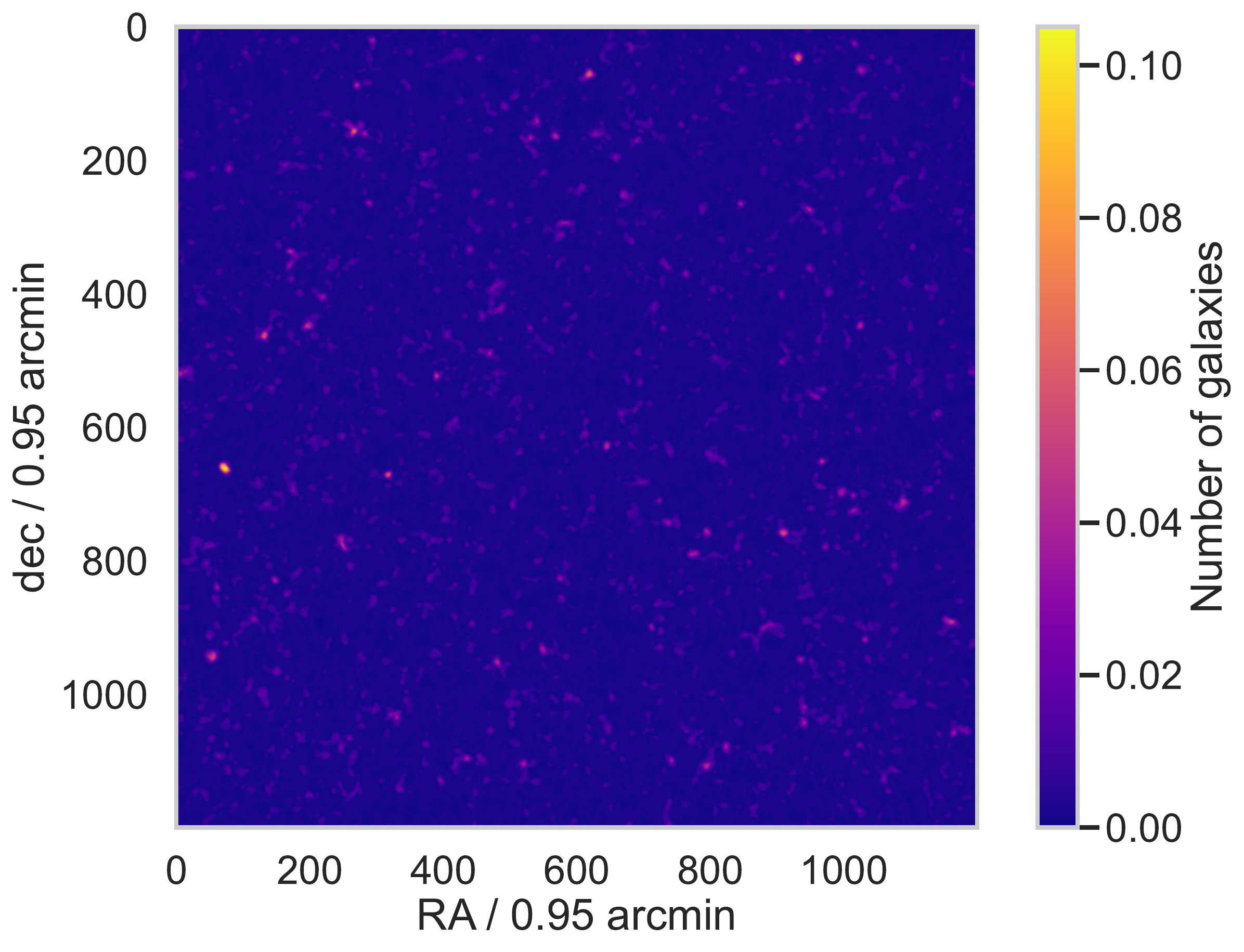}
  \caption{A map of the expected number of galaxies in some projected redshift slice per projected grid cell. \textit{Top:} galaxies with mass $M_*>10^{10}\,\Msun$ between $4.5<z\leq5.5$. \textit{Bottom:} galaxies with mass $M_*>10^{9}\,\Msun$ between $9.5<z\leq10.5$}
  \label{fig:map_mstar}
\end{figure}

\Fig{map_mstar} shows the expected number of galaxies per projected grid cell in two redshift slices: $M_*>10^{10}\,\Msun$, $4.5<z\leq5.5$ in the top panel; and $M_*>10^{9}\,\Msun$, $9.5<z\leq10.5$ in the lower panel.  These have been chosen, somewhat arbitrarily, to represent relatively abundant and sparse sources, respectively.  In the upper panel, it can be seen quite clearly that there is significant clustering of the galaxies at $z\sim5$; this is also true, but less obvious, in the lower panel at $z\sim10$.

\begin{figure*}
  \includegraphics[width=8.7cm]{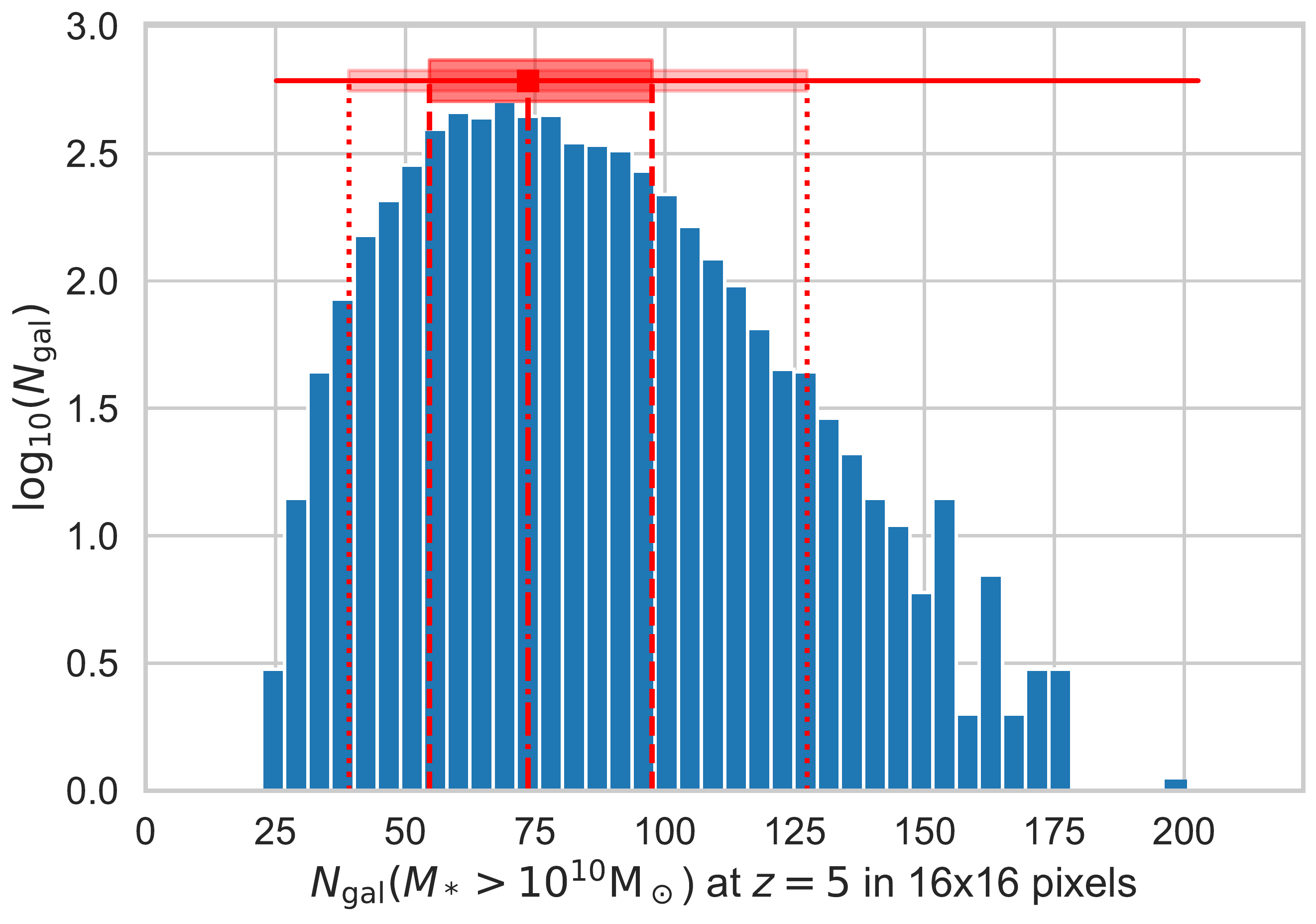}\includegraphics[width=8.7cm]{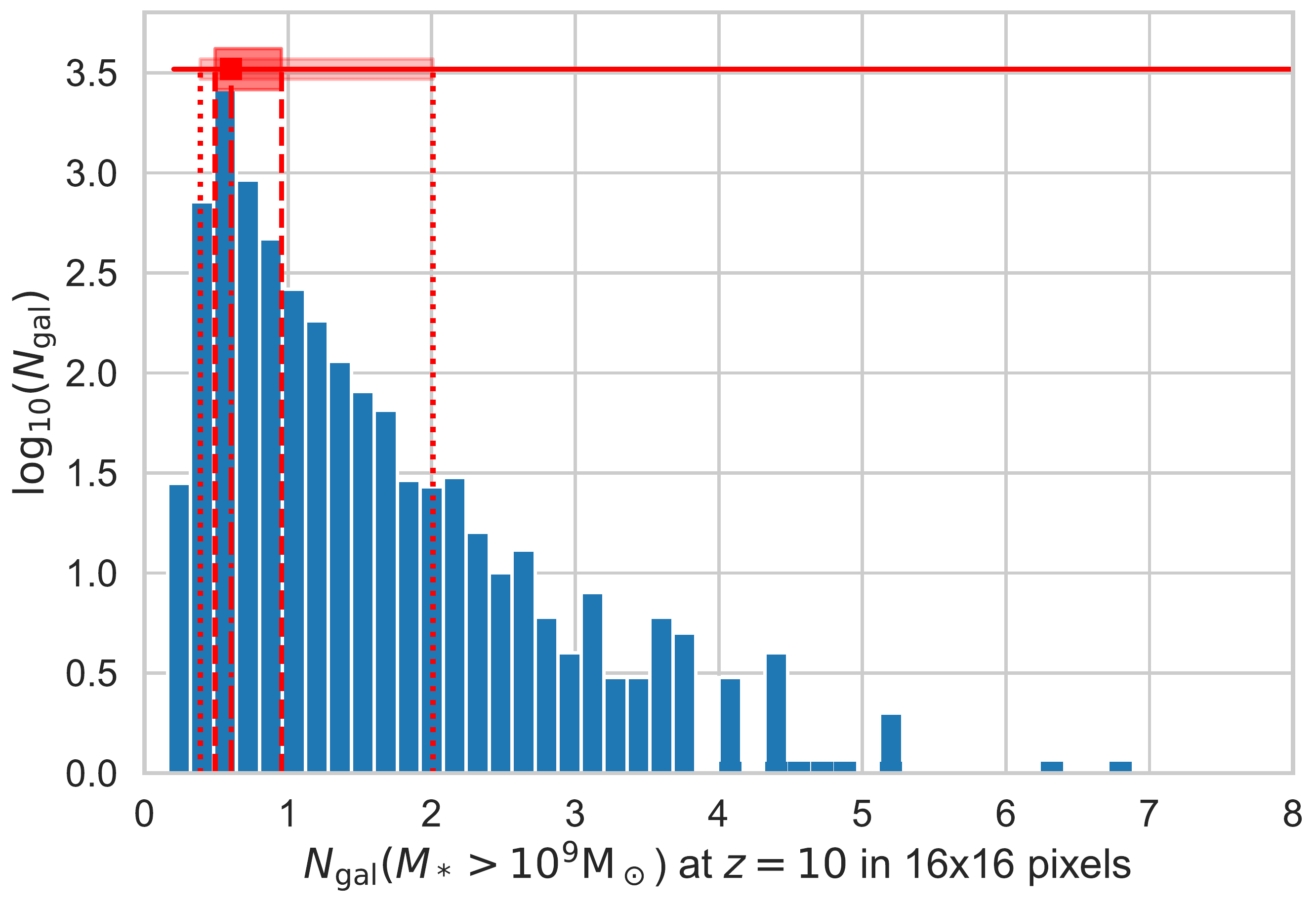}
  \includegraphics[width=8.7cm]{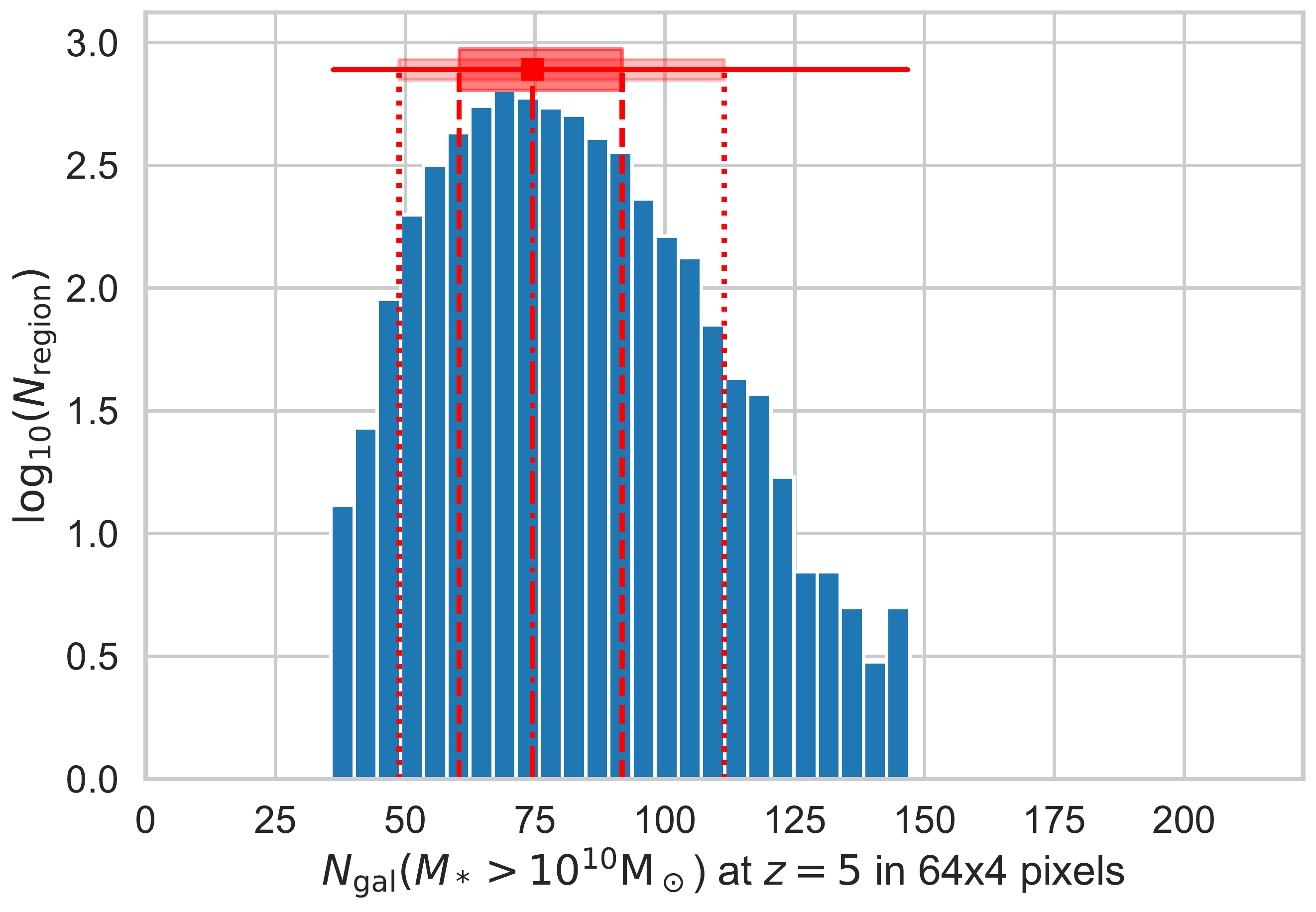}\includegraphics[width=8.7cm]{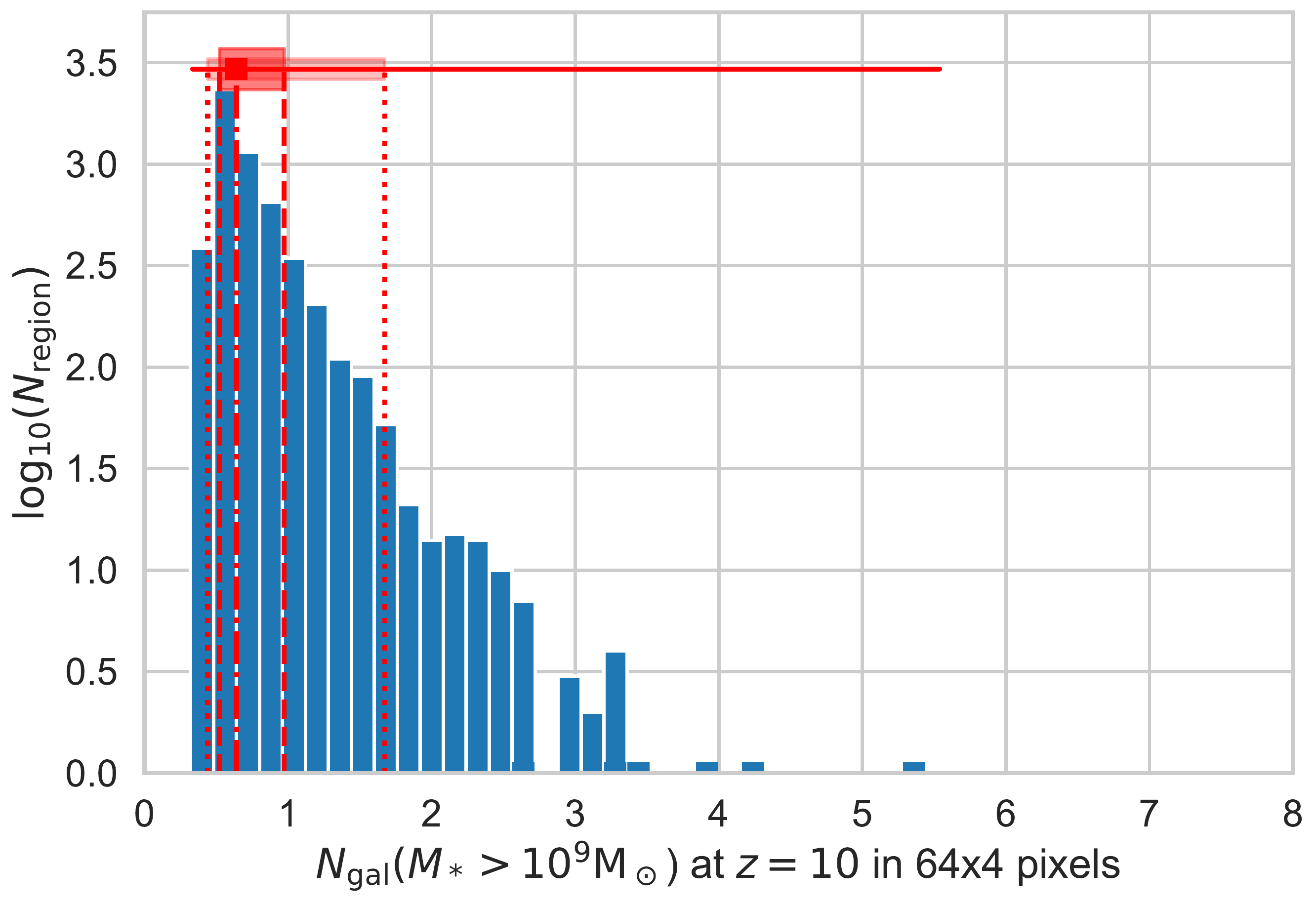}
  \includegraphics[width=8.7cm]{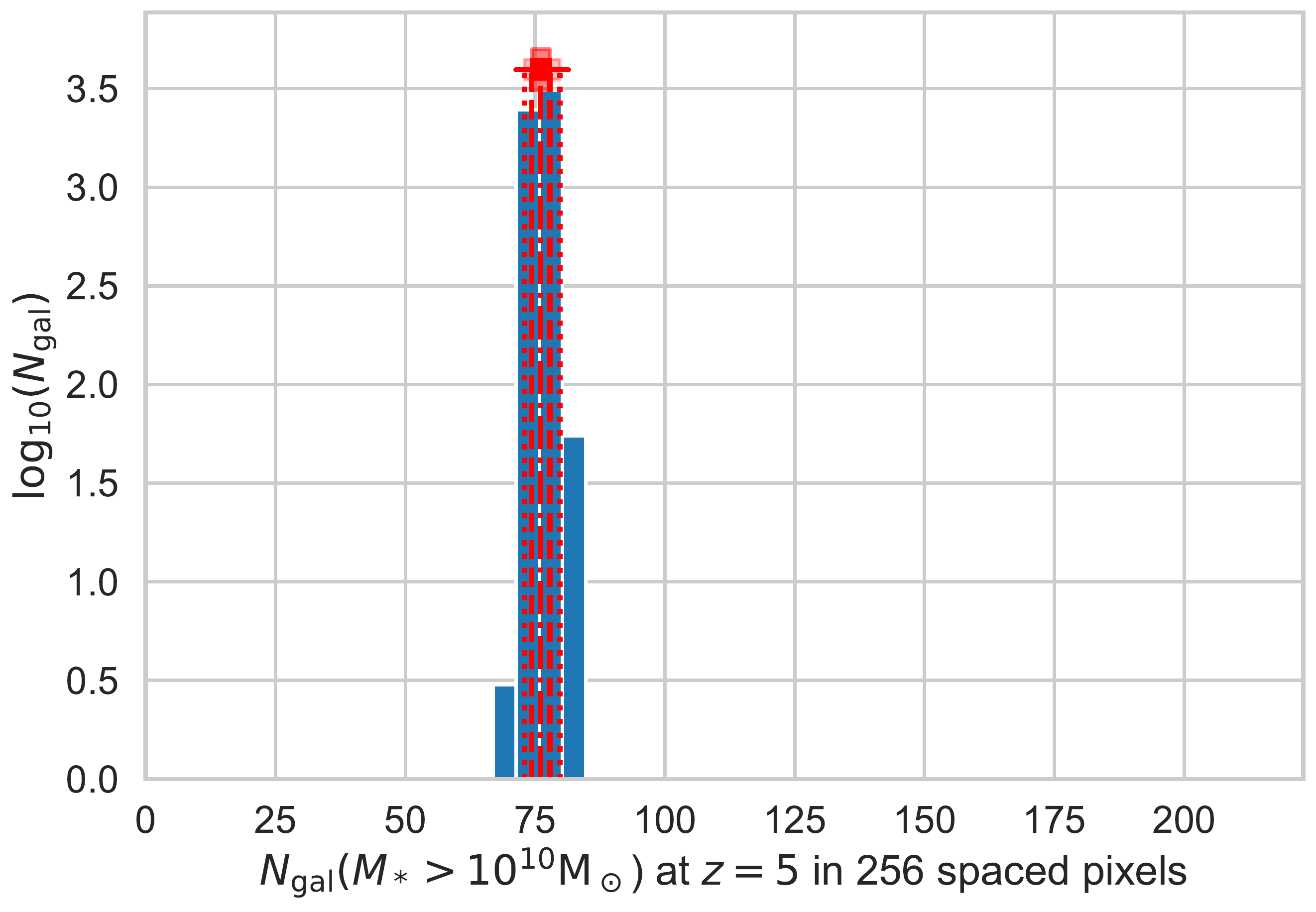}\includegraphics[width=8.7cm]{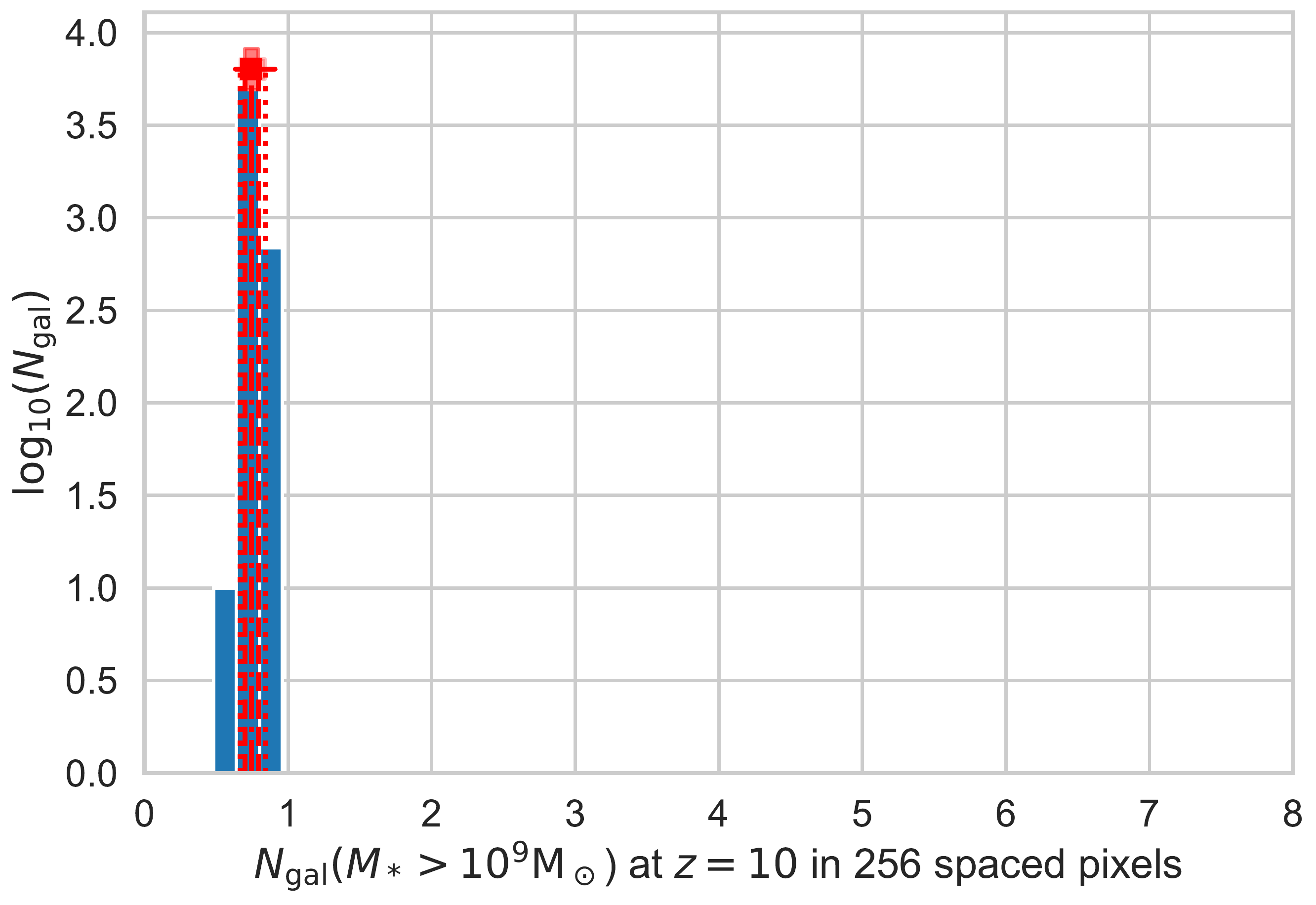}
  \caption{Histograms of the number of galaxies within a 256 grid cell ($\sim300\,$arcmin$^2$) survey region, above a particular mass and in a given redshift slice, according to the geometry of the survey: left column -- $M_*>10^{10}\,\Msun$, $4.5<z\leq5.5$; right column -- $M_*>10^{9}\,\Msun$, $9.5<z\leq10.5$; upper row -- 16\,x\,16; middle row -- 64\,x\,4; lower row -- 256 widely spaced grid cells. The dot-dashed, dashed and dotted lines show the median, one-sigma and two-sigma ranges, respectively; the box-plot shows the full extent of the data, plus the one and two-sigma ranges.  In the top, right-hand panel a single point with $N=12.3$ has been omitted, for clarity.}
  \label{fig:hist_mstar}
\end{figure*}

To show what effect this might have on the variance of galaxy numbers detected in surveys, we plot in \Fig{hist_mstar} the galaxy counts in an area of approximately $\sim300\,$arcmin$^2$, corresponding to 256 projected grid cells, for 3 different survey designs: the upper row is a square survey region of 16\,x\,16 grid cells, approximately (17\,arcmin)$^2$; the middle row is a long strip of 64\,x\,4 grid cells, approximately 1\,deg\,x\,4\,arcmin; and the lower row is 256 separate, widely-spaced and hence uncorrelated grid cells.  There are 5625 separate samples in the upper and lower rows; slightly fewer in the middle row because of the shape of the region and a desire not to sample the same grid cell twice.

From the bottom row, we can see that we have sampled a sufficient number of independent regions that the expected number of galaxies in each mock survey lies close to the mean.  For the square survey regions, however, there is a large variation in the expected number of detected galaxies, by a factor of 8 (top-left) for the most abundant sources at $z\sim5$, to 60 (top-right) for the rarer sources at $z\sim10$; the 2-sigma ranges for these are factors of 3.3 and 5.2, respectively.  The long, thin surveys shown in the middle row, reduce this variance a little but still show considerable spread about the mean value.

We compare our results from that of the Cosmic Variance Calculator\footnote{\protect{\tt https://www.ph.unimelb.edu.au/~mtrenti/cvc/CosmicVariance.html}} \citep{Trenti08} in \app{variance:cvc}.  The latter show a similar trend but with reduced variance.  The main reasons for this are likely to be the limited volume of the dark matter simulations which underlie their results, and the use of halo occupation models extrapolated to high redshift that may well not capture the extreme biases that we see for rare objects.  The ratio of the standard deviation of the \flares{} predictions to that of the CVC is slowly varying with both the survey area, $A_\mathrm{survey}$, and the expected number of galaxies in the survey, $N_\mathrm{gal}$.   For a square survey area it is reasonably well fit by the relation
\begin{equation}
  \label{eq:variance:cvc}
        {\sigma\over\sigma_\mathrm{CVC}}=
        \left(A_\mathrm{survey}\over\mathrm{arcmin}^2\right)^{0.092}N_\mathrm{gal}^{-0.038}.
\end{equation}

Although we have presented here results for galactic stellar mass, those for star formation rate, shown in \App{variance:SFR} are similar.  Moreover, we would expect the same to hold for flux-limited surveys also, as we expect a strong correlation between mass/SFR and observable fluxes.  That is not to say that there won't be some environmental dependence in that correlation. We will explore this in future work, where we generate mock surveys in different bands.

\begin{figure}
  \includegraphics[width=8.7cm]{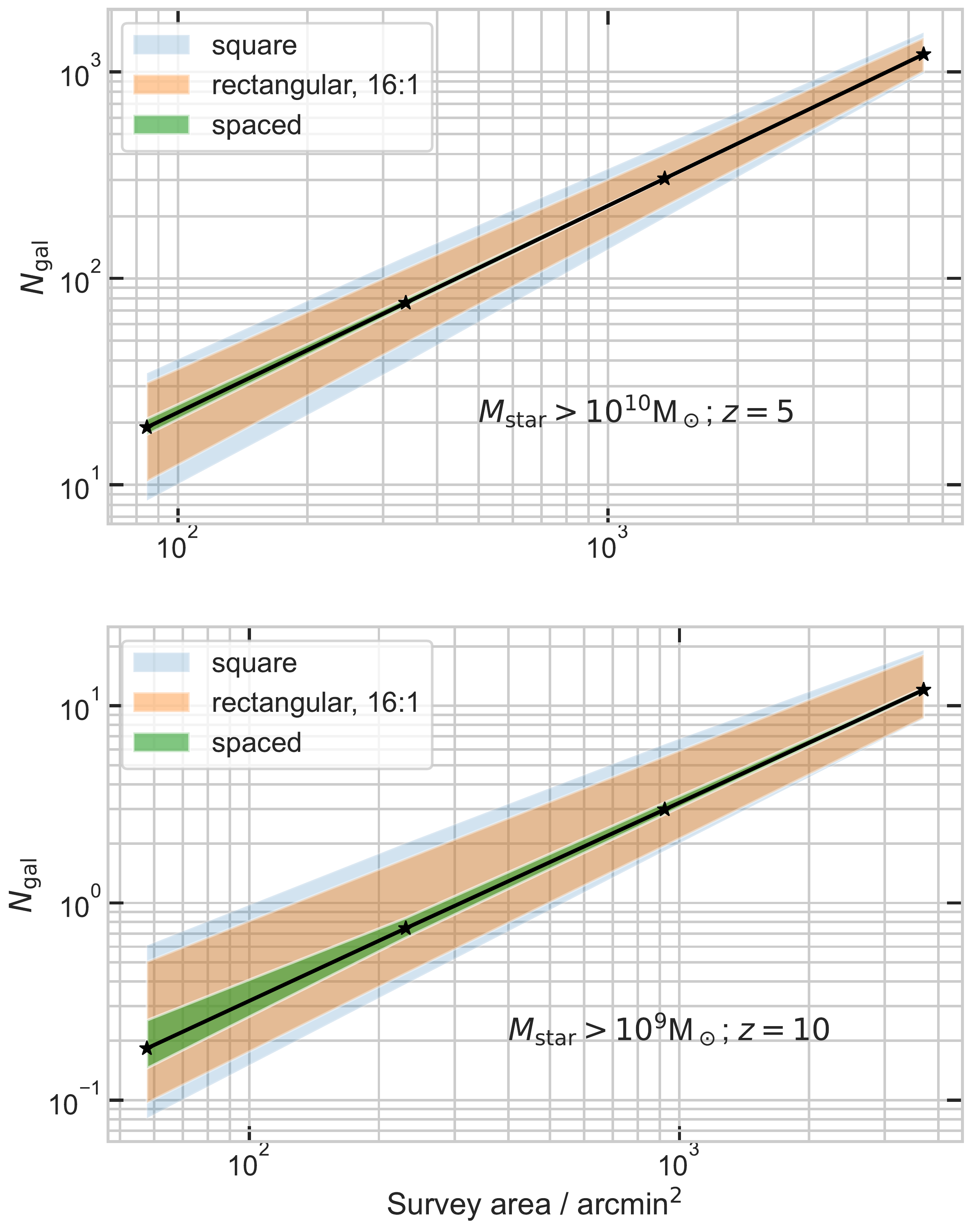}
  \caption{The mean and 2-sigma (2.3$-$97.7 percentile) range for expected galaxy counts as a function of survey area and shape.  Measured values are taken for survey areas corresponding to 64, 256, 1024 and 4096 pixels and results interpolated between these points.  We report results for stellar mass: the upper/lower panels are for a large/low number count; similar results are found for star formation rates.}
  \label{fig:variance_mass}
\end{figure}

\Fig{variance_mass} shows the mean and 2-sigma spread for the number counts as a function of the survey area, for both high and low number counts.  The variances are reduced as the survey area is increased.  We show only our two mass selections here; results for the star formation selection are very similar.  Measured values are taken for survey areas corresponding to 64, 256, 1024 and 4096 pixels and results interpolated between these points; in \App{variance:large} we show histograms of number counts for the largest survey area of 4096 pixels, or approximately 1.4\,deg$^2$.  

\subsection{Application to existing and proposed surveys}
\label{sec:variance:surveys}

The implications for the interpretation of galaxy surveys at these redshifts are clear: in any survey of limited spatial extent, the variance in the number of detected galaxies is likely to be large, and one should take that into account when making any measurement of the number density of sources.

Figure~8 of \flaresone\ showed galactic stellar mass functions at $z=5$ from a range of observations \citep{Gonzalez11,Duncan14,Song16,Stefanon17}, varying in survey area from 50\,arcmin$^2$ to 1\,deg$^2$.  These show a variation in normalisation of a factor of 3 at low masses to 10 at high masses.  Now, while some of this difference will be due to the different observational bands and analysis, a significant fraction may be due to sampling variation across different survey areas.

Cycle 1 of \jwst\ has a number of large area survey programs. One of the many aims of these surveys is to investigate galaxies in the EoR.  JADES \citep{Jades20} imaging will cover 2 fields, each roughly square and 100\,arcmin$^2$ in area.  The predicted galaxy numbers per unit redshift interval \citep{Williams18} vary from many thousand at $z=5$ to a few hundred at $z=10$: the survey variance will therefore roughly correspond to that in \Fig{hist_mstar}, i.e.~a 2-sigma range of about a factor of just over 3.  \citet{Robertson22} report first results for galaxies at $z>10$ in JADES \citep[spectroscopcally confirmed by][]{Curtis-Lake22}, finding 4 galaxies in a survey area of 65\,arcmin$^2$, for which the 2-sigma survey variance is about a factor of 6.

CEERS \citep{Ceers22} is undertaking imaging and spectroscopy of the EGS HST legacy field, in an area of approximately 100 (20\,x\,5)\,arcmin$^2$, and PRIMER \citep{Primer21} is providing imaging of the CANDELS/COSMOS and CANDELS/UDS fields, each of order 100\,arcmin$^2$.  They too will suffer survey variance similar to that shown in \Fig{hist_mstar}.  Recently CEERS published preliminary results in a survey area of 35.5 arcmin$^2$ \citep{Finkelstein23}.  They split their sample of high-redshift galaxies into 15, 9 \& 2 objects at redshifts of 8.5--10, 10--12 \& $z>12$, respectively.  Our results suggest that the 2-sigma range in number counts caused by survey variance in the two lower redshift bins is likely to extend over a factor of around 10.

On a larger scale, the COSMOS-Web imaging survey \citep{CosmosWeb22} will have an area of 0.54\,deg$^2$ with an estimated galaxy count of several thousand per unit redshift interval at $z=6$ and 30$-$70 at $z=10$.  The survey area is approximately square in shape and so the 2-sigma range for galaxy counts will be around a factor of 2.5, as seen in \Fig{variance_mass}.

\citet{Castellano22} report a detection of 7 galaxies at $z\approx10$ in a survey area of 37\,arcmin$^2$ from JWST/NIRCam imaging data, 3-10 times higher than previously reported results.  They suggest that survey variance may be contributing and call for more pencil-beam surveys to confirm their results.  For the same mean number count in a similar field of view, we find a 2-sigma percentile range of between 2 and 15, thus reinforcing that need.

\citet{Harikane23} summarises the high-redshift ,$z\gtrsim9$, photometric observations of galaxy candidates to-date.  At $z\sim9$ the majority, 8/12, of their sample comes from CEERS with an effective survey area of just over 200\,arcmin$^2$, which corresponds in our results to a 2-sigma spread of a factor of approximately 4.  Other reported results come from surveys that currently have much smaller areas and which will therefore have even greater survey variances.

\citet{Adams23b} determine the UV luminosity function to date from PEARLS and other public surveys, predominantly CEERS.  The total survey area is about 110\,arcmin$^2$ split over several fields, with the majority, 64\,arcmin$^2$ coming from CEERS.  They find slightly lower luminosity functions at high redshift than \citet{Finkelstein23}, which they attribute to a greater area and survey depth.

\citet{McLeod23} combine results from 12 different JWST surveys with a total effective area of 260\,arcmin$^2$.  This large area plus large number of pointings is likely to result in the lowest survey variance of any of the high-redshift JWST studies to date.  They find similar results at $z>9$ to previous studies, but with reduced error bars, and mean star formation rates that are consistent with \flares{} (but higher than other simulations that have restricted volumes that do not sample regions of the highest overdensity).

Finally we note that, as stressed by \citet{Adams23} and many others, the photometric redshifts used in many preliminary studies are highly uncertain and currently provide the biggest uncertainty in number count estimates at the upper end of the EoR, $z\gtrsim9$.  However, as spectroscopic redshifts become available, survey variance is likely to dominate.

\subsection{Future work: more realistic mocks}

There are a number of enhancements that we intend to make to this study in order to make more accurate predictions of survey variance:
\begin{shortitem}
\item The use of the mean density field in dark matter grid cells to predict star and galaxy formation rates is fairly crude and leaves a lot of residual scatter, as seen in \Fig{stars_DM_od_raw}, that we have struggled to reduce.  Future work will use a higher resolution background dark matter simulation that will resolve halos and allow a better mapping from dark matter to galaxies.
\item By resolving halos, we will be able to project onto light-cones centred on an observer without the need for smoothing, rather than projecting parallel to the simulation grid.
\item We will use galaxies from the high-resolution, hydrodynamic simulations to make mock images of the sky in various bands, utilising the known star formation and metal enrichment histories, and applying realistic dust absorption.
\item We will then make mock observations of those images, reproducing the selection criteria of the different surveys.
\end{shortitem}
This is a substantial undertaking that will take some time to come to fruition, which is why we have given in this paper crude estimates of the magnitude of the survey variance that we expect to see that will still be of significant value as a qualitative estimate of the effect of survey variance for a given survey geometry.

\section{Conclusions}
\label{sec:conc}

In this paper we investigate the variation of star formation and galaxy properties with environment in the \flares{} simulations of galaxy formation in the early Universe.  Those simulations are designed to sample the full range of overdensities, averaged on a scale of 14\,$h^{-1}\,$cMpc, within a (3.2\,Gpc)$^3$ box.  For the most part we look at properties averaged within cubical grid cells of edge 2.67\,cMpc as a function of the overdensity averaged within a sphere of radius 14\,$h^{-1}$\,cMpc, $\delta_{14}$. We reach the following conclusions:
\begin{shortitem}
\item The ratio of stellar density, $\rho_\mathrm{star}$, to dark matter, $\rho_\mathrm{DM}$, density within each grid cell is highly biased, varying by a factor of 100 between the lowest and highest values of $\delta_{14}$ (\Fig{stars_DM_od_raw}).
\item Moreover, even at a fixed value of $\delta_{14}$, the scatter in $\rho_\mathrm{star}$/$\rho_\mathrm{DM}$ is enormous and the distribution is highly skewed such that the mean is 10 times the median.
\item This bias remains constant across all redshifts between $z=5$ and $z=10$ (\Fig{stars_DM_od_mean_snaps}).
\item For $\delta_{14}<0.9$ more than half the grid cells have no stars whatsoever within them; whereas in the highest overdensity cell roughly 10 per cent of the baryons have been turned into stars (\Fig{stars_skewness_od})
\item The bias seen in the stellar distribution is replicated in star-forming gas, metals and black holes, and in the star formation and black hole accretion rates; that in non-star-forming gas is, however, much lower and similar to that of the dark matter (Figures~\ref{fig:various_od_mean} and \ref{fig:various_od_rates})
\item The mean galactic stellar mass function (GSMF) follows a similar form to that for the grid cells of mean matter density ($\delta_{14}\approx0$) in the mass range where they overlap, but has a slightly higher normalisation due to the strong bias towards extra star formation in overdense regions (\Fig{GSMF_od_bins}).  Only the highest overdensity regions contribute to the high-mass end of the GSMF and these are very rare, which gives rise the the exponential decline in the GSMF.
\item Because the highest mass galaxies are only found in the most overdense regions, we note that resimulation of such regions within large volumes, such as undertaken in \flares, is the only way to capture them in simulations.
\item The star formation rate function (SFRF) shows a similar behaviour to the GSMF, with the largest star formation rates being dominated by galaxies in the highest overdensity bins (\Fig{SFRF_od_bins}).
\item Maps of unit redshift slices show significant clustering of galaxies at all redshifts and at both high and low number densities (\Fig{map_mstar}).
\item \Fig{hist_mstar} illustrates the effect of clustering by looking at the variation in number counts in a region consisting of 256 grid cells (approximately $\sim300\,$arcmin$^2$) in different configurations.  If the cells are widely-spaced then the variance is small, as would be expected.  However, for a square survey area of 16x16 grid cells, then the 2-sigma variation in number counts is more like a factor of 4 (slightly less for high number counts and higher for low number counts).
\item Using rectangular survey volumes improves the survey variance slightly, but to reduce it significantly requires multiple survey areas over wide separations.
\item We compare our results with those from the Cosmic Variance Calculator of \citet{Trenti08} and find larger standard deviations by a factor of 1.3--1.8 (\fig{variance_cvc_mass}).  We provide a scaling formula to convert between the two (\eq{variance:cvc}).
\item Very similar results hold for maps of galaxies exceeding a particular star formation rate (Figures~\ref{fig:map_SFR} and \ref{fig:hist_SFR}).
  \item For larger survey areas, the variance is reduced, dropping to a factor of about 2 for an area of 1.4\,deg$^2$ (Figures~\ref{fig:hist_mstar_large} and \ref{fig:hist_SFR_large}).
\end{shortitem}

Although we have presented results for physical rather than observable properties of galaxies, we would expect similar results to hold for flux-limited surveys also, as we expect a strong correlation between stellar mass / star formation rate and observable fluxes.  We will explore this in future work, where we generate mock surveys in different bands.

The implications for the interpretation of galaxy surveys at these redshifts are clear: in any flux-limited survey of limited spatial extent, the variance in the number of detected galaxies is likely to be large.  It should not be surprising to find number densities from different survey areas that differ by a factor of 2--4.  Multiple widely-spaced regions will need to be combined to beat down sample variance.  Number densities obtained from (a large number of) background regions in targeted observations of unrelated, compact, low-redshift sources would be one way to do that.  Our conclusions thus reinforce those of previous studies such as \citet{Trenti08} and \citet{Moster11}.


\section*{Acknowledgements}

We thank the \eagle\ team for their efforts in developing the \eagle\ simulation code.
We also wish to acknowledge the following open source software packages used in the analysis: \textsf{Scipy} \citep{SciPy20}, \textsf{Astropy} \citep{AstroPy13}, and \textsf{Matplotlib} \citep{Matplotlib07}.

This work used the DiRAC@Durham facility managed by the Institute for Computational Cosmology on behalf of the STFC DiRAC HPC Facility (www.dirac.ac.uk).
The equipment was funded by BEIS capital funding via STFC capital grants ST/K00042X/1, ST/P002293/1, ST/R002371/1 and ST/S002502/1, Durham University and STFC operations grant ST/R000832/1.
DiRAC is part of the National e-Infrastructure. The \eagle\ simulations were performed using the DiRAC-2 facility at Durham, managed by the ICC, and the PRACE facility Curie based in France at TGCC, CEA, Bruyeres-le-Chatel.

CCL acknowledges support from a Dennis Sciama fellowship funded by the University of Portsmouth for the Institute of Cosmology and Gravitation.
DI acknowledges support by the European Research Council via ERC Consolidator Grant KETJU (no. 818930) and the CSC – IT Center for Science, Finland.
APV acknowledges support from the Carlsberg Foundation (grant no CF20-0534). The Cosmic Dawn Center (DAWN) is funded by the Danish National Research Foundation under grant No. 140.

We list here the roles and contributions of the authors according to the Contributor Roles Taxonomy (CRediT)\footnote{\url{https://credit.niso.org/}}.
\textbf{Peter Thomas}: Conceptualization, Data curation, Methodology, Investigation, Formal Analysis, Visualization, Writing - original draft.
\textbf{Christopher C. Lovell, Aswin P. Vijayan}: Data curation, Writing - review \& editing.
\textbf{Maxwell Maltz}: Methodology, Writing - review \& editing.
\textbf{Stephen M. Wilkins}: Conceptualization, Writing - review \& editing.
\textbf{Dimitrios Irodotou, Louise Seeyave, Will Roper}: Writing - review \& editing.

We would like to thank the anonymous referee whose comments significantly improved the paper.

\section*{Data Availability}
A portion of the data used to produce this work can be found online: \href{https://flaresimulations.github.io/\#data.html}{flaresimulations.github.io/\#data}. Much of the analysis used the raw data produced by the simulation which can be made available upon request.


\bibliographystyle{mnras}
\bibliography{bias}

\begin{thebibliography}{}
\makeatletter
\relax
\def\mn@urlcharsother{\let\do\@makeother \do\$\do\&\do\#\do\^\do\_\do\%\do\~}
\def\mn@doi{\begingroup\mn@urlcharsother \@ifnextchar [ {\mn@doi@}
  {\mn@doi@[]}}
\def\mn@doi@[#1]#2{\def\@tempa{#1}\ifx\@tempa\@empty \href
  {http://dx.doi.org/#2} {doi:#2}\else \href {http://dx.doi.org/#2} {#1}\fi
  \endgroup}
\def\mn@eprint#1#2{\mn@eprint@#1:#2::\@nil}
\def\mn@eprint@arXiv#1{\href {http://arxiv.org/abs/#1} {{\tt arXiv:#1}}}
\def\mn@eprint@dblp#1{\href {http://dblp.uni-trier.de/rec/bibtex/#1.xml}
  {dblp:#1}}
\def\mn@eprint@#1:#2:#3:#4\@nil{\def\@tempa {#1}\def\@tempb {#2}\def\@tempc
  {#3}\ifx \@tempc \@empty \let \@tempc \@tempb \let \@tempb \@tempa \fi \ifx
  \@tempb \@empty \def\@tempb {arXiv}\fi \@ifundefined
  {mn@eprint@\@tempb}{\@tempb:\@tempc}{\expandafter \expandafter \csname
  mn@eprint@\@tempb\endcsname \expandafter{\@tempc}}}

\bibitem[\protect\citeauthoryear{{Adams} et~al.,}{{Adams}
  et~al.}{2023a}]{Adams23b}
{Adams} N.~J.,  et~al., 2023a, arXiv e-prints, p. arXiv:2304.13721

\bibitem[\protect\citeauthoryear{{Adams} et~al.,}{{Adams}
  et~al.}{2023b}]{Adams23}
{Adams} N.~J.,  et~al., 2023b, \mn@doi [\mnras] {10.1093/mnras/stac3347}, 518,
  4755

\bibitem[\protect\citeauthoryear{{Angulo}, {Springel}, {White}, {Jenkins},
  {Baugh}  \& {Frenk}}{{Angulo} et~al.}{2012}]{Angulo12}
{Angulo} R.~E.,  {Springel} V.,  {White} S.~D.~M.,  {Jenkins} A.,  {Baugh}
  C.~M.,   {Frenk} C.~S.,  2012, \mn@doi [\mnras]
  {10.1111/j.1365-2966.2012.21830.x}, 426, 2046

\bibitem[\protect\citeauthoryear{{Bagley} et~al.,}{{Bagley}
  et~al.}{2022}]{Ceers22}
{Bagley} M.~B.,  et~al., 2022, \mn@doi [arXiv:2211.02495]
  {10.48550/arXiv.2211.02495}

\bibitem[\protect\citeauthoryear{Bahé et~al.,}{Bahé et~al.}{2017}]{Bahe17}
Bahé Y.~M.,  et~al., 2017, \mn@doi [MNRAS] {10.1093/mnras/stx1403}, 470, 4186

\bibitem[\protect\citeauthoryear{{Bardeen}, {Bond}, {Kaiser}  \&
  {Szalay}}{{Bardeen} et~al.}{1986}]{Bardeen86}
{Bardeen} J.~M.,  {Bond} J.~R.,  {Kaiser} N.,   {Szalay} A.~S.,  1986, \mn@doi
  [\apj] {10.1086/164143}, 304, 15

\bibitem[\protect\citeauthoryear{{Barnes} et~al.,}{{Barnes}
  et~al.}{2017}]{Barnes17}
{Barnes} D.~J.,  et~al., 2017, \mn@doi [MNRAS] {10.1093/mnras/stx1647}, 471,
  1088

\bibitem[\protect\citeauthoryear{{Bhatawdekar}, {Conselice},
  {Margalef-Bentabol}  \& {Duncan}}{{Bhatawdekar} et~al.}{2019}]{Bhatawdekar19}
{Bhatawdekar} R.,  {Conselice} C.~J.,  {Margalef-Bentabol} B.,   {Duncan} K.,
  2019, \mn@doi [\mnras] {10.1093/mnras/stz866}, 486, 3805

\bibitem[\protect\citeauthoryear{{Casey} et~al.,}{{Casey}
  et~al.}{2022}]{CosmosWeb22}
{Casey} C.~M.,  et~al., 2022, \mn@doi [arXiv:2211.07865]
  {10.48550/arXiv.2211.07865}

\bibitem[\protect\citeauthoryear{{Castellano} et~al.,}{{Castellano}
  et~al.}{2022}]{Castellano22}
{Castellano} M.,  et~al., 2022, \mn@doi [arXiv e-prints]
  {10.48550/arXiv.2212.06666}, p. arXiv:2212.06666

\bibitem[\protect\citeauthoryear{Crain et~al.,}{Crain et~al.}{2009}]{Crain09}
Crain R.~A.,  et~al., 2009, \mn@doi [MNRAS] {10.1111/j.1365-2966.2009.15402.x},
  399, 1773

\bibitem[\protect\citeauthoryear{{Curtis-Lake} et~al.,}{{Curtis-Lake}
  et~al.}{2022}]{Curtis-Lake22}
{Curtis-Lake} E.,  et~al., 2022, \mn@doi [arXiv:2212.04568]
  {10.48550/arXiv.2212.04568}

\bibitem[\protect\citeauthoryear{Davé, Anglés-Alcázar, Narayanan, Li,
  Rafieferantsoa  \& Appleby}{Davé et~al.}{2019}]{Dave19}
Davé R.,  Anglés-Alcázar D.,  Narayanan D.,  Li Q.,  Rafieferantsoa M.~H.,
  Appleby S.,  2019, \mn@doi [MNRAS] {10.1093/mnras/stz937}, 486, 2827

\bibitem[\protect\citeauthoryear{{Desjacques}, {Jeong}  \&
  {Schmidt}}{{Desjacques} et~al.}{2018}]{Desjacques18}
{Desjacques} V.,  {Jeong} D.,   {Schmidt} F.,  2018, \mn@doi [\physrep]
  {10.1016/j.physrep.2017.12.002}, 733, 1

\bibitem[\protect\citeauthoryear{Donnan et~al.,}{Donnan
  et~al.}{2022}]{Donnan22}
Donnan C.~T.,  et~al., 2022, \mn@doi [arXiv:2207.12356]
  {10.48550/arXiv.2207.12356}

\bibitem[\protect\citeauthoryear{{Duncan} et~al.,}{{Duncan}
  et~al.}{2014}]{Duncan14}
{Duncan} K.,  et~al., 2014, \mn@doi [\mnras] {10.1093/mnras/stu1622}, 444, 2960

\bibitem[\protect\citeauthoryear{{Dunlop} et~al.,}{{Dunlop}
  et~al.}{2021}]{Primer21}
{Dunlop} J.~S.,  et~al., 2021, {PRIMER: Public Release IMaging for
  Extragalactic Research}, JWST Proposal. Cycle 1, ID. \#1837

\bibitem[\protect\citeauthoryear{{Einasto}, {Liivam{\"a}gi}  \&
  {Einasto}}{{Einasto} et~al.}{2023}]{Einasto23}
{Einasto} J.,  {Liivam{\"a}gi} L.~J.,   {Einasto} M.,  2023, \mn@doi [\mnras]
  {10.1093/mnras/stac3181}, 518, 2164

\bibitem[\protect\citeauthoryear{Feng, Di-Matteo, Croft, Bird, Battaglia  \&
  Wilkins}{Feng et~al.}{2015}]{Feng15}
Feng Y.,  Di-Matteo T.,  Croft R.~A.,  Bird S.,  Battaglia N.,   Wilkins S.,
  2015, \mn@doi [MNRAS] {10.1093/mnras/stv2484}, 455, 2778

\bibitem[\protect\citeauthoryear{{Finkelstein} et~al.,}{{Finkelstein}
  et~al.}{2023}]{Finkelstein23}
{Finkelstein} S.~L.,  et~al., 2023, \mn@doi [\apjl] {10.3847/2041-8213/acade4},
  946, L13

\bibitem[\protect\citeauthoryear{Genel et~al.,}{Genel et~al.}{2014}]{Genel14}
Genel S.,  et~al., 2014, \mn@doi [MNRAS] {10.1093/mnras/stu1654}, 445, 175

\bibitem[\protect\citeauthoryear{{Gonz{\'a}lez}, {Labb{\'e}}, {Bouwens},
  {Illingworth}, {Franx}  \& {Kriek}}{{Gonz{\'a}lez} et~al.}{2011}]{Gonzalez11}
{Gonz{\'a}lez} V.,  {Labb{\'e}} I.,  {Bouwens} R.~J.,  {Illingworth} G.,
  {Franx} M.,   {Kriek} M.,  2011, \mn@doi [\apjl]
  {10.1088/2041-8205/735/2/L34}, 735, L34

\bibitem[\protect\citeauthoryear{Harikane et~al.,}{Harikane
  et~al.}{2022}]{Harikane22}
Harikane Y.,  et~al., 2022, \mn@doi [arXiv:2208.01612]
  {10.48550/arXiv.2208.01612}

\bibitem[\protect\citeauthoryear{{Harikane}, {Nakajima}, {Ouchi}, {Umeda},
  {Isobe}, {Ono}, {Xu}  \& {Zhang}}{{Harikane} et~al.}{2023}]{Harikane23}
{Harikane} Y.,  {Nakajima} K.,  {Ouchi} M.,  {Umeda} H.,  {Isobe} Y.,  {Ono}
  Y.,  {Xu} Y.,   {Zhang} Y.,  2023, \mn@doi [arXiv e-prints]
  {10.48550/arXiv.2304.06658}, p. arXiv:2304.06658

\bibitem[\protect\citeauthoryear{Hunter}{Hunter}{2007}]{Matplotlib07}
Hunter J.~D.,  2007, \mn@doi [Computing in Science \& Engineering]
  {10.1109/MCSE.2007.55}, 9, 90

\bibitem[\protect\citeauthoryear{{Kaiser}}{{Kaiser}}{1984}]{Kaiser84}
{Kaiser} N.,  1984, \mn@doi [\apjl] {10.1086/184341}, 284, L9

\bibitem[\protect\citeauthoryear{{Katz} \& {White}}{{Katz} \&
  {White}}{1993}]{Katz93}
{Katz} N.,  {White} S. D.~M.,  1993, \mn@doi [\apj] {10.1086/172935}, 412, 455

\bibitem[\protect\citeauthoryear{{Kim}, {Park}, {Gott}  \& {Dubinski}}{{Kim}
  et~al.}{2009}]{Kim09}
{Kim} J.,  {Park} C.,  {Gott} J.~Richard I.,   {Dubinski} J.,  2009, \mn@doi
  [\apj] {10.1088/0004-637X/701/2/1547}, 701, 1547

\bibitem[\protect\citeauthoryear{Kitzbichler \& White}{Kitzbichler \&
  White}{2007}]{Kitzbichler07}
Kitzbichler M.~G.,  White S. D.~M.,  2007, \mn@doi [\mnras]
  {10.1111/j.1365-2966.2007.11458.x}, 376, 2

\bibitem[\protect\citeauthoryear{{Knebe} et~al.,}{{Knebe}
  et~al.}{2018}]{Knebe18}
{Knebe} A.,  et~al., 2018, \mn@doi [\mnras] {10.1093/mnras/stx3274}, 475, 2936

\bibitem[\protect\citeauthoryear{Labbe et~al.,}{Labbe et~al.}{2022}]{Labbe22}
Labbe I.,  et~al., 2022, \mn@doi [arXiv:2207.12446] {10.48550/arXiv.2207.12446}

\bibitem[\protect\citeauthoryear{{Lovell}, {Vijayan}, {Thomas}, {Wilkins},
  {Barnes}, {Irodotou}  \& {Roper}}{{Lovell} et~al.}{2021}]{Flares1}
{Lovell} C.~C.,  {Vijayan} A.~P.,  {Thomas} P.~A.,  {Wilkins} S.~M.,  {Barnes}
  D.~J.,  {Irodotou} D.,   {Roper} W.,  2021, MNRAS, 500, 2127

\bibitem[\protect\citeauthoryear{Maksimova, Garrison, Eisenstein, Hadzhiyska,
  Bose  \& Satterthwaite}{Maksimova et~al.}{2021}]{Maksimova21}
Maksimova N.~A.,  Garrison L.~H.,  Eisenstein D.~J.,  Hadzhiyska B.,  Bose S.,
   Satterthwaite T.~P.,  2021, \mn@doi [MNRAS] {10.1093/mnras/stab2484}, 508,
  4017

\bibitem[\protect\citeauthoryear{{McLeod} et~al.,}{{McLeod}
  et~al.}{2023}]{McLeod23}
{McLeod} D.~J.,  et~al., 2023, \mn@doi [arXiv e-prints]
  {10.48550/arXiv.2304.14469}, p. arXiv:2304.14469

\bibitem[\protect\citeauthoryear{Moster, Somerville, Newman  \& Rix}{Moster
  et~al.}{2011}]{Moster11}
Moster B.~P.,  Somerville R.~S.,  Newman J.~A.,   Rix H.-W.,  2011, \mn@doi
  [ApJ] {10.1088/0004-637X/731/2/113}, 731, 113

\bibitem[\protect\citeauthoryear{Naidu et~al.,}{Naidu et~al.}{2022}]{Naidu22}
Naidu R.~P.,  et~al., 2022, \mn@doi [arXiv:2207.09434]
  {10.48550/arXiv.2207.09434}

\bibitem[\protect\citeauthoryear{Nelson et~al.,}{Nelson
  et~al.}{2017}]{Nelson17}
Nelson D.,  et~al., 2017, \mn@doi [MNRAS] {10.1093/mnras/stx3040}, 475, 624

\bibitem[\protect\citeauthoryear{Newman \& Davis}{Newman \&
  Davis}{2002}]{Newman02}
Newman J.~A.,  Davis M.,  2002, \mn@doi [ApJ] {10.1086/324148}, 564, 567

\bibitem[\protect\citeauthoryear{Pillepich et~al.,}{Pillepich
  et~al.}{2017}]{Pillepich17}
Pillepich A.,  et~al., 2017, \mn@doi [MNRAS] {10.1093/mnras/stx3112}, 475, 648

\bibitem[\protect\citeauthoryear{{Rieke}}{{Rieke}}{2020}]{Jades20}
{Rieke} M.,  2020, in {da Cunha} E.,  {Hodge} J.,  {Afonso} J.,  {Pentericci}
  L.,   {Sobral} D.,  eds,  Proceedings of the International Astronomical Union
  Vol. 352, Uncovering Early Galaxy Evolution in the ALMA and JWST Era. pp
  337--341, \mn@doi{10.1017/S1743921319008950}

\bibitem[\protect\citeauthoryear{{Robertson} et~al.,}{{Robertson}
  et~al.}{2022}]{Robertson22}
{Robertson} B.~E.,  et~al., 2022, \mn@doi [arXiv:2212.04480]
  {10.48550/arXiv.2212.04480}

\bibitem[\protect\citeauthoryear{Robitaille et~al.,}{Robitaille
  et~al.}{2013}]{AstroPy13}
Robitaille T.~P.,  et~al., 2013, \mn@doi [A\&A] {10.1051/0004-6361/201322068},
  558, A33

\bibitem[\protect\citeauthoryear{Rodighiero, Bisigello, Iani, Marasco, Grazian,
  Sinigaglia, Cassata  \& Gruppioni}{Rodighiero et~al.}{2022}]{Rodighiero22}
Rodighiero G.,  Bisigello L.,  Iani E.,  Marasco A.,  Grazian A.,  Sinigaglia
  F.,  Cassata P.,   Gruppioni C.,  2022, \mn@doi [MNRAS]
  {10.1093/mnrasl/slac115}, 518, L19

\bibitem[\protect\citeauthoryear{{Roper}, {Lovell}, {Vijayan}, {Marshall},
  {Irodotou}, {Kuusisto}, {Thomas}  \& {Wilkins}}{{Roper}
  et~al.}{2022}]{Roper22}
{Roper} W.~J.,  {Lovell} C.~C.,  {Vijayan} A.~P.,  {Marshall} M.~A.,
  {Irodotou} D.,  {Kuusisto} J.~K.,  {Thomas} P.~A.,   {Wilkins} S.~M.,  2022,
  \mn@doi [\mnras] {10.1093/mnras/stac1368}, 514, 1921

\bibitem[\protect\citeauthoryear{{Schaller}, {Dalla Vecchia}, {Schaye},
  {Bower}, {Theuns}, {Crain}, {Furlong}  \& {McCarthy}}{{Schaller}
  et~al.}{2015}]{Schaller15}
{Schaller} M.,  {Dalla Vecchia} C.,  {Schaye} J.,  {Bower} R.~G.,  {Theuns} T.,
   {Crain} R.~A.,  {Furlong} M.,   {McCarthy} I.~G.,  2015, \mn@doi [MNRAS]
  {10.1093/mnras/stv2169}, 454, 2277

\bibitem[\protect\citeauthoryear{Schaye et~al.,}{Schaye et~al.}{2015}]{Eagle15}
Schaye J.,  et~al., 2015, \mn@doi [MNRAS] {10.1093/mnras/stu2058}, 446, 521

\bibitem[\protect\citeauthoryear{Somerville, Lee, Ferguson, Gardner, Moustakas
  \& Giavalisco}{Somerville et~al.}{2004}]{Somerville04}
Somerville R.~S.,  Lee K.,  Ferguson H.~C.,  Gardner J.~P.,  Moustakas L.~A.,
  Giavalisco M.,  2004, \mn@doi [ApJ] {10.1086/378628}, 600, L171

\bibitem[\protect\citeauthoryear{{Song} et~al.,}{{Song} et~al.}{2016}]{Song16}
{Song} M.,  et~al., 2016, \mn@doi [\apj] {10.3847/0004-637X/825/1/5}, 825, 5

\bibitem[\protect\citeauthoryear{{Springel} et~al.,}{{Springel}
  et~al.}{2005}]{Springel05}
{Springel} V.,  et~al., 2005, \mn@doi [\nat] {10.1038/nature03597}, 435, 629

\bibitem[\protect\citeauthoryear{Stark, Loeb  \& Ellis}{Stark
  et~al.}{2007}]{Stark07}
Stark D.~P.,  Loeb A.,   Ellis R.~S.,  2007, \mn@doi [ApJ] {10.1086/520947},
  668, 627

\bibitem[\protect\citeauthoryear{{Stefanon}, {Bouwens}, {Labb{\'e}}, {Muzzin},
  {Marchesini}, {Oesch}  \& {Gonzalez}}{{Stefanon} et~al.}{2017}]{Stefanon17}
{Stefanon} M.,  {Bouwens} R.~J.,  {Labb{\'e}} I.,  {Muzzin} A.,  {Marchesini}
  D.,  {Oesch} P.,   {Gonzalez} V.,  2017, \mn@doi [\apj]
  {10.3847/1538-4357/aa72d8}, 843, 36

\bibitem[\protect\citeauthoryear{{Trapp} \& {Furlanetto}}{{Trapp} \&
  {Furlanetto}}{2020}]{Trapp20}
{Trapp} A.~C.,  {Furlanetto} S.~R.,  2020, \mn@doi [\mnras]
  {10.1093/mnras/staa2828}, 499, 2401

\bibitem[\protect\citeauthoryear{Trapp, Furlanetto  \& Yang}{Trapp
  et~al.}{2022}]{Trapp21}
Trapp A.~C.,  Furlanetto S.~R.,   Yang J.,  2022, \mn@doi [MNRAS]
  {10.1093/mnras/stab3801}, 510, 4844

\bibitem[\protect\citeauthoryear{Trenti \& Stiavelli}{Trenti \&
  Stiavelli}{2008}]{Trenti08}
Trenti M.,  Stiavelli M.,  2008, \mn@doi [The Astrophysical Journal]
  {10.1086/528674}, 676, 767

\bibitem[\protect\citeauthoryear{{Treu} et~al.,}{{Treu} et~al.}{2022}]{Treu22}
{Treu} T.,  et~al., 2022, \mn@doi [\apj] {10.3847/1538-4357/ac8158}, 935, 110

\bibitem[\protect\citeauthoryear{{Vijayan}, {Lovell}, {Wilkins}, {Thomas},
  {Barnes}, {Irodotou}, {Kuusisto}  \& {Roper}}{{Vijayan}
  et~al.}{2021}]{Vijayan21}
{Vijayan} A.~P.,  {Lovell} C.~C.,  {Wilkins} S.~M.,  {Thomas} P.~A.,  {Barnes}
  D.~J.,  {Irodotou} D.,  {Kuusisto} J.,   {Roper} W.~J.,  2021, \mn@doi
  [\mnras] {10.1093/mnras/staa3715}, 501, 3289

\bibitem[\protect\citeauthoryear{{Virtanen} et~al.,}{{Virtanen}
  et~al.}{2020}]{SciPy20}
{Virtanen} P.,  et~al., 2020, \mn@doi [Nature Methods]
  {https://doi.org/10.1038/s41592-019-0686-2}, 17, 261

\bibitem[\protect\citeauthoryear{{Williams} et~al.,}{{Williams}
  et~al.}{2018}]{Williams18}
{Williams} C.~C.,  et~al., 2018, \mn@doi [\apjs] {10.3847/1538-4365/aabcbb},
  236, 33

\bibitem[\protect\citeauthoryear{Zeldovich, Einasto  \& Shandarin}{Zeldovich
  et~al.}{1982}]{Zeldovich82}
Zeldovich Y.~B.,  Einasto J.,   Shandarin S.~F.,  1982, \mn@doi [Nature]
  {10.1038/300407a0}, 300, 407

\makeatother
\end{thebibliography}


\appendix

\section{Variance in mean number count predictions}
\label{sec:variance:ngal}

In this section we investigate whether, for rare galaxies, the limited number of grid cells which are populated can lead to excess variance in number count predictions, over and above that expected from Poisson variation.

\fig{stars_DM_od_raw} shows that we have a fair sample of star formation density within grid cells at all overdensities. However, the matter under question here is whether the same is true of the number counts of galaxies.  To test this we use bootstrap resampling to estimate the variance in our results that might arise from the rarity of massive galaxies.  We perform that test on a sample with $M>10^9\Msun$ at $z=10$ which has a mean number count of just 1 in a 300\,arcmin$^2$ area.

\begin{figure}
  \includegraphics[width=8.7cm]{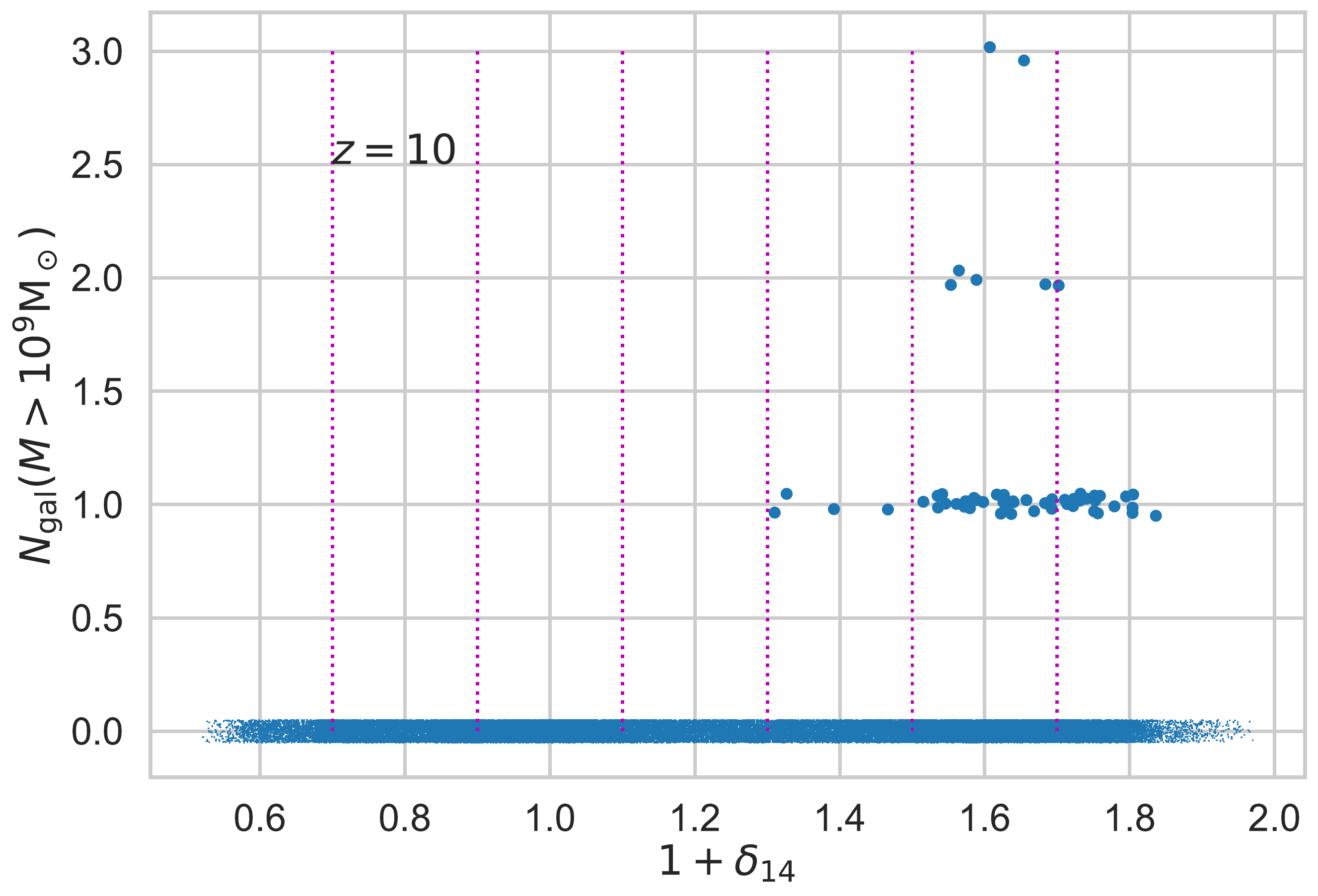}
  \caption{The number of galaxies of galaxies in each grid cell versus the overdensity of that grid cell, for $M>10^9\Msun$ at $z=10$.  The number of galaxies is integral but has been given a small dispersion so that individual points are clearly visible on the plot.}
  \label{fig:ngal_od}
\end{figure}

\fig{ngal_od} shows the number of galaxies within a single grid cell versus the overdensity of the cell in the DM-only run.  The number of galaxies is integral but has been given a small dispersion so that they are clearly visible on the plot; likewise the size of the symbols has been reduced to a single pixel for those grid cells with 0 galaxies.  It can be seen that we have chosen an example where almost all grid cells have zero galaxies and so the number counts are dominated by a few grid cells: there are 51, 5 \& 2 cells out of 56,960 that have 1, 2 \& 3 galaxies, respectively.

\begin{figure}
  \includegraphics[width=8.7cm]{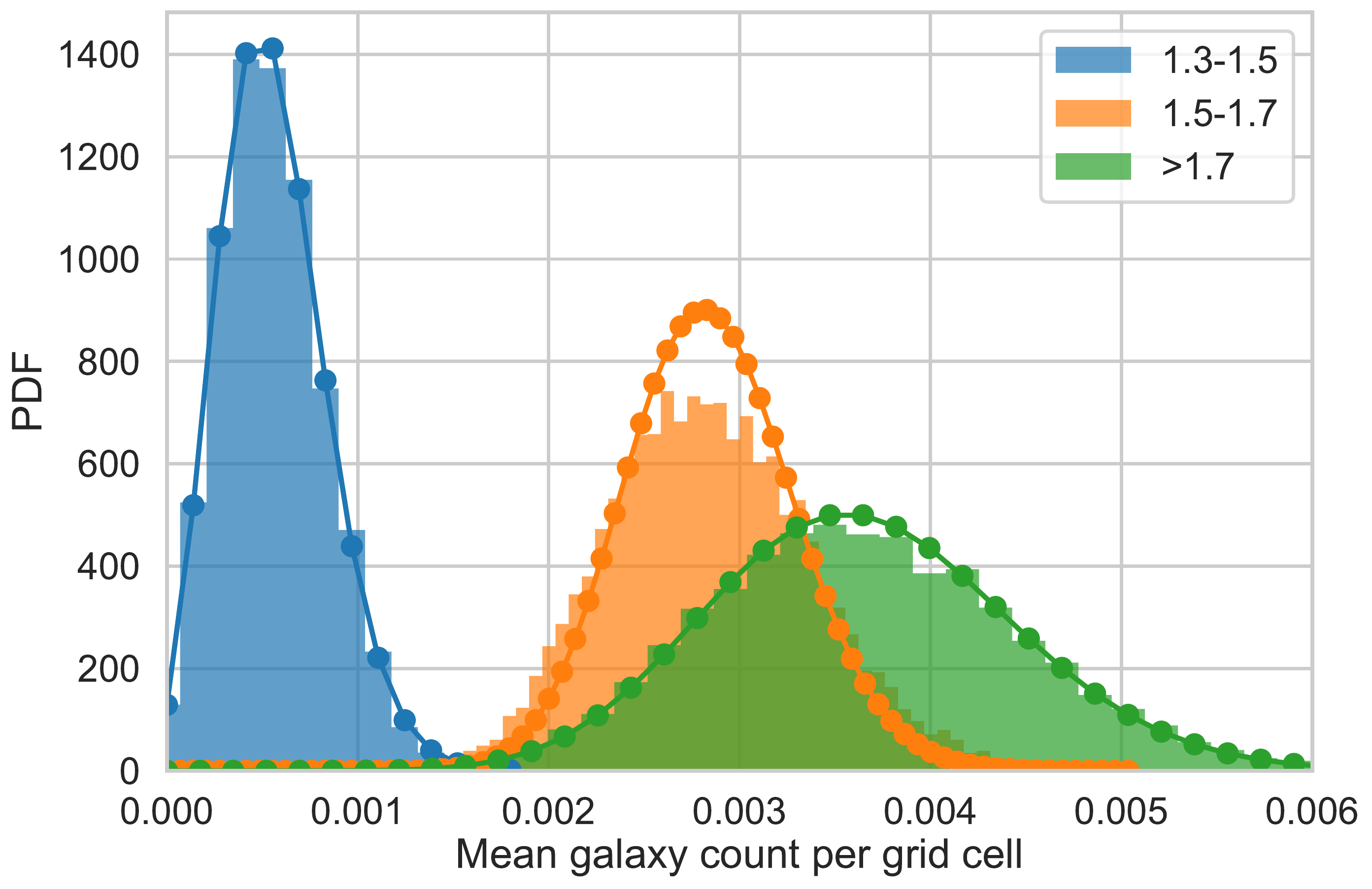}
  \caption{The histograms show the probability distribution functions of the expected number of counts per grid cell within different density bins.  The solid lines and points show the Poisson distributions with the same mean number of galaxies.}
  \label{fig:ngal_pdf}
\end{figure}

We estimate the sample variance that this would introduce by performing a bootstrap analysis within each density bin: we draw 10,000 samples of equal size to the original one, but including replacement, with the distribution of galaxy counts shown in \fig{ngal_pdf}. The mean within each density bin is unchanged and matches the value used in the paper: it is zero for low density bins and 4, 41 and 21, respectively, for the three highest density bins.  

The solid lines and data points show the equivalent Poisson distributions.  As can be seen, the data for the density bins $1.3< 1+\delta_{14}< 1.5$ and $1+\delta_{14}>1.7$ almost exactly match that of the Poisson distribution.  That for the bin $1.5< 1+\delta_{14}< 1.7$ is a little wider: this presumably reflects the influence of the two grid cells with 3 galaxies within them.  However, even there the difference is not overly large with a measured standard deviation of 7.8 galaxies (summed over all grid cells in the density bin) compared to the Poisson value of 6.4.

When we move to much larger galaxy numbers, then the measured and Poisson predictions become indistinguishable.

We conclude that the finite sampling is not significantly affecting our results.

\section{Comparison to cosmic variance calculator}
\label{sec:variance:cvc}

This section compares our results to that of the Cosmic Variance Calculator (CVC) of \citet{Trenti08}.  That uses analytic estimates via the two-point correlation function in extended Press-Schechter theory, as well as synthetic catalogs extracted from N-body cosmological simulations of structure formation.  However the largest box is just 160\,$h^{-1}\,$cMpc on a side and thus lacking the volume required to sample extreme density fluctuations, especially at high redshift.  Moreover they use a simple Halo Occupation Distribution model to populate the halos with galaxies, rather than galaxy formation simulations as are used in \flares{}.  We might therefore expect a larger variance in this current work than in the CVC.

\begin{figure*}
  \includegraphics[width=17.4cm]{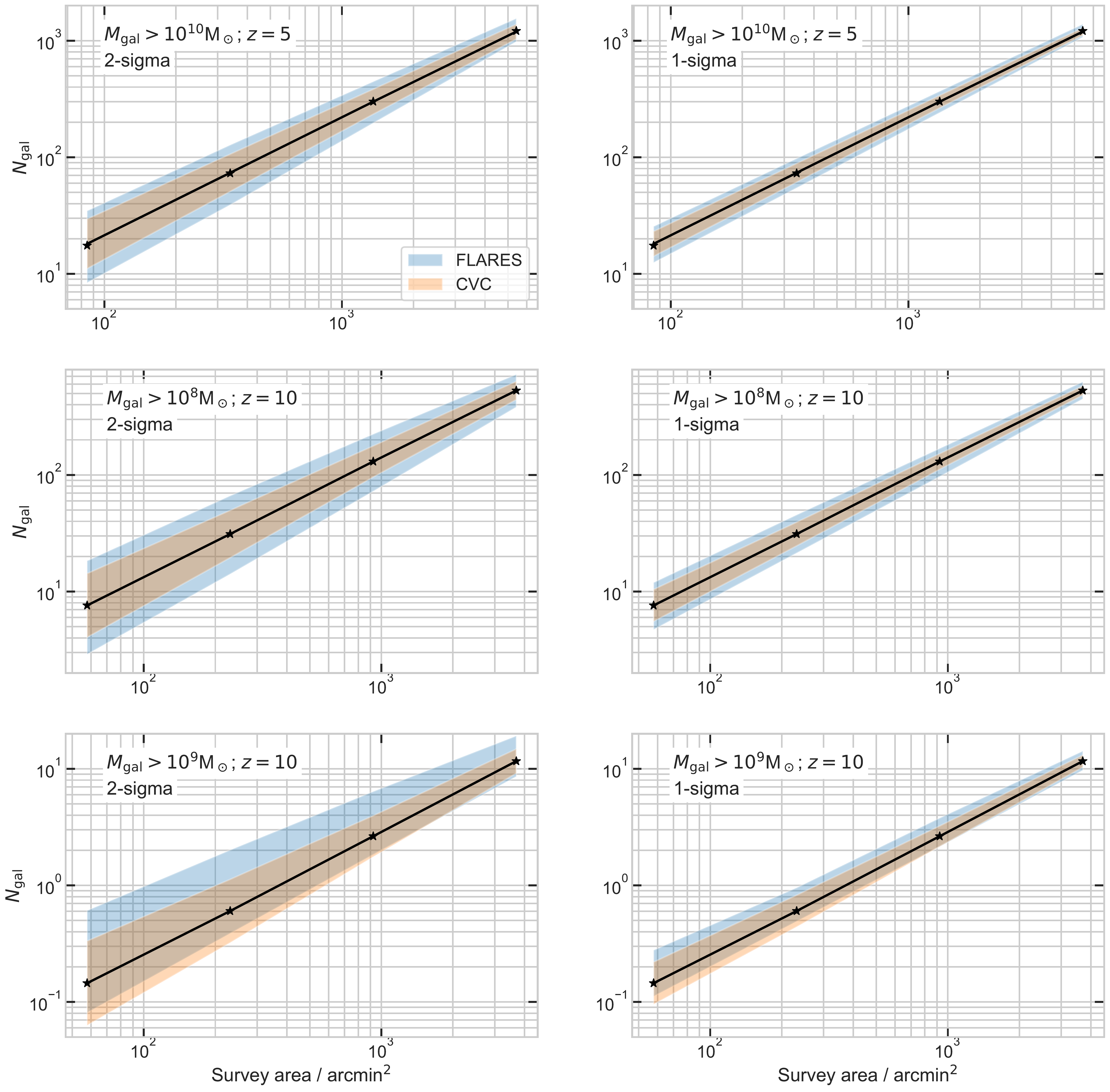}
  \caption{The distribution of predicted galaxy number counts in different square survey areas for this study (\flares{}) and the cosmic variance calculator (CVC).  The left/right columns show the 2/1-sigma dispersions; the rows give examples of galaxy selections with differing mean number counts and redshifts.}
  \label{fig:variance_cvc_mass}
\end{figure*}

\Fig{variance_cvc_mass} contrasts the variance of the predictions from \flares{} with that from the CVC for square survey area.  At all redshifts (greater than 5) and number counts, \flares{} has a higher variance.  A similar plot for the rectangular (16x1 aspect ratio) survey areas, not shown here, is similar but with slightly reduced spread.  The ratio of the standard deviation of the \flares{} predictions to that of the CVC varies from about 1.3 (large number counts) to 1.8 (low number counts).

The CVC just returns a value for the standard deviation of the distribution, although we note that Figure~1 of \citet{Trenti08} shows that the distribution is skewed.  In \flares{} this skewness is apparent both in the number count histograms of \fig{hist_mstar}\footnote{and of Figures~\ref{fig:hist_SFR}, \ref{fig:hist_mstar_large} \& \ref{fig:hist_SFR_large} below} and also in \Fig{variance_cvc_mass}, especially in the bottom row.  We note that this skewness is less apparent in the middle row of the latter figure, which suggests that it the low number counts rather than high redshift which is the main driver.

\section{Survey variance in SFR}
\label{sec:variance:SFR}

\begin{figure}
  \includegraphics[width=8.7cm]{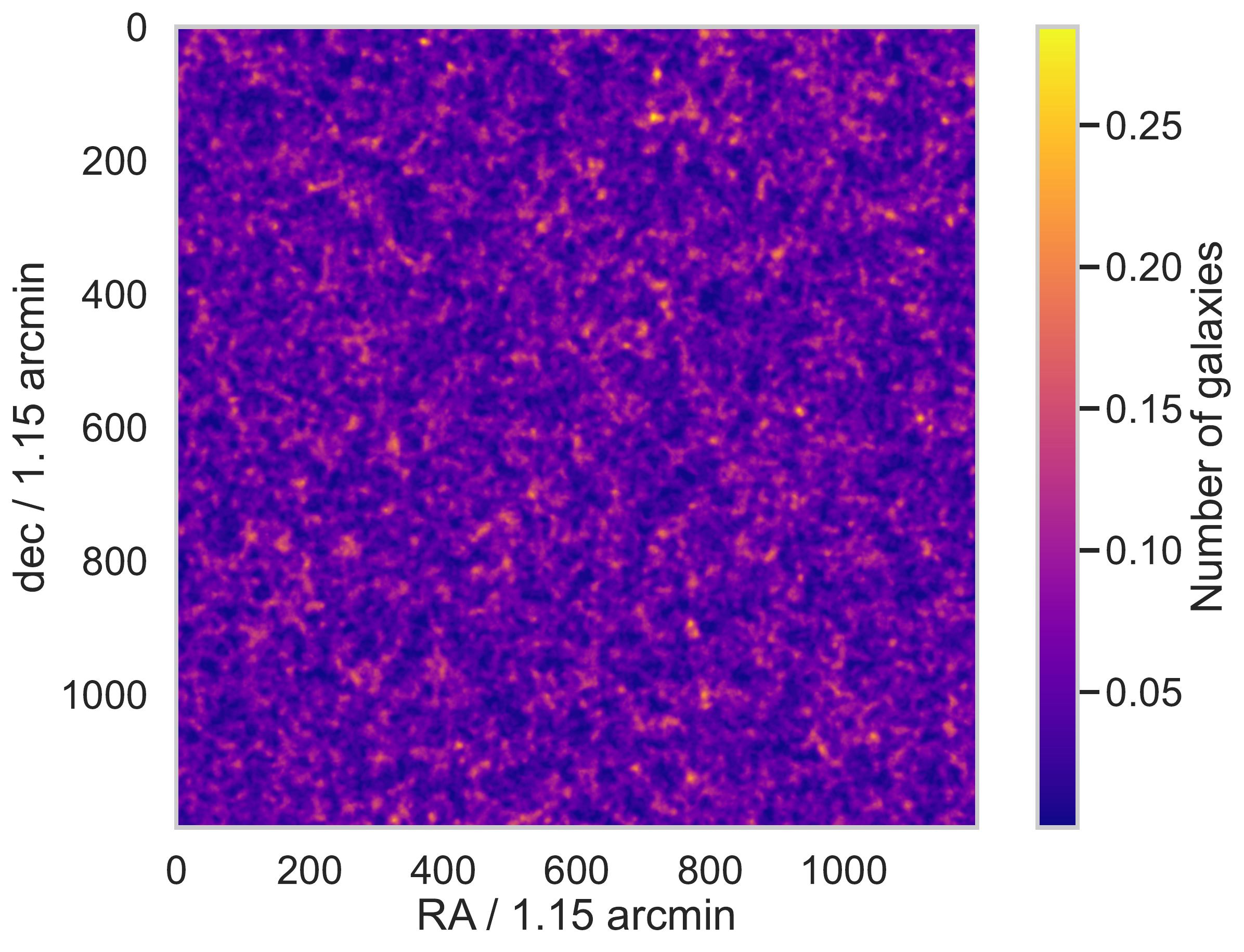}
  \includegraphics[width=8.7cm]{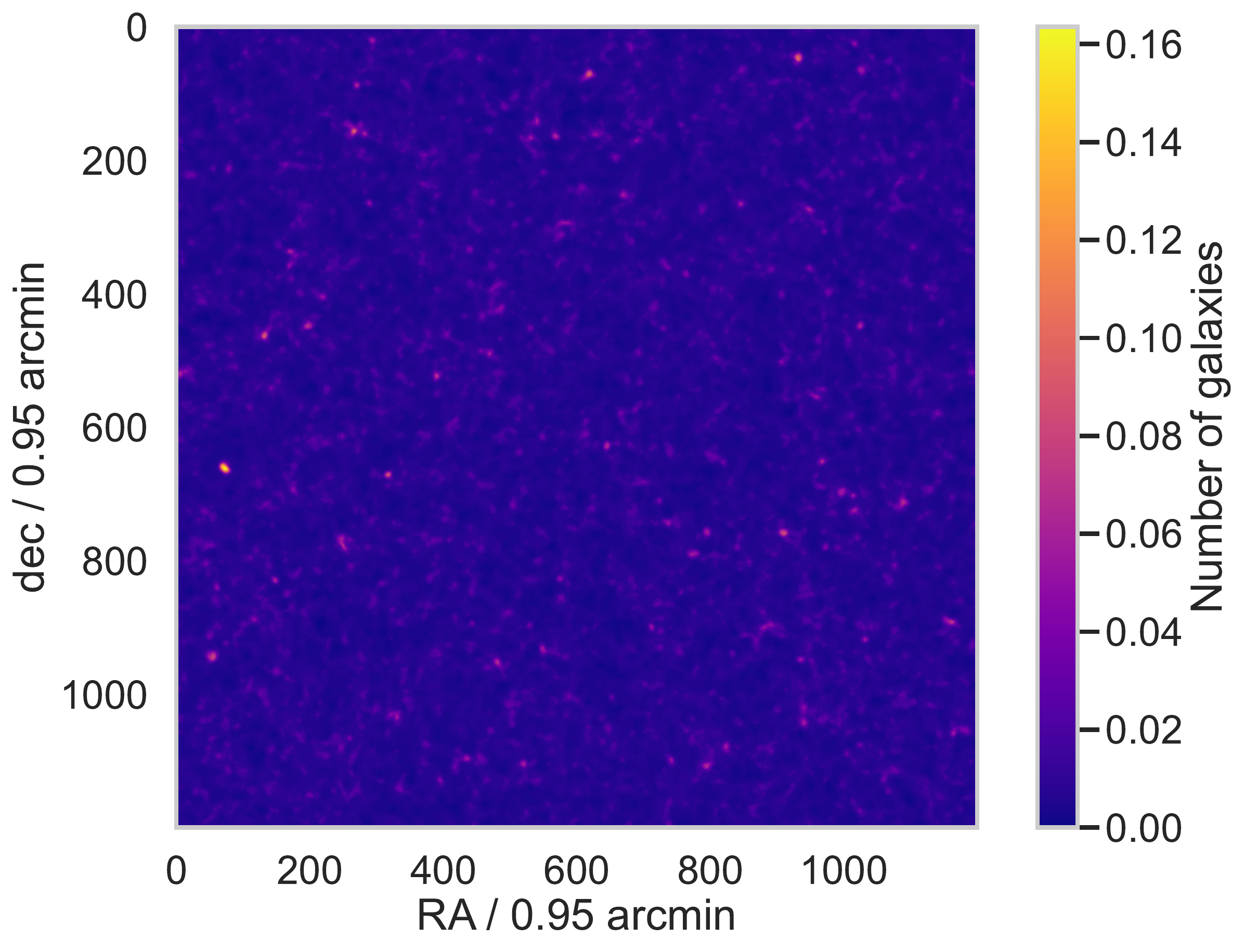}
  \caption{The number of galaxies of star formation rate SFR\,$>100\,\Msun\,$yr$^{-1}$ in the redshift slice $4.5<z\leq5.5$ (upper panel), or SFR\,$>10\,\Msun\,$yr$^{-1}$ in the redshift slice $9.5<z\leq10.5$ (lower panel) per projected grid cell.}
  \label{fig:map_SFR}
\end{figure}

\begin{figure*}
  \includegraphics[width=8.7cm]{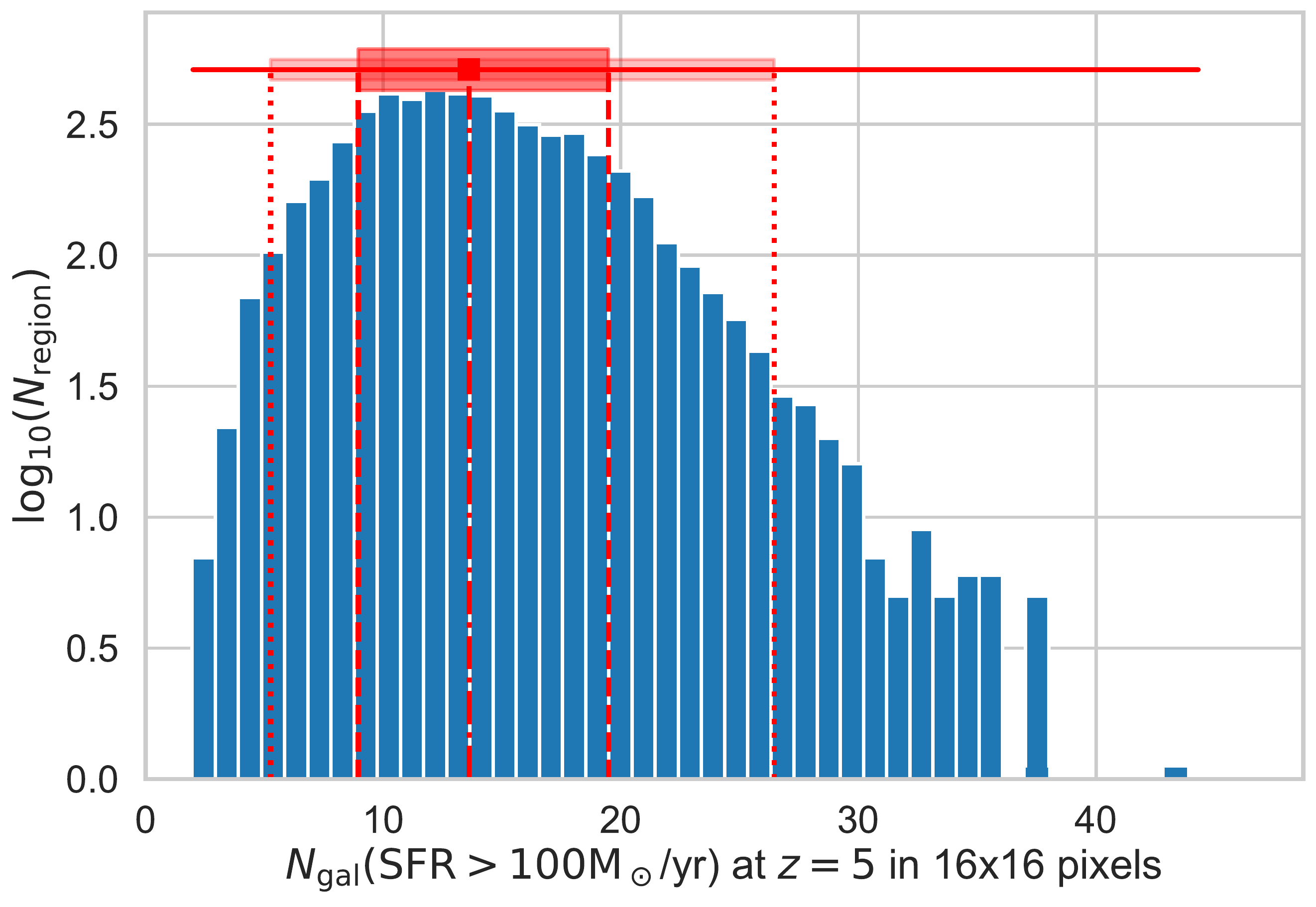}\includegraphics[width=8.7cm]{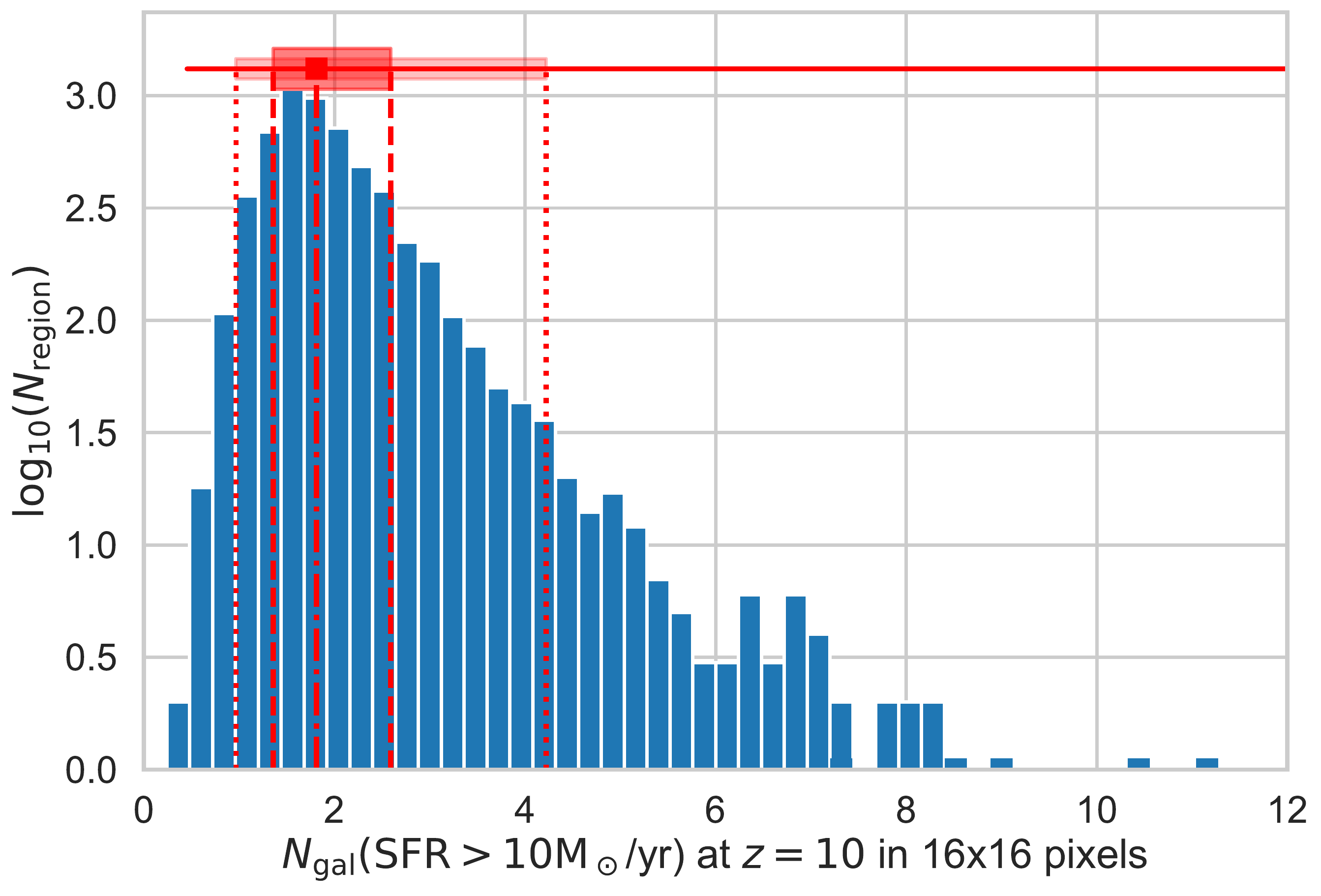}
  \includegraphics[width=8.7cm]{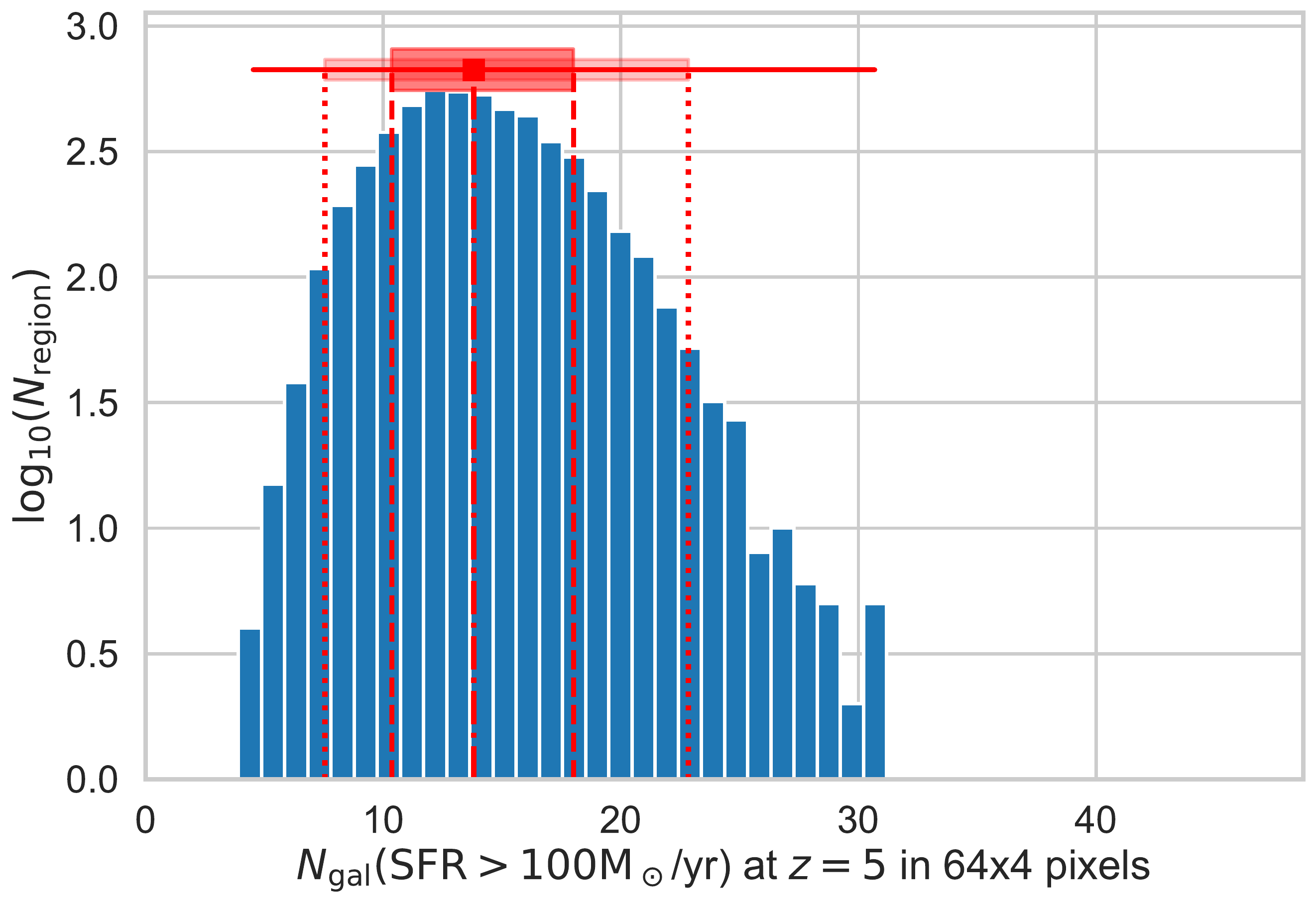}\includegraphics[width=8.7cm]{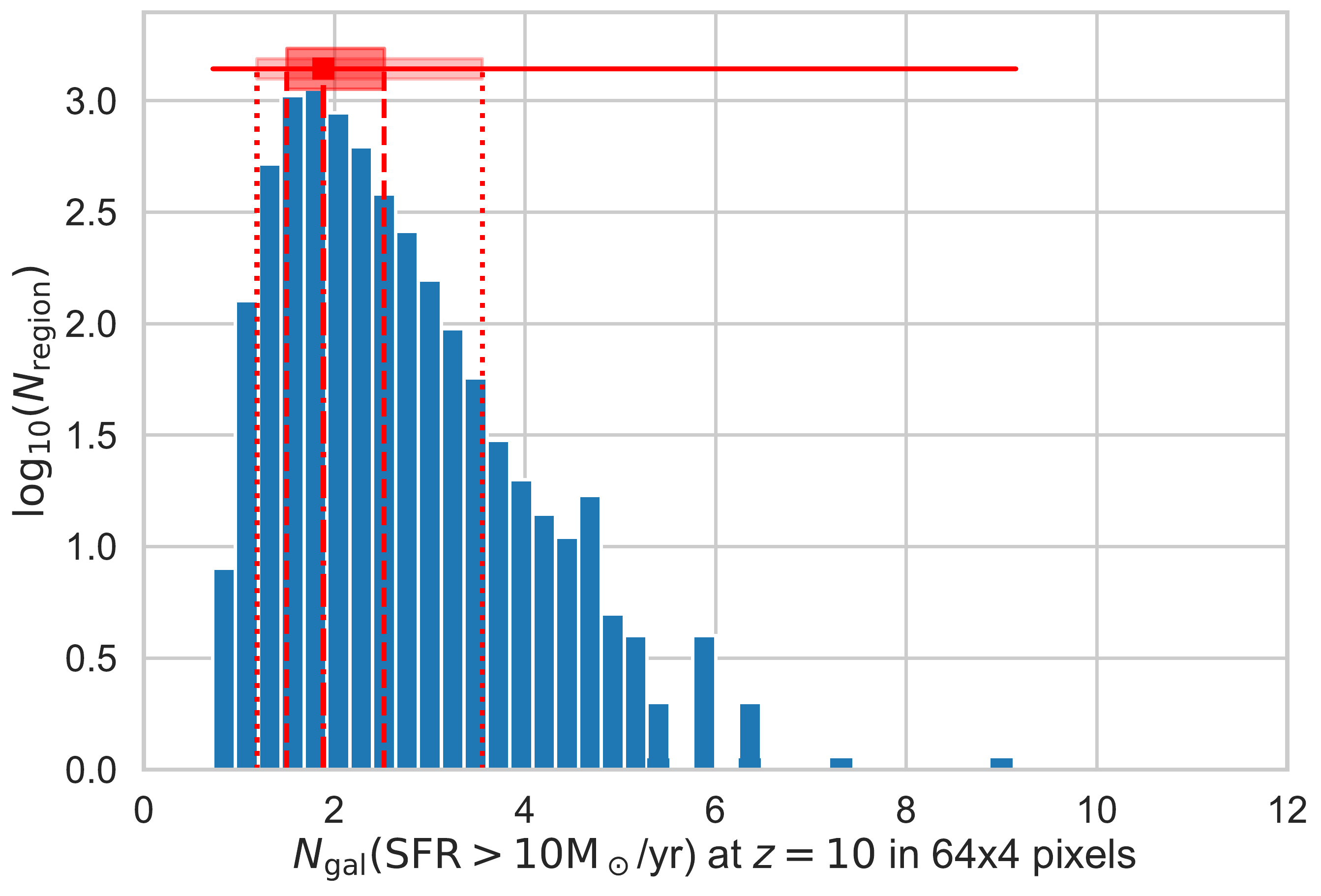}
  \includegraphics[width=8.7cm]{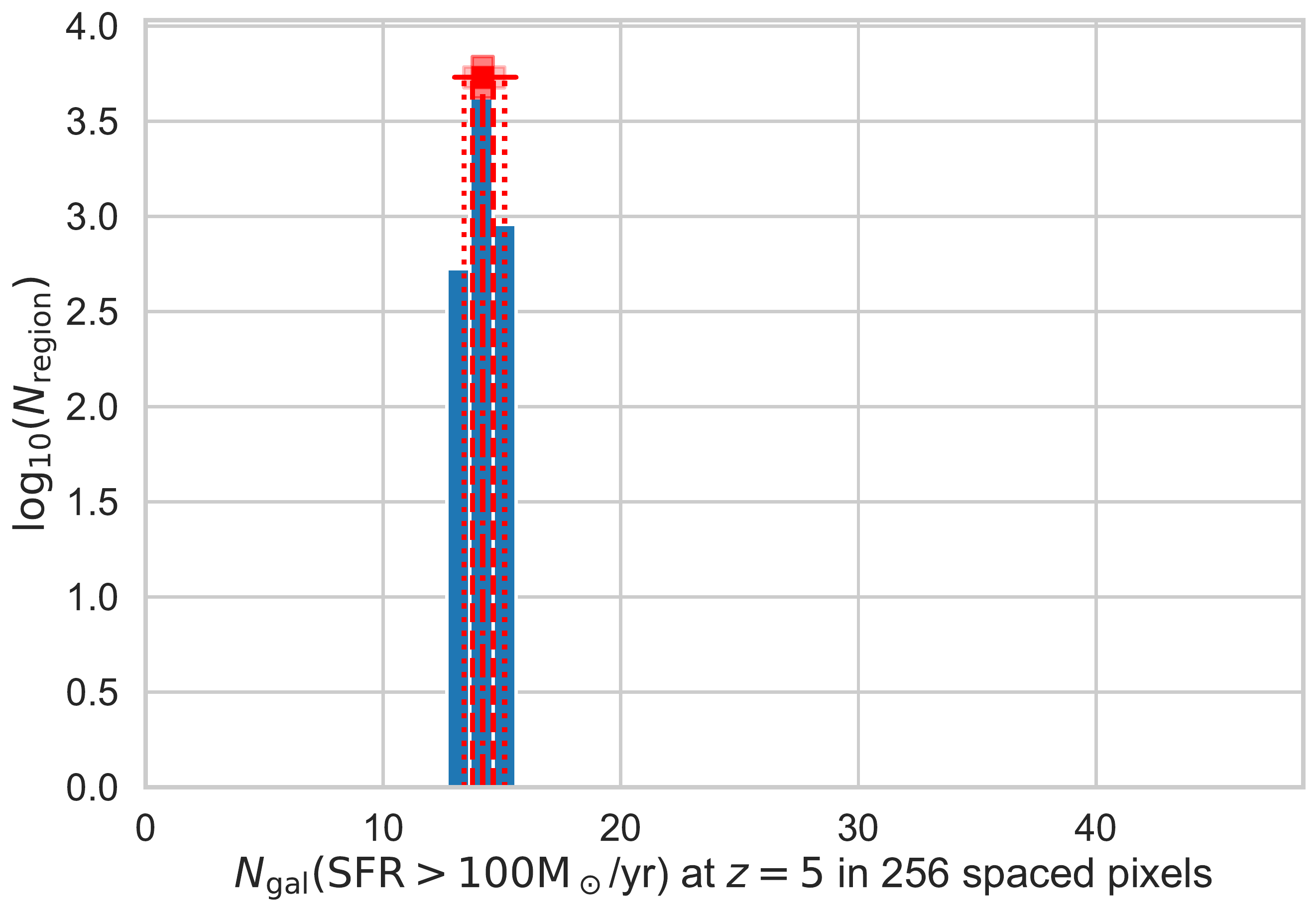}\includegraphics[width=8.7cm]{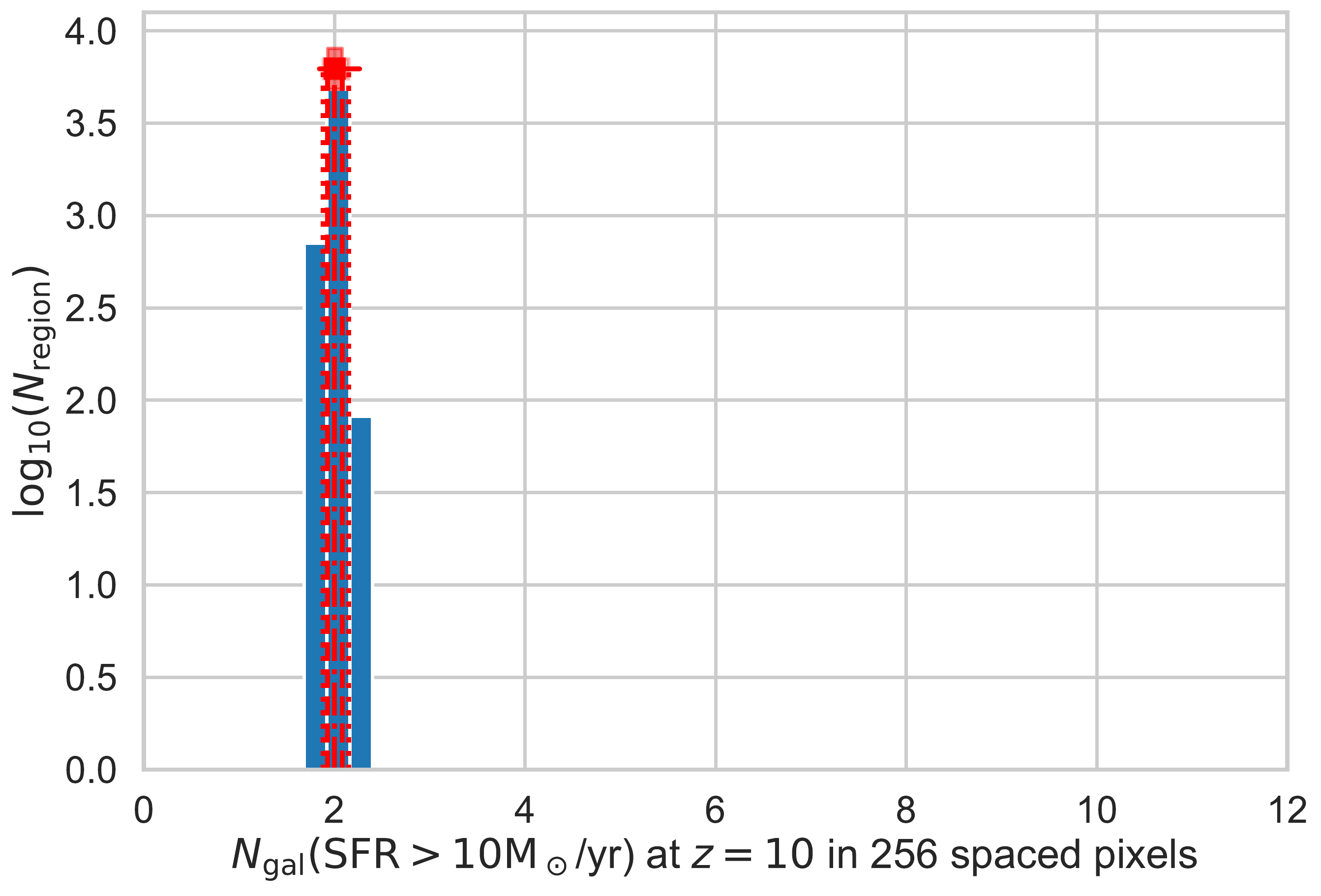}
  \caption{Histograms of the number of galaxies within a 256 grid cell ($\sim300\,$arcmin$^2$) survey region, above a particular star formation rate and in a given redshift slice, according to the geometry of the survey: left column -- $M_*>100\,\Msun\,$yr$^{-1}$, $4.5<z\leq5.5$; right column -- $M_*>10\,\Msun\,$yr$^{-1}$, $9.5<z\leq10.5$; upper row -- 16\,x\,16; middle row -- 256\,x\,1; lower row -- 256 widely spaced grid cells. The dot-dashed, dashed and dotted lines show the median, one-sigma and two-sigma ranges, respectively; the box-plot shows the full extent of the data, plus the one and two-sigma ranges.  In the top, right-hand panel a single point with $N=19.2$ has been omitted, for clarity.}
  \label{fig:hist_SFR}
\end{figure*}

Here we repeat the results of \Sec{variance:results} but for galaxies above a particular star formation rate rather than stellar mass.  The results presented in \Fig{map_SFR} and \Fig{hist_SFR} are qualitatively very similar to those seen in \Fig{map_mstar} and \Fig{hist_mstar}, respectively.

\section{Variance for larger area surveys}
\label{sec:variance:large}

\begin{figure*}
  \includegraphics[width=8.7cm]{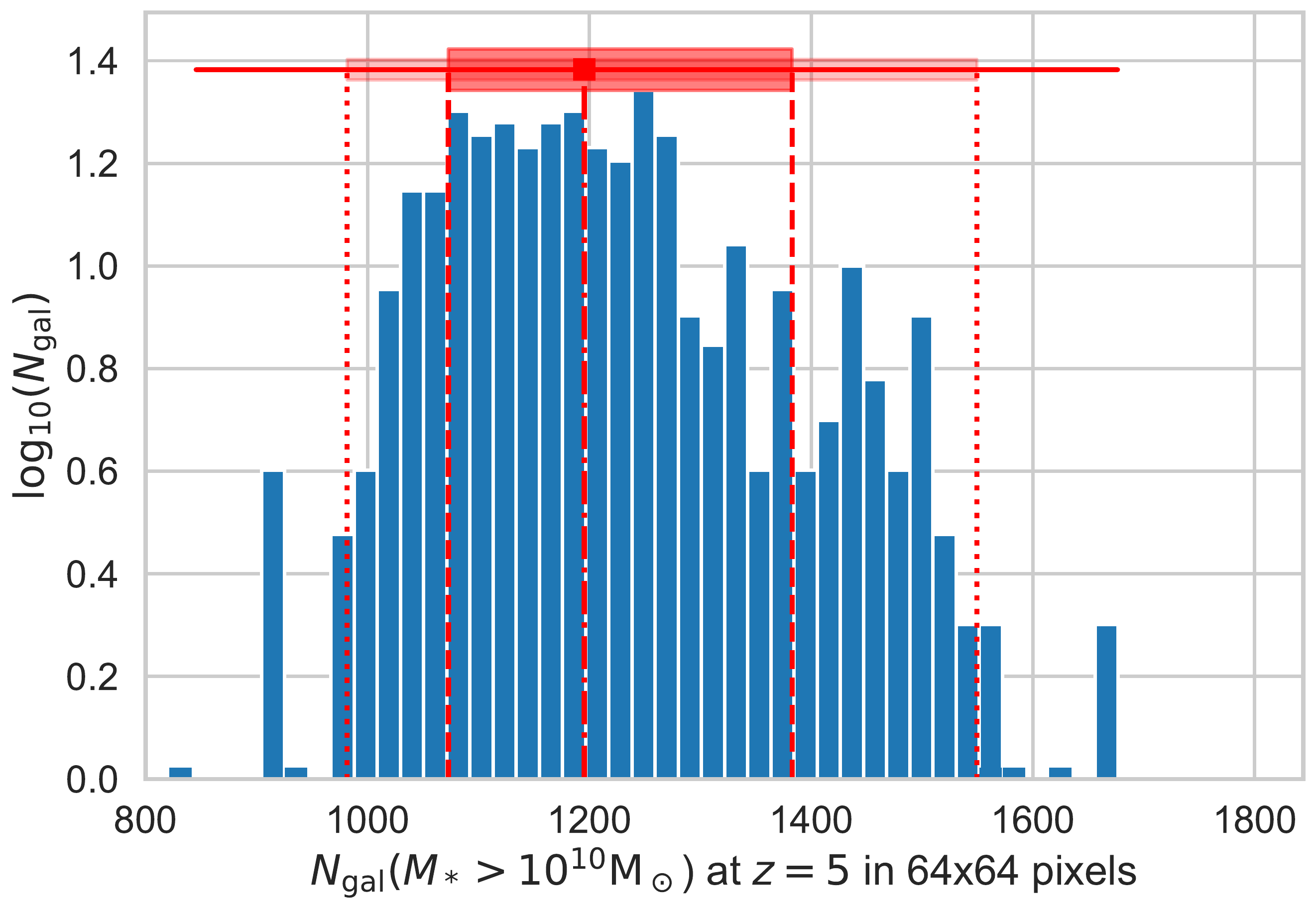}\includegraphics[width=8.7cm]{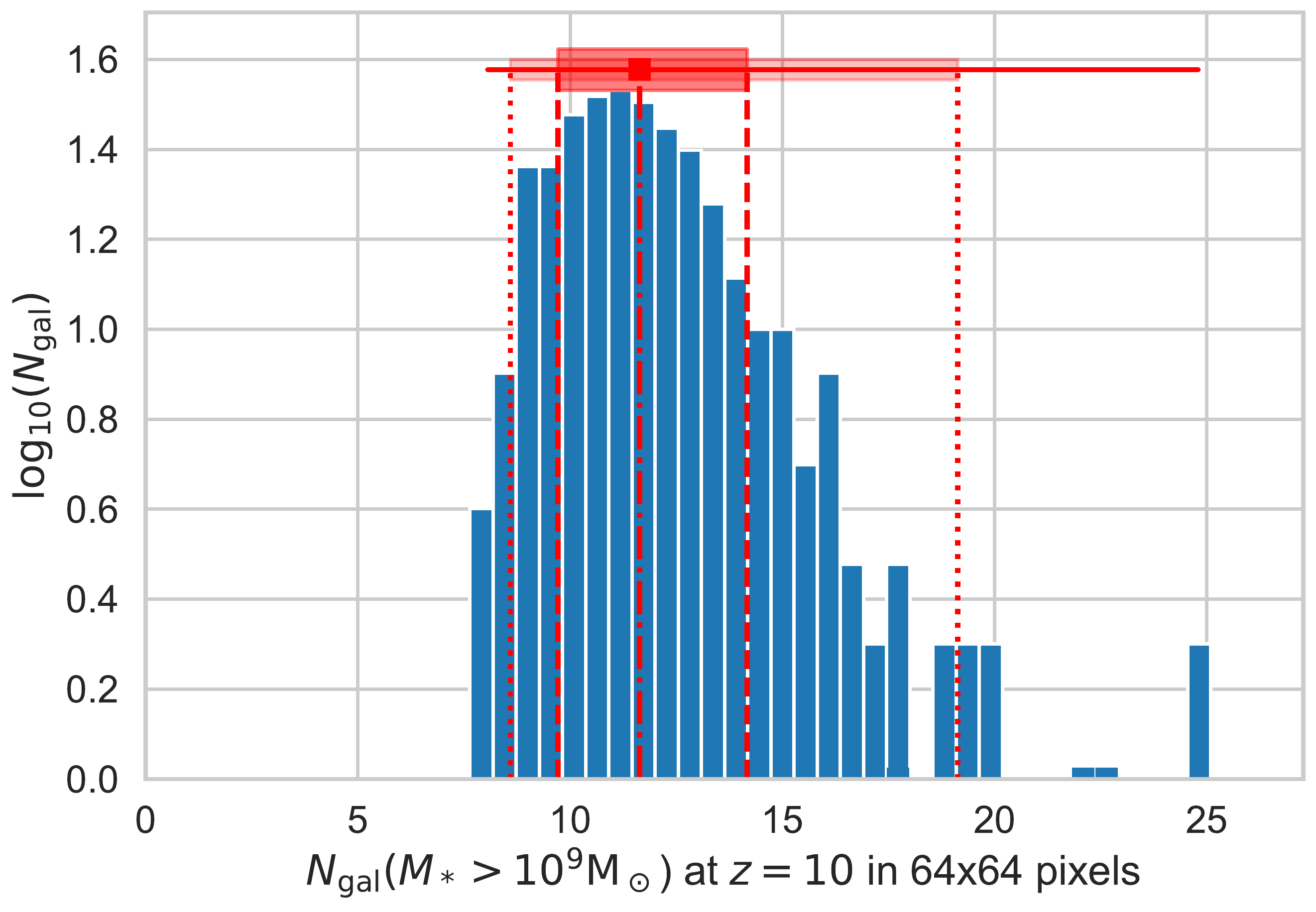}
  \includegraphics[width=8.7cm]{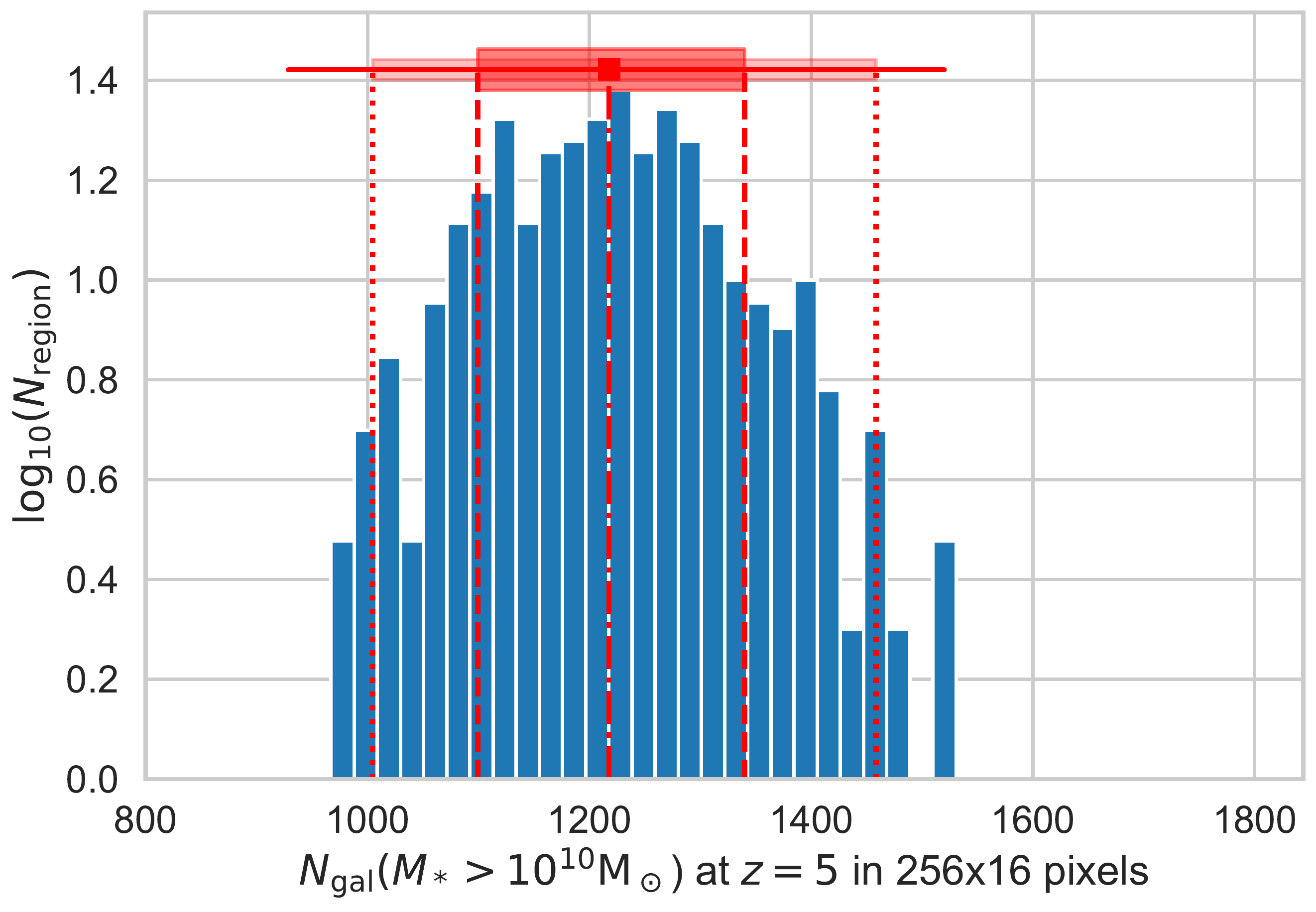}\includegraphics[width=8.7cm]{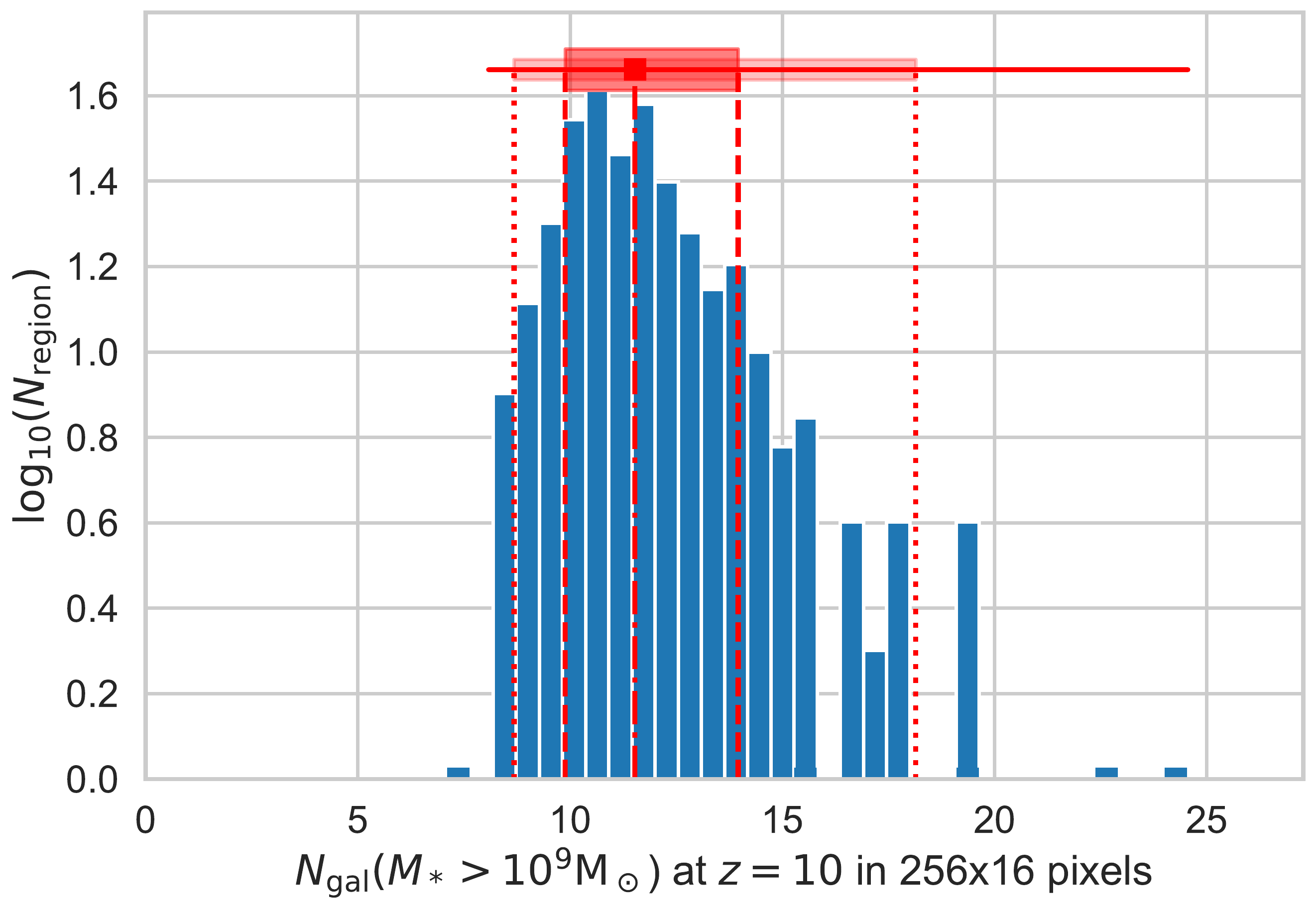}
  \includegraphics[width=8.7cm]{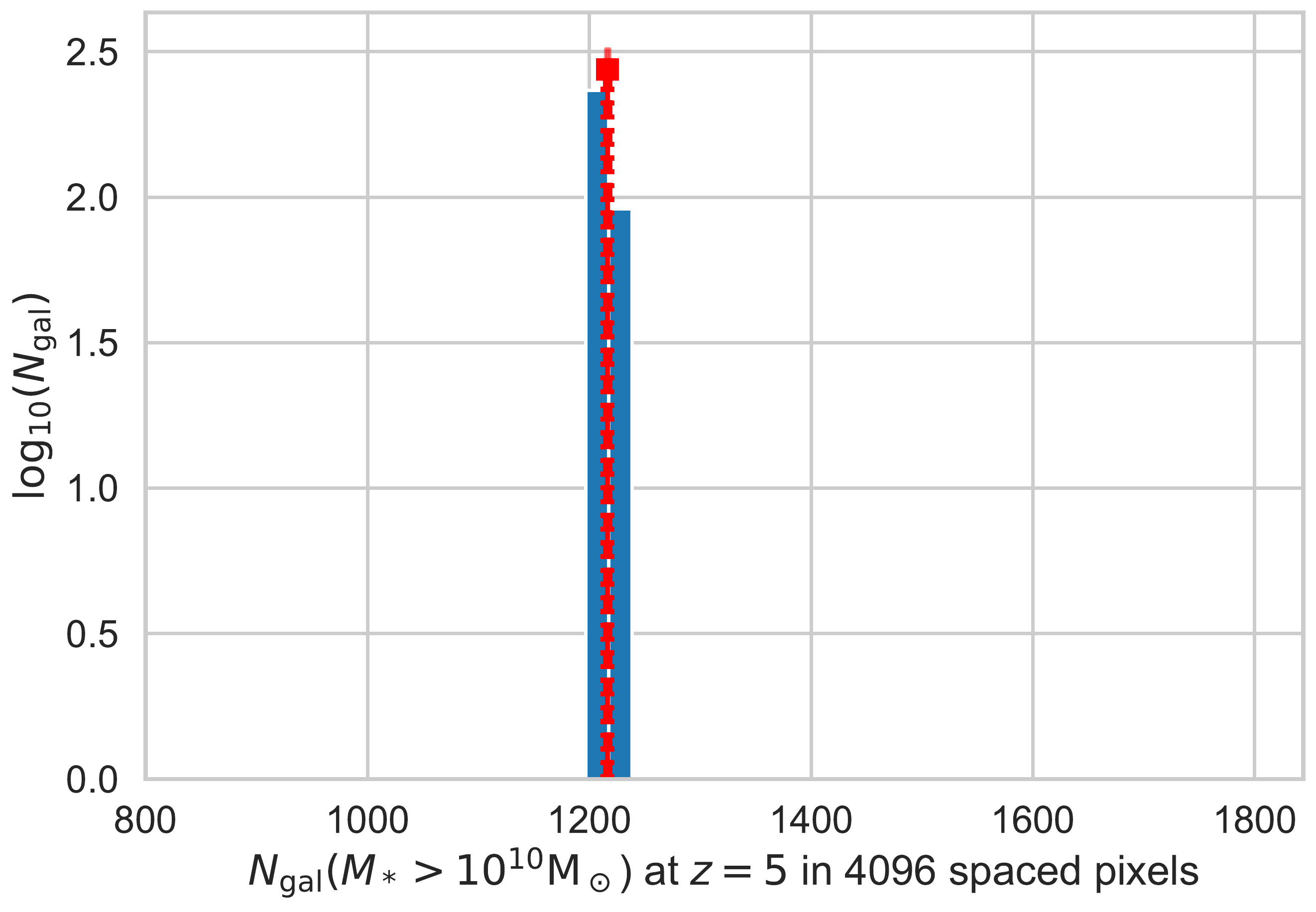}\includegraphics[width=8.7cm]{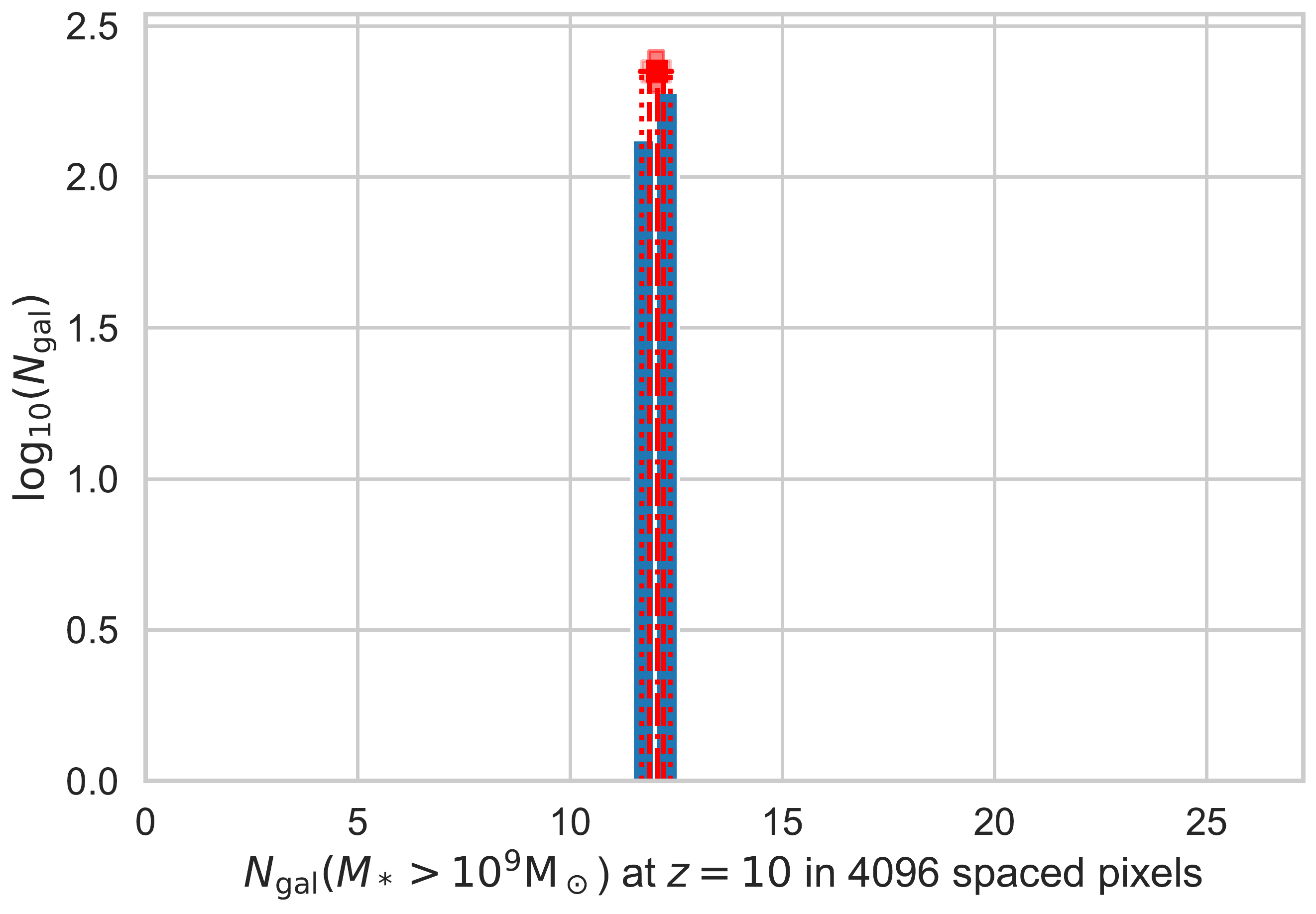}
  \caption{Histograms of the number of galaxies within a 4096 grid cell ($\sim1.4\,$deg$^2$) survey region, above a particular mass and in a given redshift slice, according to the geometry of the survey: left column -- $M_*>10^{10}\,\Msun$, $4.5<z\leq5.5$; right column -- $M_*>10^{9}\,\Msun$, $9.5<z\leq10.5$; upper row -- 64\,x\,64; middle row -- 1024\,x\,4; lower row -- 4096 widely spaced grid cells. The dot-dashed, dashed and dotted lines show the median, one-sigma and two-sigma ranges, respectively; the box-plot shows the full extent of the data, plus the one and two-sigma ranges.}
  \label{fig:hist_mstar_large}
\end{figure*}

\begin{figure*}
  \includegraphics[width=8.7cm]{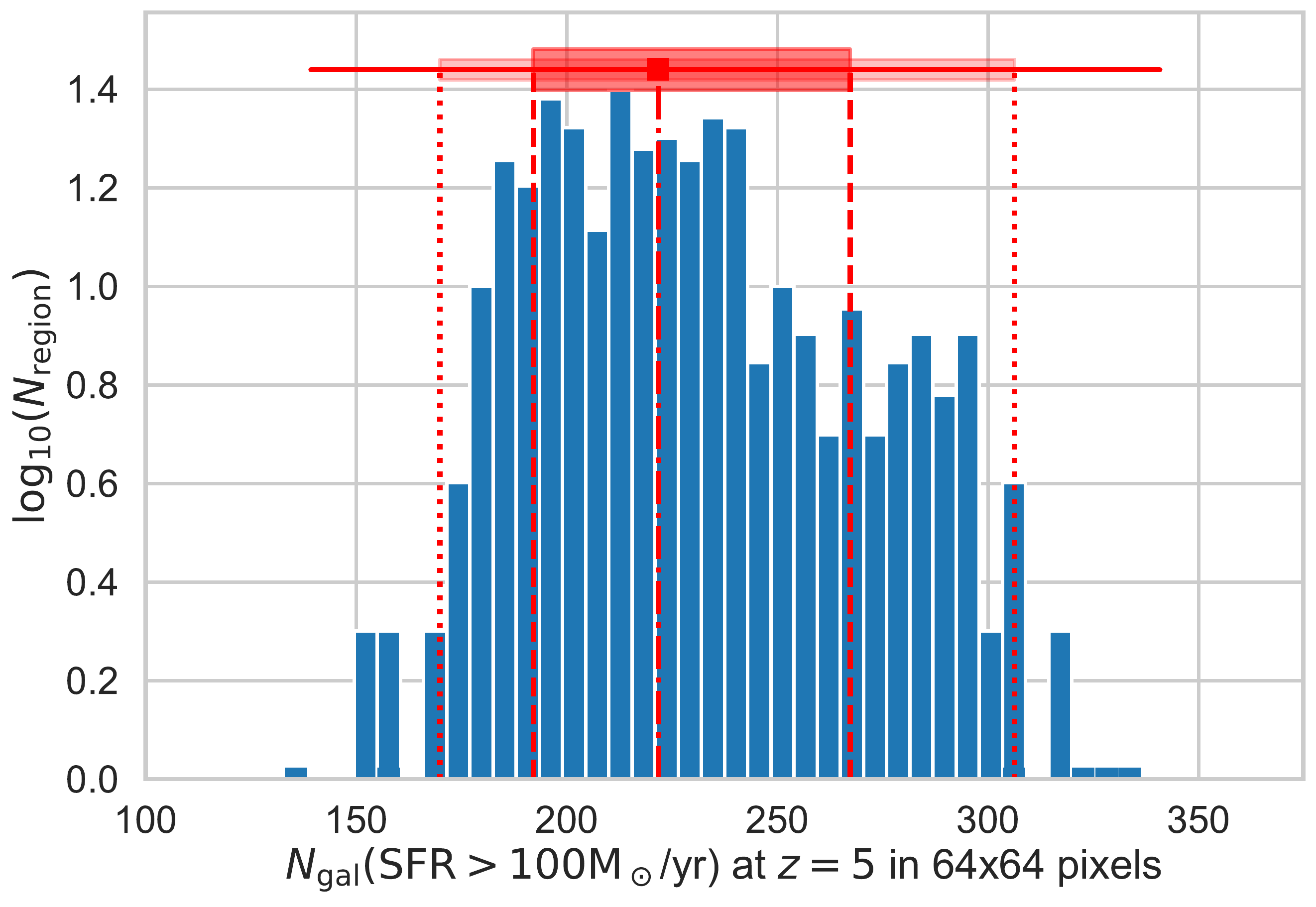}\includegraphics[width=8.7cm]{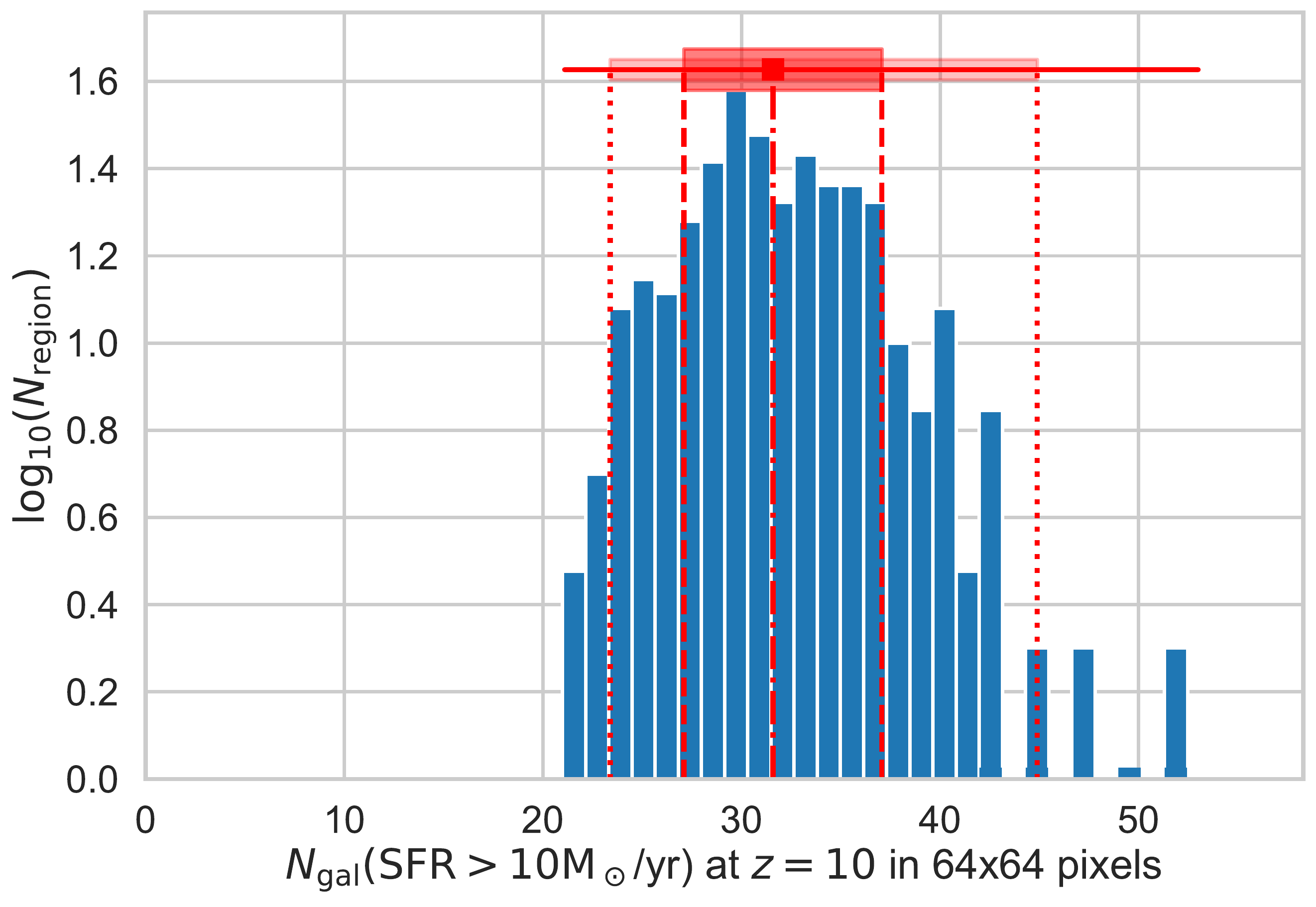}
  \includegraphics[width=8.7cm]{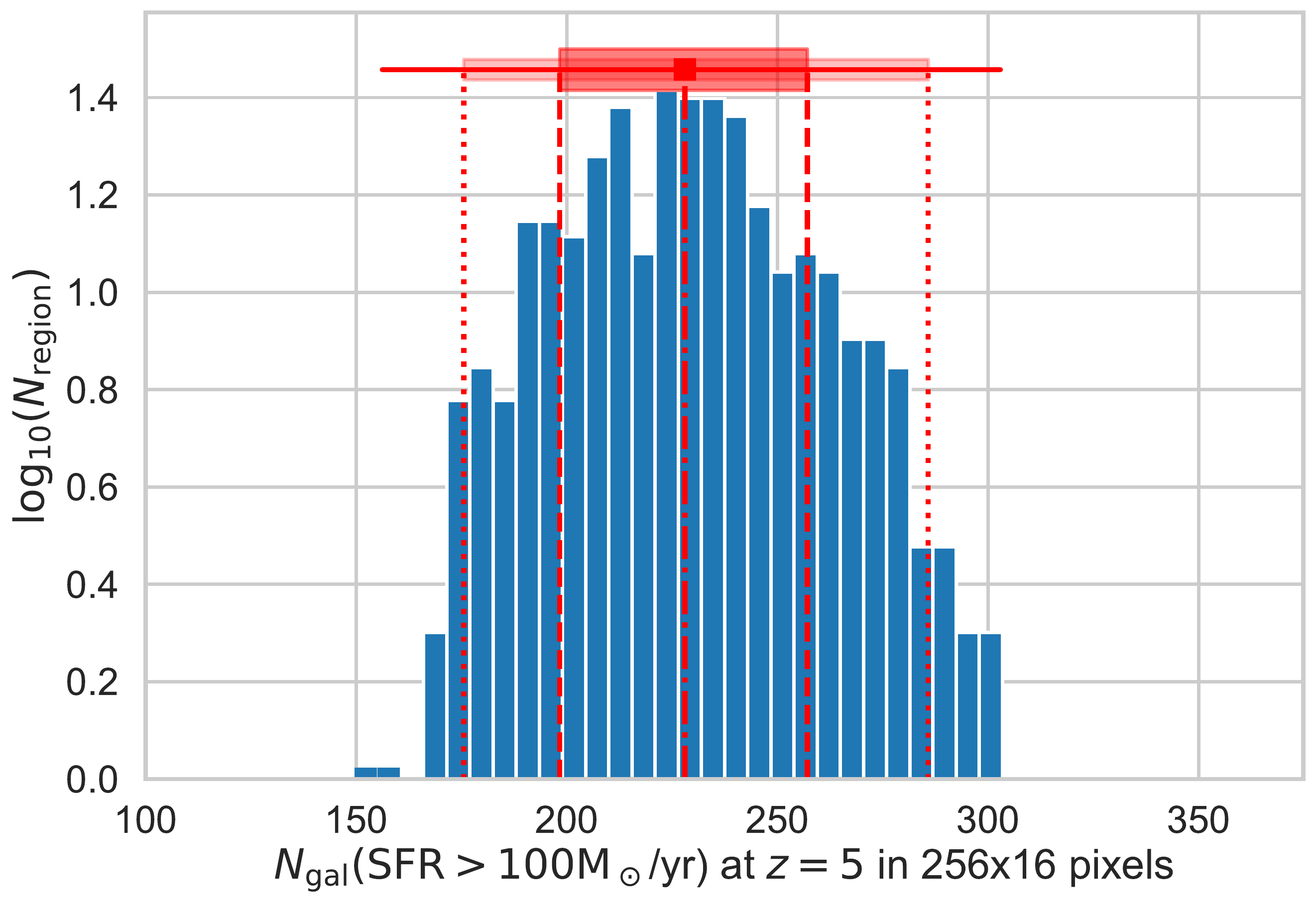}\includegraphics[width=8.7cm]{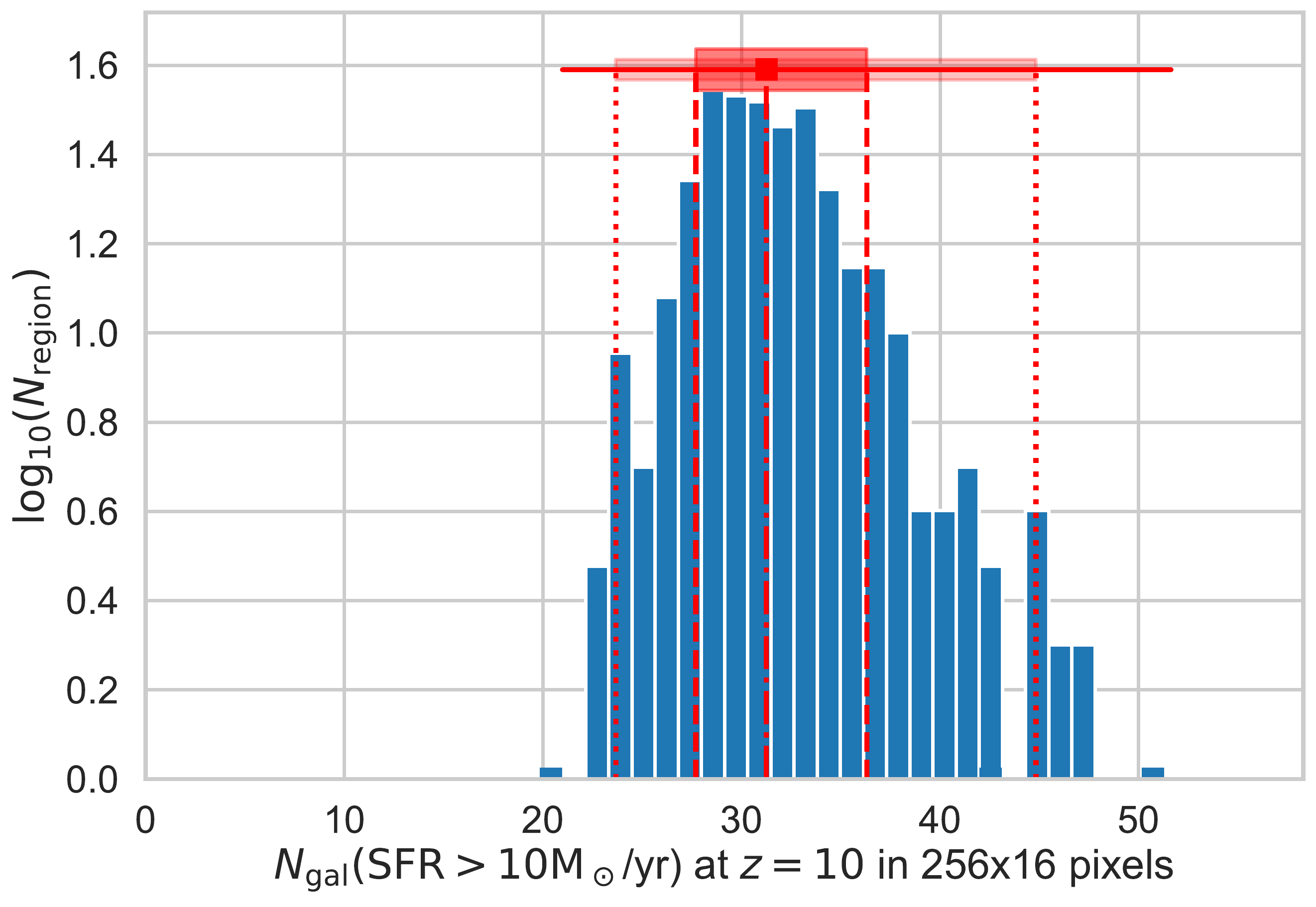}
  \includegraphics[width=8.7cm]{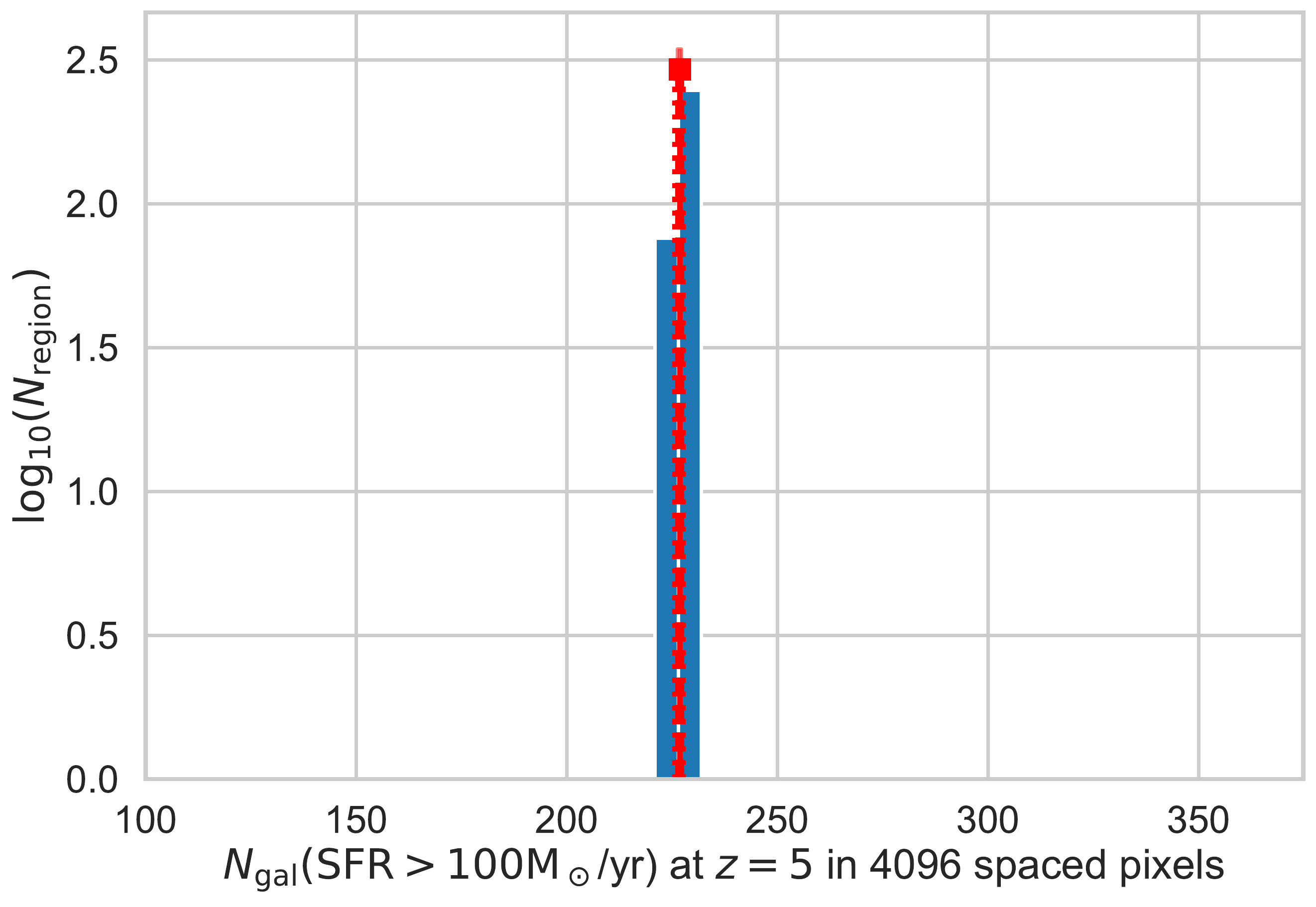}\includegraphics[width=8.7cm]{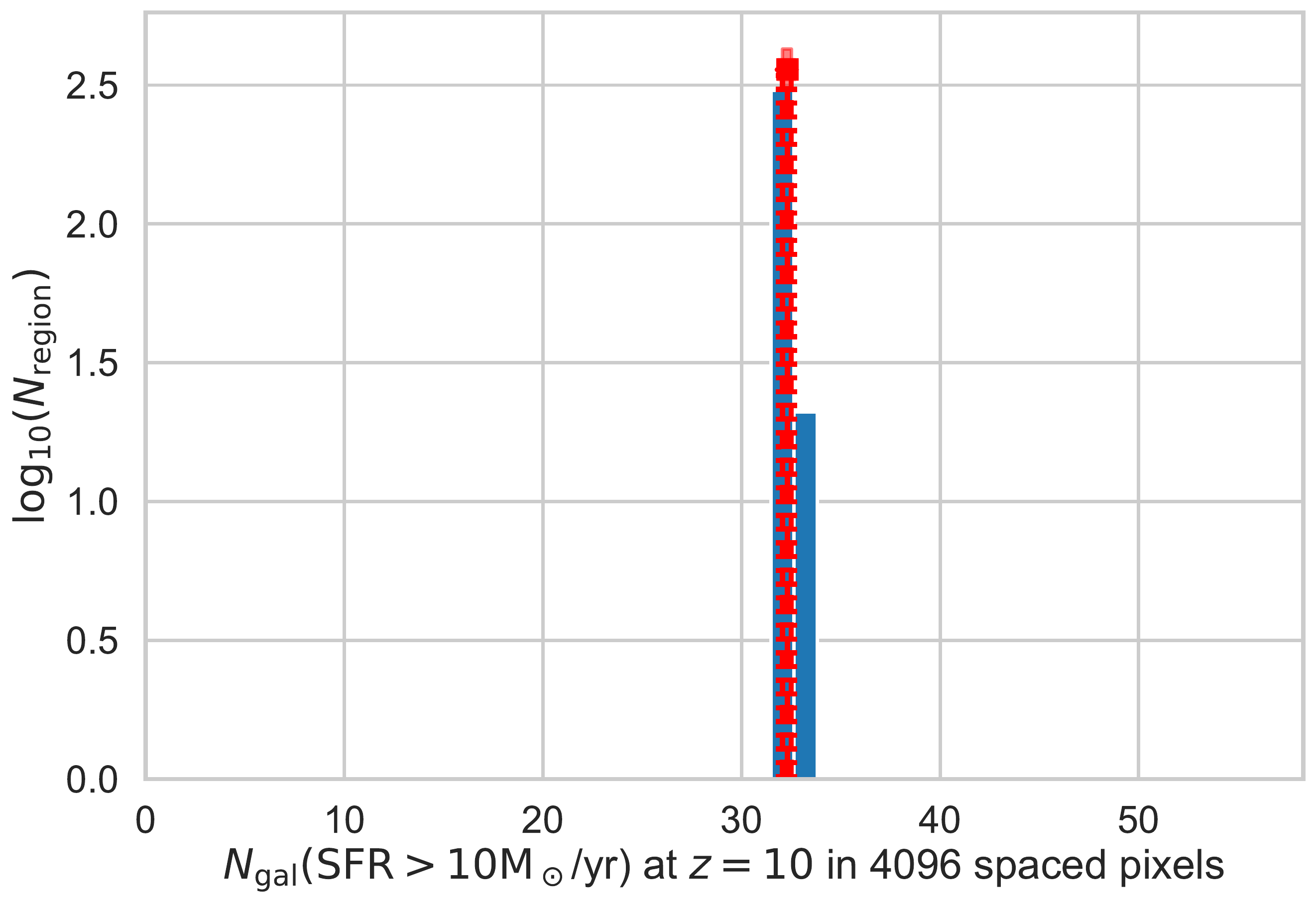}
  \caption{Histograms of the number of galaxies within a 4096 grid cell ($\sim1.4\,$deg$^2$) survey region, above a particular star formation rate and in a given redshift slice, according to the geometry of the survey: left column -- $M_*>100\,\Msun\,$yr$^{-1}$, $4.5<z\leq5.5$; right column -- $M_*>10\,\Msun\,$yr$^{-1}$, $9.5<z\leq10.5$; upper row -- 64\,x\,64; middle row -- 1024\,x\,4; lower row -- 4096 widely spaced grid cells. The dot-dashed, dashed and dotted lines show the median, one-sigma and two-sigma ranges, respectively; the box-plot shows the full extent of the data, plus the one and two-sigma ranges.}
  \label{fig:hist_SFR_large}
\end{figure*}

Figures~\ref{fig:hist_mstar_large} and ~\ref{fig:hist_SFR_large} show histograms of the expected number of galaxies exceeding a particular mass or star formation threshold, respecitvely, for a survey region consisting of 4096 pixels, approximately 1.4\,deg$^2$.


\bsp	
\label{lastpage}
\end{document}